\documentclass[twocolumn,letterpaper,superscriptaddress,prl]{revtex4-1}
\usepackage[usenames]{color}
\usepackage{graphicx}
\usepackage{multirow}
\usepackage{verbatim}
\usepackage{amsmath}
\usepackage{amsfonts}
\usepackage{amssymb}
\usepackage{mathrsfs}
\usepackage{url}
\usepackage{color}      
\usepackage{hyperref}   
\usepackage{caption}
\usepackage{float}
\usepackage{subcaption}
\usepackage{amsmath}
\usepackage{braket}
\begin{document}

\title{Theory of Thermal Relaxation of Electrons in Semiconductors}
\author{Sridhar~Sadasivam}
\email{sadasivam@anl.gov}
\affiliation{Center for Nanoscale Materials, Argonne National Laboratory, Argonne IL 60439, USA}
\author{Maria~K.~Y.~Chan}
\affiliation{Center for Nanoscale Materials, Argonne National Laboratory, Argonne IL 60439, USA}
\author{Pierre~Darancet}
\email{pdarancet@anl.gov}
\affiliation{Center for Nanoscale Materials, Argonne National Laboratory, Argonne IL 60439, USA}


\begin{abstract}
We compute the transient dynamics of phonons in contact with high energy ``hot'' charge carriers in 12 polar and non-polar semiconductors, using a first-principles Boltzmann transport framework. 
For most materials, we find that the decay in electronic temperature departs significantly from a single-exponential model at times ranging from 1 ps to 15 ps after electronic excitation, a phenomenon concomitant with the appearance of non-thermal vibrational modes. 
We demonstrate that these effects result from the slow thermalization within the phonon subsystem, caused by the large heterogeneity in the timescales of electron-phonon and phonon-phonon interactions in these materials. 
We propose a generalized 2-temperature model accounting for the phonon thermalization as a limiting step of electron-phonon thermalization, which captures the full thermal relaxation of hot electrons and holes in semiconductors. 
A direct consequence of our findings is that, for semiconductors, information about the \emph{spectral distribution} of  electron-phonon and phonon-phonon coupling can be extracted from the multi-exponential behavior of the electronic temperature.    
\end{abstract}

\maketitle

Following the seminal works of Kaganov et al.~\cite{kaganov1957relaxation} and Allen~\cite{allen1987theory}, the thermalization of a system of highly energetic charge carriers with a lattice is frequently understood as an electron-phonon mediated, temperature equilibration process with a single characteristic timescale  $\tau_\textrm{el-ph}$. 
Such description, referred to as the two temperature (2T) model, relies on the central assumption that both electrons and phonons remain in distinct thermal equilibria and can therefore be described by time-dependent temperatures $T_\textrm{el}(t)$ and $T_\textrm{ph}(t)$ during the thermal equilibration process. 
In metals, due to the relative homogeneity of the electron-phonon interactions and the  rates of thermalization within the electronic and phononic subsystems, the hypothesis of subsystem-wide thermal equilibrium is generally accurate, and the 2T model has been successful in modeling ultra-fast laser heating~\cite{qiu1992short,tien1993heat,ivanov2003combined}, despite some notable deviations from the 2T predictions  in graphene and aluminum~\cite{vallabhaneni2016reliability,sullivan2017optical,waldecker2016electron}. 
In semiconductors, the highly heterogeneous electron-phonon interactions (e.g. in polar semiconductors with Fr\"{o}hlich interactions~\cite{frohlich1954electrons}) and, in some cases, the higher lattice thermal conductivity in comparison to metals weaken the hypothesis of a thermalized phononic subsystem \cite{yang2017novel,waldecker2017momentum}, hence calling for the reexamination of the 2T physical picture in semiconductors. 
%

In this context, the advent of first-principles techniques able to predict the mode- and energy-resolved electron-phonon~\cite{giustino2017electron,ponce2016epw,verstraete2013ab} and phonon-phonon interactions~\cite{esfarjani2008method,togo2015distributions} provides an important opportunity: In their modern implementations~\cite{ponce2016epw,li2014shengbte,togo2015distributions}, these methods have been able to predict lattice thermal conductivities~\cite{garg2011role,lindsay2013first,romero2015thermal,liao2015significant}, the temperature- and pressure- dependence of the electronic bandgap~\cite{noffsinger2012phonon,antonius2014many,kawai2014electron,monserrat2014electron,giustino2010electron,antonius2016temperature,monserrat2016temperature},  electrical conductivities \cite{park2014electron,liu2017first}, and  hot carrier dynamics~\cite{bernardi2014abinitio,bernardi2015abinitio}. However, to the best of our knowledge and despite these early successes, these approaches have yet to be applied to the computation of electron-induced, non-equilibrium phonon distributions and their effects on thermal relaxation of electrons. 

% ... 

In this work, we combine first-principles calculations of electron-phonon and third-order phonon-phonon interactions within the semi-classical Boltzmann transport equation (BTE) for predicting the joint time-evolution of electron and  phonon populations  after hot carrier excitation. 
For 12 polar and non-polar cubic semiconductors, we show that the resulting phonon and electron dynamics departs qualitatively from the 2T physical picture over timescales of 1-15 ps after excitation. 
We demonstrate that this disagreement stems from the breakdown of the hypothesis of thermal equilibrium within the lattice subsystem, caused by the wide range of timescales associated with electron-phonon and phonon-phonon interactions in these systems. 
We generalize the 2T model of Allen to account for the slow phonon thermalization as a limiting step of electron-phonon thermalization, show that our generalized 2T model captures the transient dynamics for all compounds, and discuss its implication for time-resolved spectroscopy experiments. 
We anticipate our findings to apply to any material with broad spectral distributions of electron-phonon interactions (e.g. polar materials) and weak phonon-phonon interactions (in comparison to bulk metals).

\begin{figure*}[!t]
	\begin{center}
	\includegraphics[width=0.90\textwidth]{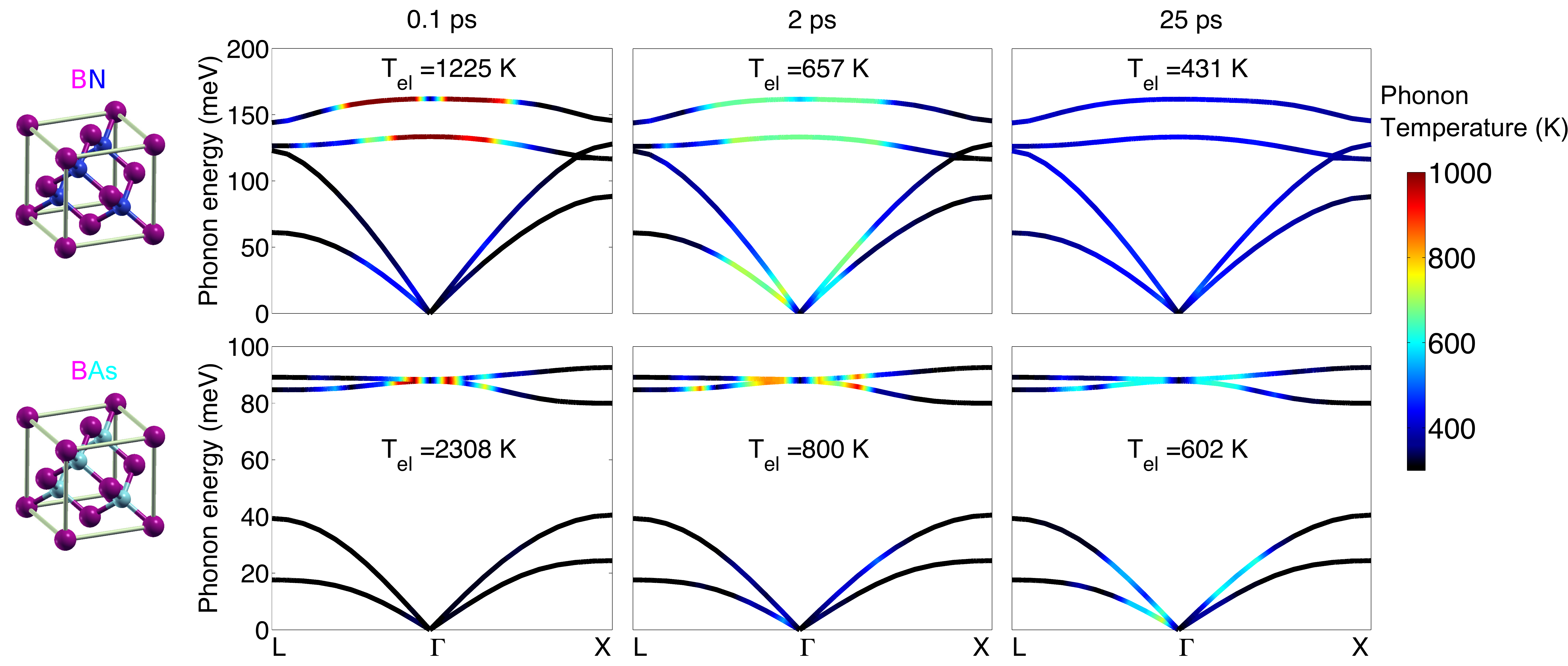}
	\end{center}
    \caption{Temperature maps of phonon modes in cubic boron nitride (BN, top row) and boron arsenide (BAs, bottom row) along the L-$\Gamma$-X directions as a function of time starting from a hot equilibrium electron distribution at 3000 K. In each panel, the phonon modes with the largest temperature are observed to be nearly in equilibrium with electrons (temperature color bar is saturated for T$>$1000K).}
    \label{fig1}
\end{figure*}

We compute the mode- and time-resolved phonon occupation function $n_{\textbf{q},\nu}(t)$  in the presence of an electron occupation function $f_{n\mathbf{k}}(t)$ solving a coupled system of equations parametrized using density functional theory (DFT)-based approaches. The time-evolution of the phonon occupation function is obtained by solving the BTE:  $\frac{d n_{\textbf{q}\nu}(t)}{d t} = \left(\frac{\partial n_{\textbf{q}\nu} (t) }{\partial t}\right)_{ep} \left[n_{\textbf{q}\nu}(t) , f_{n\mathbf{k}} (t)\right]+\left(\frac{\partial n_{\textbf{q}\nu} (t)}{\partial t}\right)_{pp}\left[ n_{\textbf{q}\nu}(t) \right]$, where the drift term has been neglected due to the lack of spatial temperature gradient, and $\left[ \dots \right]$ indicates the functional dependence. The two terms on the right denote the time-dependent scattering potentials due to electron-phonon (EPI) and phonon-phonon interactions (PPI), both computed using first-principles methods, as detailed below.  Importantly, we make the assumption that charge carriers are in thermal equilibrium and that $ f_{n\mathbf{k}} (t)$ can be approximated by a time-dependent Fermi-Dirac function centered near the top of the valence band for holes and near the bottom of the conduction band for electrons at the temperature $T_\textrm{el}(t)$. Depending on the material and the nature of charge carriers, the timescale of the phonon-mediated carrier thermalization to the band edges was found to range from 0.1 to 1 ps~\cite{bernardi2014abinitio,jhalani2017asymmetry} which also corresponds to limits of validity of the semi-classical description. Hence, we expect our simulation method and the approximation of $ f_{n\mathbf{k}} (t)$ to be quantitative at subsequent times. 

Specifically, we define the EPI scattering potential as an explicit functional of the phonon \emph{and} electron occupation functions at time $t$, and compute it using Fermi's golden rule: $\left(\frac{\partial n_{\textbf{q}\nu} (t) }{\partial t}\right)_{ep} = \frac{4\pi}{\hbar}\sum\limits_{\textbf{k},m,n} |g_{\textbf{q}\nu}(m\textbf{k}+\textbf{q},n\textbf{k})|^2 \mathcal{M}_{mn\nu\textbf{k}\textbf{q}}(t)$, in which  $|g_{\textbf{q}\nu}(m\textbf{k}+\textbf{q},n\textbf{k})|$ is the time-independent electron-phonon matrix elements involving electronic states $\vert n\mathbf{k}\rangle$ and $\vert m\mathbf{k}+\textbf{q}\rangle$ and vibrational state $\vert\textbf{q}\nu\rangle$ evaluated using Wannier interpolation with the EPW code~\cite{ponce2016epw}. $\mathcal{M}_{mn\nu\textbf{k}\textbf{q}}(t)$ is the time-dependent joint density of states computed from $n_{\textbf{q}\nu}(t) $, $f_{n\mathbf{k}} (t)$, $f_{m\mathbf{k}+\textbf{q}} (t)$, and the electron and phonon spectral densities (detailed formulas are given in Supplemental Material).  Similarly, we evaluate the scattering caused by PPI $\left(\frac{\partial n_{\textbf{q}\nu} (t) }{\partial t}\right)_{pp}$ from Fermi's golden rule, using the time-independent 3-phonon scattering matrix elements $|\Psi_{\textbf{q}\textbf{q}'\textbf{q}\pm\textbf{q}'+\textbf{G}}^{\nu\nu'\nu''}|^2$ computed with DFT~\cite{esfarjani2008method} and the time-dependent density of final states computed from $n_{\textbf{q},\nu}(t)$, $n_{\textbf{q}',\nu'}(t)$, $n_{\textbf{q}\pm\textbf{q}',\nu''}(t)$. At each time step, the net energy transfer $Q_{ep}$ between electrons and phonons is computed and a new electronic temperature is derived as $T_\textrm{el}(t+\Delta t) = T_\textrm{el}(t)-Q_{ep}(t)/C_\textrm{el}(T_\textrm{el})$ where $C_\textrm{el}(T_\textrm{el})$ is the instantaneous electronic heat capacity at temperature $T_\textrm{el}$. The BTE is solved for 48000 phonon modes using an explicit time-stepping scheme with a time-step of 0.5 fs and a total simulation time of 25 ps for 12 cubic semiconducting compounds (BN, BP, BAs, BSb,  AlP, AlAs, AlSb, GaN, GaP, GaAs, diamond, Si). All the simulations discussed below were initialized with an equilibrium phonon distribution at 300 K and a Fermi-Dirac distribution of electrons at 3000 K with the Fermi level set at 0.3 eV below the valence band maximum (other choices of initial temperatures and Fermi energies are shown to lead to similar conclusions in Supplemental Material). 

The electronic structure was computed with DFT in the local density approximation, using norm-conserving pseudopotentials, a 10$\times$10$\times$10 $\mathbf{k}$-grid and the Quantum Espresso package~\cite{QE-2009}. The phonon dispersion was computed using density functional perturbation theory~\cite{baroni2001phonons} and a 5$\times$5$\times$5 $\mathbf{q}$-grid. Third-order force constants were computed in real space using finite differences on a  6$\times$6$\times$6 supercell~\footnote{Decay of real-space third-order force constants for all 12 semiconductors is provided in Supplemental Information} and Fourier transformed to obtain phonon-phonon interaction matrix elements $|\Psi_{\textbf{q}\textbf{q}'\textbf{q}\pm\textbf{q}'+\textbf{G}}^{\nu\nu'\nu''}|^2$ on a 20$\times$20$\times$20 $\textbf{q}-$grid~\cite{lindsay2012thermal}. The present approach neglects the temperature dependence of the third-order force constants~\cite{hellman2011lattice,hellman2013temperature}, as this simplification has been shown to accurately predict the temperature-dependent lattice thermal conductivity for cubic semiconductors~\cite{lindsay2013first,zhou2014lattice,lindsay2012thermal,luo2013gallium}. Electron-phonon interactions were evaluated on  20$\times$20$\times$20 and 40$\times$40$\times$40 grids for phonons and electrons, respectively.  Convergence studies  are provided in the Supplemental Material.

In Fig.~\ref{fig1}, we show the time-dependent phonon occupations along high symmetry directions of the Brillouin zone for BN and BAs (snapshots for all materials can be found in  Supplemental Material). 
At short times $t<1$ ps, the electronic energy is transferred to long-wavelength optical phonons, an effect originating from the larger electron-phonon scattering phase space associated with low-momentum phonons near the top (bottom) of the valence (conduction) bands in all 12 compounds, and further magnified by the $1/\textbf{q}$ divergence in the Fr\"{o}hlich coupling in polar semiconductors~\cite{frohlich1954electrons,verdi2015frohlich}:  Accordingly, we  observe that more energy is transferred to the LO and TO modes of BN than to the modes of BAs, as expected from the Born effective charges (1.86 for BN vs 0.56 for BAs) and polarity, which also lead to a larger electron-phonon coupling and LO-TO splitting.  Surprisingly, these ``hot'' phonon modes are found for all compounds to achieve near-thermal equilibrium with the \emph{electrons} rather than with the rest of the phonons, a strong departure from the hypothesis of local thermal equilibrium within the lattice.
At longer times, $1<t<10$ ps, long wavelength LO and TO phonons in BN remain in near-thermal equilibrium with electrons, while transferring their energy to acoustic modes via 3-phonon processes, through Klemens (decay to two acoustic phonons)~\cite{klemens1966anharmonic} and Ridley (decay to one optical \& one acoustic  phonons) mechanisms~\cite{ridley1996phonon}. The ``hot phonon'' cooling in BAs is slower in comparison to BN as the large acoustic-optical phonon band gap~\cite{lindsay2013first,ma2016boron} (originating from the mass mismatch)  truncates the Klemens scattering phase space, while the Ridley decay is reduced by the small LO-TO splitting. 
Near-thermalization within the phonon subsystem (and, concomitantly, between electrons and phonons) is achieved in BN at  $t\simeq25$ ps, with an electronic temperature 50 K away from the average lattice temperature $T_{\textrm{ph}}=380$ K. In stark contrast to BN, for BAs both electrons (602 K) and hot phonons remain in near equilibrium with each other, but far from the average lattice temperature $T_{\textrm{ph}}=344$ K.
Importantly, the same two trends are observed for all simulated materials: (1) electrons first achieve near thermal equilibrium with a small number of high energy phonon modes; (2) full electron-lattice thermalization and intra-phonon thermalization are always achieved \emph{simultaneously}; both trends implying that electron cooling is limited by thermalization within the phonon-subsystem.

\begin{figure}[t]
	\begin{center}
    \includegraphics[width=0.45\textwidth]{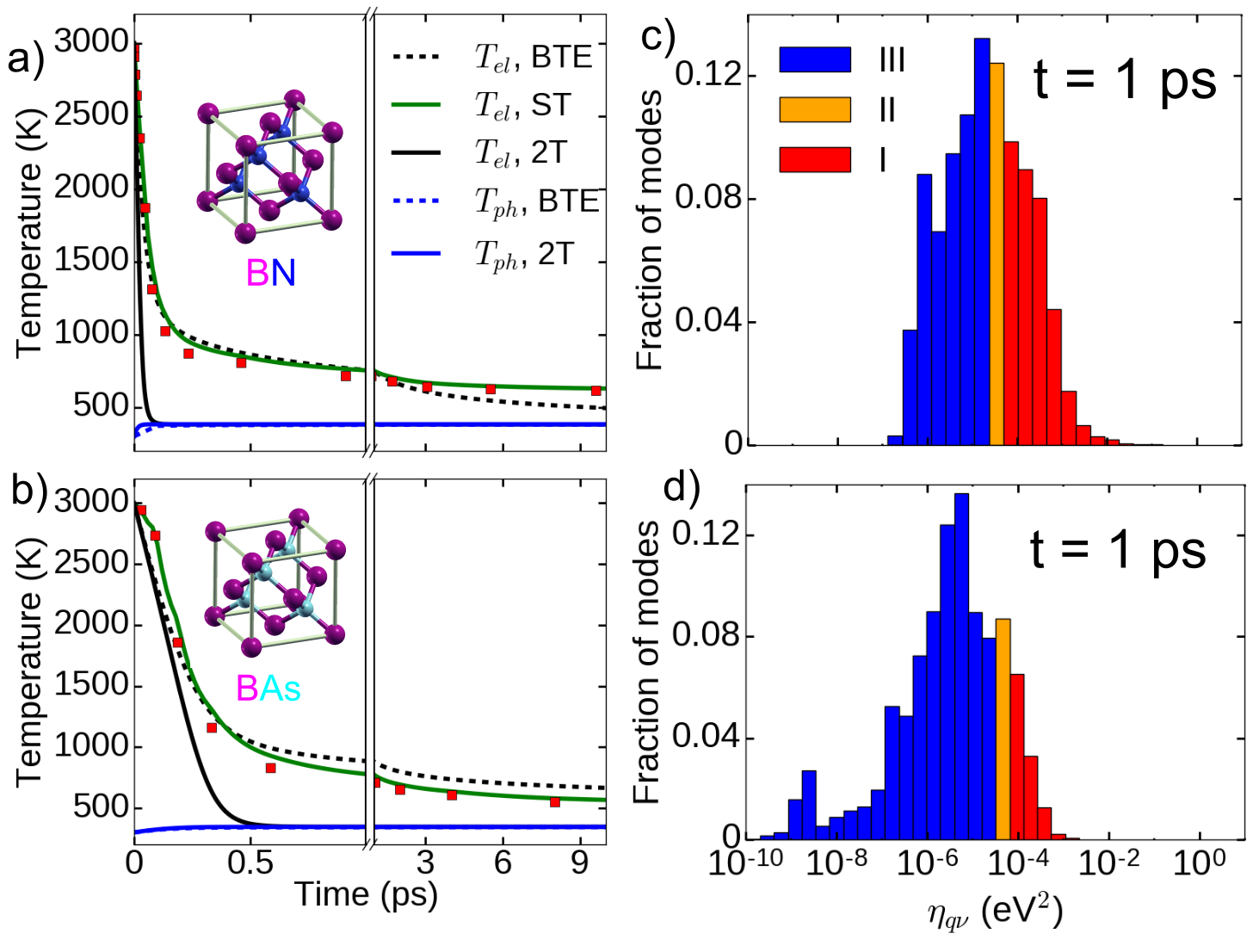}
  \end{center}
    \caption{Electronic and lattice temperatures in BN (a) and BAs (b) obtained from the 2T model, BTE and a constrained ``successive thermalization'' (ST) simulation using the 2T model on a subset of phonons (see main text). The red squares indicate the times (and corresponding equilibration temperatures) at which subspace thermalization is achieved and a new set of modes is introduced in the ST simulation. Histograms of the distribution of interaction strength $\eta_{\textbf{q}\nu}$ (Eq. \ref{Eq:Eta}) for BN (c) and BAs (d), showing the partition scheme and the phonons included in the ST simulation at $t=1$ ps. Phonon modes I are the phonons thermalized with electrons, II are the phonons undergoing thermalization, and III are the phonons non-interacting with electrons.}
    \label{fig2}
\end{figure}


Further illustrating the non-equilibrium between phonon modes, we see in Fig. \ref{fig2} that the agreement between the BTE simulation and a 2T model parametrized from first-principles (see Supplemental Material for details) is good at times $t<0.05$ ps and $t<0.2$ ps for BN and BAs, respectively, but quickly deteriorates afterwards. For all compounds, the 2T model predicts a thermalization that is at least an order of magnitude faster than that observed in the full BTE simulation, proving that electronic cooling becomes limited by another mechanism,  not accounted for in the  2T model~\footnote{As shown in Supplemental Material, this overly fast cooling is \emph{not} corrected by simple higher level descriptions such as the 3-temperature model proposed by Waldecker et al.~\cite{waldecker2016electron} in which the phonon branches are sub-divided into two categories (for example, optical and acoustic), depending on their coupling to electrons.}.

To test our hypothesis of a phonon-thermalization limited process, we perform a constrained simulation of electron cooling in which the phonons are partitioned into multiple subspaces defined by the strength  $\eta_{\textbf{q}\nu}$ of their interactions with electrons and phonons:  
\begin{equation}
\begin{split}
\eta_{\textbf{q}\nu} &= \hbar\omega_{\textbf{q}\nu}\Big\{\sum\limits_{\textbf{k},m,n} |g_{\textbf{q}\nu}(m\textbf{k}+\textbf{q},n\textbf{k})|^2 \delta_{\textbf{q}\nu,mn\textbf{k}} \\ &
+ \sum\limits_{\textbf{q}'\nu',\textbf{q}''\nu''}^{ \eta_{\textbf{q}'\nu'} > \eta_{\textbf{q}\nu}} |\Psi_{\textbf{q}\textbf{q}'\textbf{q}\pm\textbf{q}'+\textbf{G}}^{\nu\nu'\nu''}|^2 \delta_{\textbf{q}\nu,\textbf{q}'\nu',\textbf{q}''\nu''}\Big\},
\end{split}
\label{Eq:Eta}
\end{equation}
where the terms on the right approximate the scattering due to EPI and PPI for each mode $\vert\textbf{q}\nu\rangle$ (the PPI term only includes modes with a larger interaction strength and is computed self-consistently). $\delta_{\textbf{q}\nu,mn\textbf{k}}$ and $\delta_{\textbf{q}\nu,\textbf{q}'\nu',\textbf{q}''\nu''}$ are energy conservation delta functions for electron-phonon and phonon-phonon scattering respectively. The phonon modes in the largest $\eta_{\textbf{q}\nu}$ subspace~\footnote{Phonon modes with $\eta_{\textbf{q}\nu}>c\eta_{\textbf{q}\nu,max}$ are chosen to belong to a subspace and the predictions for $c=0.5$ are presented in the main text. Results for $c=0.1$ are reported in Supplemental Material with similar temperature decay predictions.} are a small subset (see Fig.~\ref{fig2} c,d) of the total number of phonons and primarily consists of long-wavelength optical phonons with strong electron-phonon interaction~\footnote{See Supplemental Material (which includes Refs.~\cite{lloyd2015lattice,nava1980electron,PhysRevB.72.014306}) for distributions of $\eta_{\textbf{q}\nu}$ for all 12 semiconductors considered in this study along with information on the average momentum of phonons within each subset.}. 
At time $t=0$,  only the modes belonging to the subspace with the largest $\eta_{\textbf{q}\nu}$ are allowed to interact with electrons until thermalization. Upon thermalization of the first subspace, the next subspace is introduced in the simulation along with the thermalized system of electrons and the first subspace. This constrained, ``successive thermalization'' (ST) process is continued until all modes are included~\footnote{The effective coupling coefficient between the interacting systems at each thermalization step is chosen to be proportional to the sum of all interaction strengths $\eta_{\textbf{q}\nu}$ of modes belonging to the subspace undergoing thermalization. We note that the proportionality constant is chosen to be the \textit{same for every subspace} and is \textit{independent of temperature}. }. As shown in Fig.~\ref{fig2} and Supplemental Material, this constrained ST simulation achieves quantitative agreement at all times for all materials considered, validating our central finding: \emph{Electron cooling in semiconductors is limited by intra-phonon thermalization}, a direct consequence of the order-of-magnitude heterogeneities in the mode-dependent electron-phonon interactions and slow phonon thermalization. 


\begin{figure}[t]
	\begin{center}
    \includegraphics[width=0.45\textwidth]{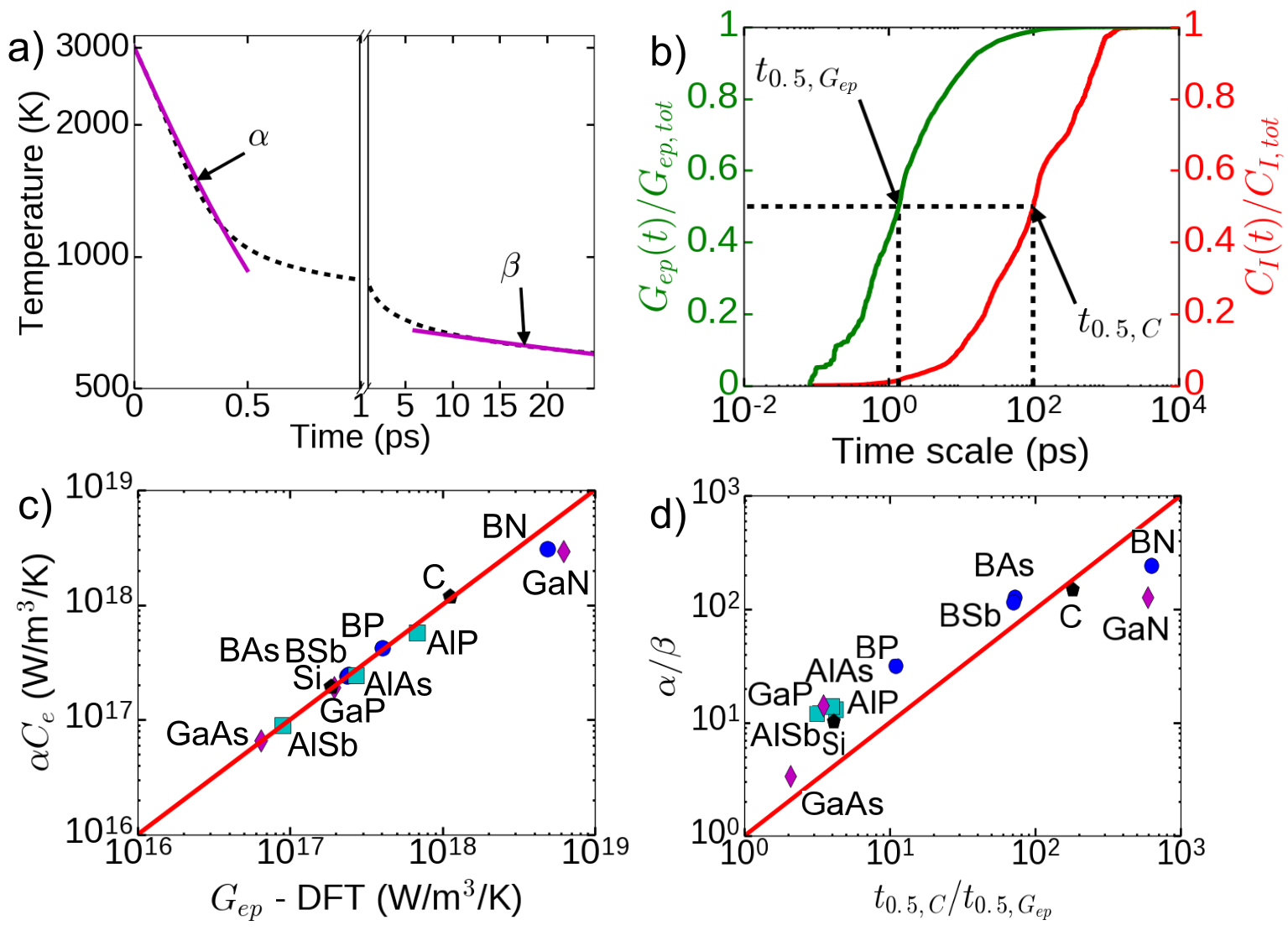}
    	\end{center}
    \caption{a) Electronic temperature decay in BAs along with the decay rates at short ($\alpha$) and long ($\beta$) time instants. b) Accumulation of lattice heat capacity $C_{\textrm{I}}(t)$ at $T_{\textrm{ph}}=300$ K and electron-phonon coupling coefficient $G_{ep}(t)$ at $T_{\textrm{el}}=3000$ K, $T_{\textrm{ph}}=300$ K as a function of phonon thermalization time-scale in BAs. c) Comparison of electron-phonon coupling coefficient obtained from the decay rate of electronic temperature at $t \rightarrow 0$ and directly from DFT for all the 12 semiconductors considered in this work. d) Comparison of the ratio between initial and long-time decay rates with the ratio of time-scales of heat capacity and electron-phonon coupling accumulation for all compounds considered in this work.}
    \label{fig3}
\end{figure}
We conclude this work by proposing a simple generalization of the 2T model based on these findings, and discuss its consequences in interpreting materials properties measured by the time-resolved decay of the electronic temperature~\cite{guo2012heat,wang2012limits}. We start by partitioning the system in an analogous way to our ST simulation, \textit{i.e.}, with 3 subsystems well described by a subsystem-wide temperature: (I) - a system containing electrons and phonons fully thermalized with each other (by definition, $T_{\textrm{I}} (t)=T_{\textrm{el}}(t) $), (II) - phonon modes in contact with electrons and/or phonons of system (I) (in the process of being thermalized) at temperature $ T_{\textrm{ph}}(t = 0) < T_{\textrm{II}} (t) < T_{\textrm{el}}(t) $, and (III) - ``cold'' phonons not in contact with (I) ($T_{\textrm{III}}  (t) = T_{\textrm{ph}}(t = 0)$). In this generalized partition scheme, the  2T model is strictly recovered by setting $(\textrm{I}) = \{\textrm{el}\}$, $(\textrm{II}) = \{\textrm{ph}\}$ and $(\textrm{III}) = \emptyset$. As more modes become thermalized with electrons as a function of time, the long-time electron-phonon thermalization can be understood as system (I) absorbing systems (II) and (III). Hence, the heat capacity of (I) becomes time-dependent with $C_{\textrm{I}}(t)$ increasing from $C_{\textrm{I}}(t=0) = C_{\textrm{el}}$ to $C_{\textrm{I}}(t\rightarrow\infty) = C_{\textrm{el}}+C_{\textrm{ph}}$ (temperature dependences were omitted for simplicity of notation). Such time-dependent heat capacity $C_{\textrm{I}}(t)$ can be understood as \emph{an accumulation function} of the phonons over the timescales of their interactions. $C_{\textrm{I}}(t)$ can be computed heuristically by defining an effective mode-dependent thermalization time $t_{\textbf{q}\nu}$  (that we set to the relaxation time) and $C_{\textrm{I}}(t) = \sum\limits_{\textbf{q}\nu} C_{\textbf{q}\nu}\Theta(t-t_{\textbf{q}\nu})$ where $\Theta(t)$ is the Heaviside function. Similarly the electron-phonon coupling accumulation can be defined as $G_{ep}(t) = \sum\limits_{\textbf{q}\nu} G_{ep,\textbf{q}\nu}\Theta(t-t_{\textbf{q}\nu})$ (see  Supplemental Material for definitions of $t_{\textbf{q}\nu}$, $C_{\textbf{q}\nu}$, $G_{ep,\textbf{q}\nu}$ and their values for all compounds). Noteworthily, for materials with large heterogeneities in their mode-dependent electron-phonon coupling rates $G_{ep,\textbf{q}\nu}$, these two accumulation functions have very different time-dependences: as seen in Fig.~\ref{fig3} b,d),  $G_{ep}(t) $ reaches 50\% of its total value 1-1000 times faster than $C_{\textrm{I}}(t)$. 

At short times (comparable to the time of accumulation of $G_{ep}(t) $), the observed electronic temperature decay rate given by this generalized 2T model can be approximated by $ G_{\textrm{I-II}} / C_{\textrm{I}}  \simeq G_{ep}(t \rightarrow \infty)/C_{\textrm{el}}$, \textit{i.e.}, the decay rate predicted by a ``standard'' 2T model. Correspondingly, in Fig.~\ref{fig3} c), we observe an excellent correlation between the initial decay rate and the electron-phonon coupling strength predicted directly from first-principles for all compounds considered in this work -- indicating that, at short time, the determination of the single-exponential decay of the electronic temperature yields the total electron-phonon coupling. At longer times, the decay rate of the electronic  temperature $ G_{\textrm{I-II}} / C_{\textrm{I}} $ is reduced by the accumulation of heat capacity in I, as $C_{\textrm{I}}>> C_{\textrm{el}}$. As shown  in Fig.~\ref{fig3} d), the reduction of the decay rate for all compounds shows a good correlation with the disparity of timescales between $C_{\textrm{I}}(t)$ and  $G_{ep}(t)$, suggesting that a measurement of the electronic temperature decay across timescales in semiconductors would yield \emph{both} the total electron-phonon coupling coefficient and information about the distribution of phonon interaction strength (and its heterogeneity) in a given material. Interestingly, as the phonon interaction strength involves both EPI and PPI (see Eq.~(\ref{Eq:Eta})), the time-dependence of the decay rates is particularly important for materials with very heterogeneous EPI (Diamond, BN, GaN), and large phonon-bandgaps (BAs, BSb), and vanishes for nearly homogeneous EPI (e.g. GaAs~\cite{bernardi2015abinitio}). 

In conclusion, we have demonstrated that electron cooling in semiconductors is limited by intra-phonon thermalization at timescales on the order of $1-20$ ps. We have proposed a generalized 2-Temperature model accounting for this effect, and shown that such a model can be used  to extract information from the measurement of the electronic temperature about both the total electron-phonon coupling and the distribution of electron-phonon and phonon-phonon interactions. More generally, we expect the phonon-limited thermalization identified in this work to have consequences on both heat and electron transport, fields in which long-lasting non-equilibrium phonon distributions have been shown to impact spectroscopic measurements \cite{vallabhaneni2016reliability,sullivan2017optical}, current-voltage characteristics \cite{steiner2009phonon}, and hot electron lifetimes \cite{conibeer2015hot,yang2016observation}. Specifically, our work offers a direct estimate of the timescales at which equilibrium models become quantitative in the presence of hot electrons (and their relationship to materials properties), and, via the tunability of the phonon-interaction strength, new pathways to control the timescales of electronic energy dissipation.

Use of the Center for Nanoscale Materials, an Office of Science user facility, was supported by the U. S. Department of Energy, Office of Science, Office of Basic Energy Sciences, under Contract No. DE-AC02-06CH11357. This material is based upon work supported by Laboratory Directed Research and Development (LDRD) funding from Argonne National Laboratory. We gratefully acknowledge the computing resources provided by the Laboratory Computing Resource Center at Argonne National Laboratory. We thank Stephen Gray, Richard Schaller, and Yi Xia for fruitful discussions. 
\bibliography{references}
\end{document}


\title{Theory of Thermal Relaxation of Electrons in Semiconductors \\ Supplemental Information}
\author{Sridhar~Sadasivam}
\email{sadasivam@anl.gov}
\affiliation{Center for Nanoscale Materials, Argonne National Laboratory, Argonne IL 60439, USA}
\author{Maria~K.~Y.~Chan}
\affiliation{Center for Nanoscale Materials, Argonne National Laboratory, Argonne IL 60439, USA}
\author{Pierre~Darancet}
\email{pdarancet@anl.gov}
\affiliation{Center for Nanoscale Materials, Argonne National Laboratory, Argonne IL 60439, USA}


\begin{abstract}
Supporting information for the manuscript: Theory of Thermal Relaxation of Electrons in Semiconductors.
\end{abstract}

\maketitle

\tableofcontents

\section{Boltzmann transport equation}
\subsection{Definition of the scattering rates}
The complete mathematical expressions for electron-phonon and phonon-phonon scattering rates in the Boltzmann transport equation are provided here.
\subsubsection{Electron-phonon scattering}
\begin{equation}
\begin{split}
\frac{\partial n_{\textbf{q}\nu}}{\partial t}\bigg\rvert_{ep} = \frac{4\pi}{\hbar}\sum\limits_{\textbf{k},m,n}\biggl\{&|g_{\textbf{q}\nu}(m\textbf{k}+\textbf{q},n\textbf{k})|^2 [f_{\textbf{mk}+\textbf{q}}(1-f_{n\textbf{k}})(n_{\textbf{q}\nu}+1)\\
&-(1-f_{m\textbf{k}+\textbf{q}})f_{n\textbf{k}}n_{\textbf{q}\nu}]\delta(E_{m\textbf{k}+\textbf{q}}-E_{n\textbf{k}}-\hbar\omega_{\textbf{q}\nu})\biggr\}
\end{split}
\label{eqn_ep}
\end{equation}
In the above equation, $E_{n\textbf{k}}$, $f_{n\textbf{k}}$ denote the energy and occupation respectively of an electron with wavevector $\textbf{k}$ and band index $n$. Similarly $\omega_{\textbf{q}\nu}$, $n_{\textbf{q}\nu}$ denote the frequency and occupation of a phonon mode with wavevector $\textbf{q}$ and branch index $\nu$. $|g_{\textbf{q}\nu}(m\textbf{k}+\textbf{q},n\textbf{k})|$ is the electron-phonon scattering matrix element for scattering of an electron from state $\ket{m\textbf{k}+\textbf{q}}$ to state $\ket{n\textbf{k}}$ due to a phonon $\ket{\textbf{q}\nu}$. The time-dependent joint density of states $\mathcal{M}_{mn\nu\textbf{k}\textbf{q}}(t)$ is given by $[f_{\textbf{mk}+\textbf{q}}(1-f_{n\textbf{k}})(n_{\textbf{q}\nu}+1)-(1-f_{m\textbf{k}+\textbf{q}})f_{n\textbf{k}}n_{\textbf{q}\nu}]\delta(E_{m\textbf{k}+\textbf{q}}-E_{n\textbf{k}}-\hbar\omega_{\textbf{q}\nu})$.

\subsubsection{Phonon-phonon scattering}

\begin{equation}
\begin{split}
\frac{\partial n_{\textbf{q}\nu}}{\partial t}\bigg\rvert_{pp} = \frac{2\pi}{\hbar^2}\sum\limits_{\textbf{q}'\nu'}\sum\limits_{\nu''}\biggl\{&|\Psi_{\textbf{q}\textbf{q}'\textbf{q}_1''}^{\nu\nu'\nu''}|^2[(n_{\textbf{q}\nu}+1)(n_{\textbf{q}'\nu'}+1)n_{\textbf{q}_1''\nu''}-n_{\textbf{q}\nu} n_{\textbf{q}'\nu'} (n_{\textbf{q}_1''\nu''}+1)]\delta(\omega_{\textbf{q}\nu}+\omega_{\textbf{q}'\nu'}-\omega_{\textbf{q}_1''\nu''}) + \\
& \frac{1}{2}|\Psi_{\textbf{q}\textbf{q}'\textbf{q}_2''}^{\nu\nu'\nu''}|^2[(n_{\textbf{q}\nu}+1)n_{\textbf{q}'\nu'}n_{\textbf{q}_2''\nu''}-n_{\textbf{q}\nu} (n_{\textbf{q}'\nu'}+1) (n_{\textbf{q}_2''\nu''}+1)]\delta(\omega_{\textbf{q}\nu}-\omega_{\textbf{q}'\nu'}-\omega_{\textbf{q}_2''\nu''})\biggr\}
\end{split}
\label{eqn_pp}
\end{equation}
where $\textbf{q}_1'' = \textbf{q}+\textbf{q}'+\textbf{G}$, $\textbf{q}_2'' = \textbf{q}-\textbf{q}'+\textbf{G}$ ($\textbf{G}$ is a reciprocal lattice vector) and $|\Psi_{\textbf{q}\textbf{q}'\textbf{q}_1''}^{\nu\nu'\nu''}|$ denotes the three-phonon scattering matrix element that is computed from a Fourier transform of the real-space third-order force constants $\Phi_{b0,b'l',b''l''}^{\alpha\beta\gamma}$:
\begin{equation}
\Psi_{\textbf{q}\textbf{q}'\textbf{q}''}^{\nu\nu'\nu''}=\frac{1}{\sqrt{N}}\left(\frac{\hbar}{2}\right)^{3/2}\sum\limits_{b}\sum\limits_{b'l'}\sum\limits_{b''l''}\sum\limits_{\alpha\beta\gamma}\Phi_{b0,b'l',b''l''}^{\alpha\beta\gamma} \times \\
\frac{e^{\alpha}_{b,\textbf{q}\nu}e^{\beta}_{b',\textbf{q}'\nu'}e^{\gamma}_{b'',\textbf{q}''\nu''}}{\sqrt{m_b\omega_{\textbf{q}\nu}m_{b'}\omega_{\textbf{q}'\nu'}m_{b''}\omega_{\textbf{q}''\nu''}}}\exp{\left(i\textbf{q}'\cdot\textbf{r}_{0l'}\right)}\exp{\left(i\textbf{q}''\cdot\textbf{r}_{0l''}\right)}
\end{equation}
where $b$, $b'$, $b''$ denote indices of atoms in the unit cell, and $l'$, $l''$ denote indices of unit cell positions with respect to a reference unit cell, and $\alpha$, $\beta$, $\gamma$ represent the Cartesian directions. The decay of real-space third-order force constants $\Phi_{b0,b'l',b''l''}^{\alpha\beta\gamma}$ with atomic distance (the maximum of distances between two atoms among the three atoms involved is plotted in the x-axis) is shown in Figs. \ref{B_series_decay}, {\ref{Al_series_decay}, \ref{Ga_series_decay}, \ref{Si_diamond_decay}. In the present work, real-space third-order force constants are obtained from finite differences of forces due to small atomic displacements in periodic supercells. The present work considers only diagonal supercells in the computation of third-order force constants; however the non-diagonal supercell technique \cite{lloyd2015lattice} could potentially be used to increase the computational efficiency of these calculations for materials with complex primitive unit cells.
\begin{figure}[H]
    \centering
   \begin{subfigure}[b]{0.4\textwidth}
    \includegraphics[width=60mm]{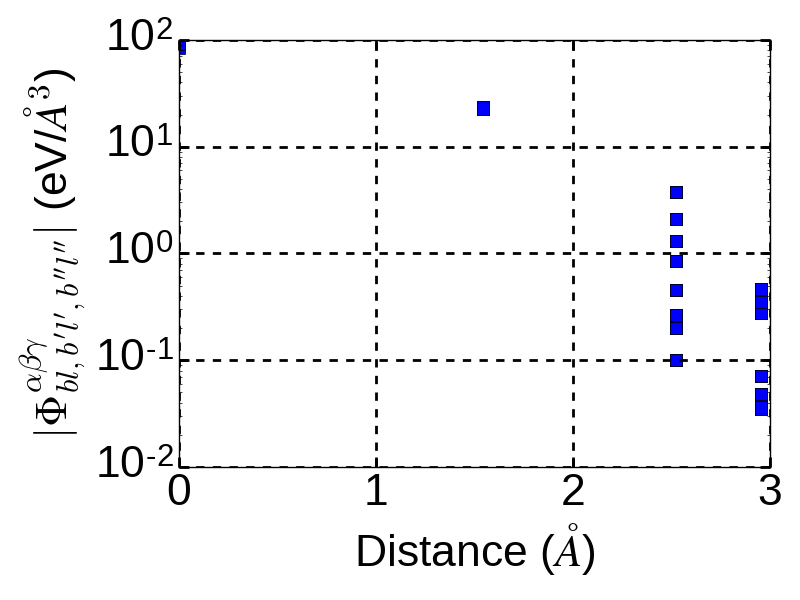}
    \caption{BN}
    \end{subfigure}\qquad\qquad
    \begin{subfigure}[b]{0.4\textwidth}
    \includegraphics[width=60mm]{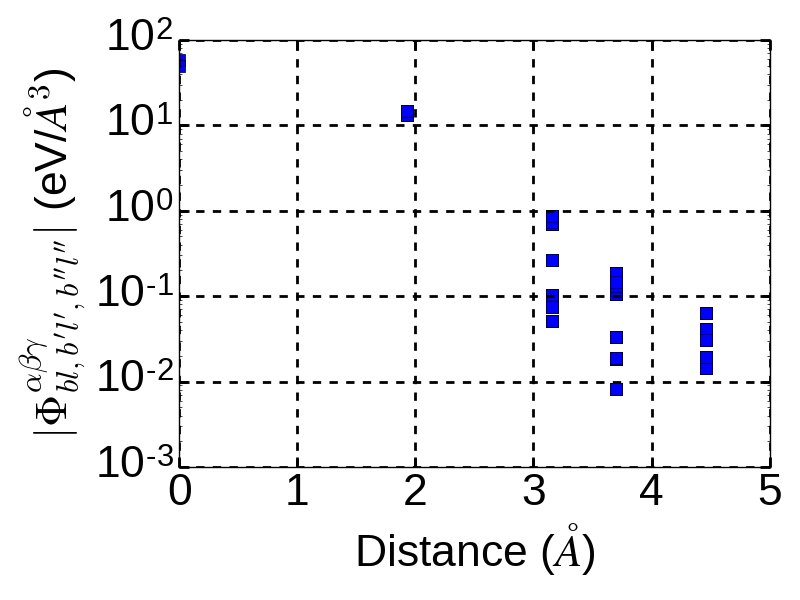}
    \caption{BP}
    \end{subfigure}\\
    \begin{subfigure}[b]{0.4\textwidth}
    \includegraphics[width=60mm]{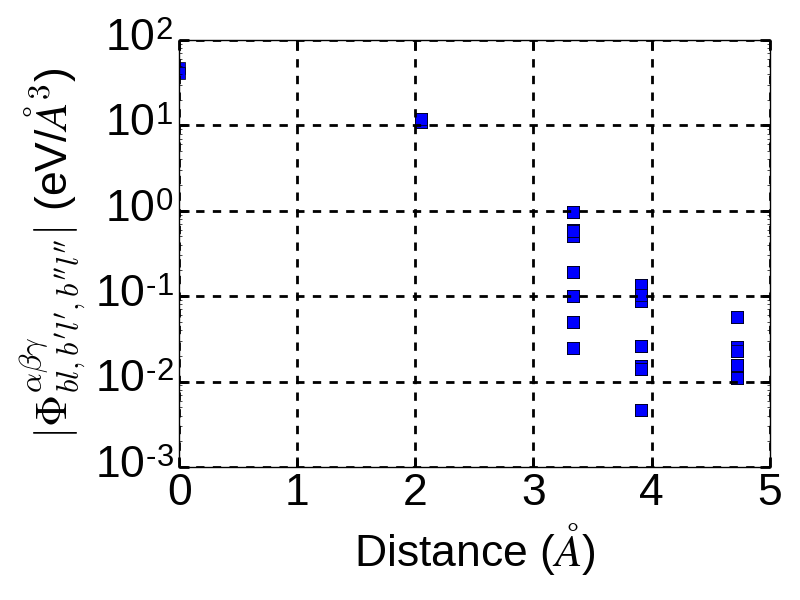}
    \caption{BAs}
    \end{subfigure}\qquad\qquad
    \begin{subfigure}[b]{0.4\textwidth}
    \includegraphics[width=60mm]{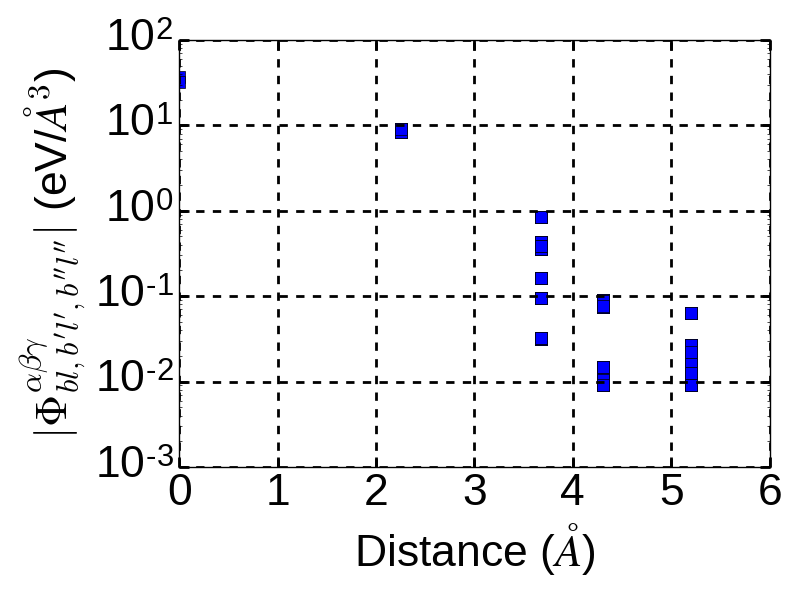}
    \caption{BSb}
    \end{subfigure}
    \caption{Decay of real-space third-order force constants with atomic distance for BN, BP, BAs and BSb.}
    \label{B_series_decay}
\end{figure}
\begin{figure}[H]
    \centering
    \begin{subfigure}[b]{0.4\textwidth}
    \includegraphics[width=60mm]{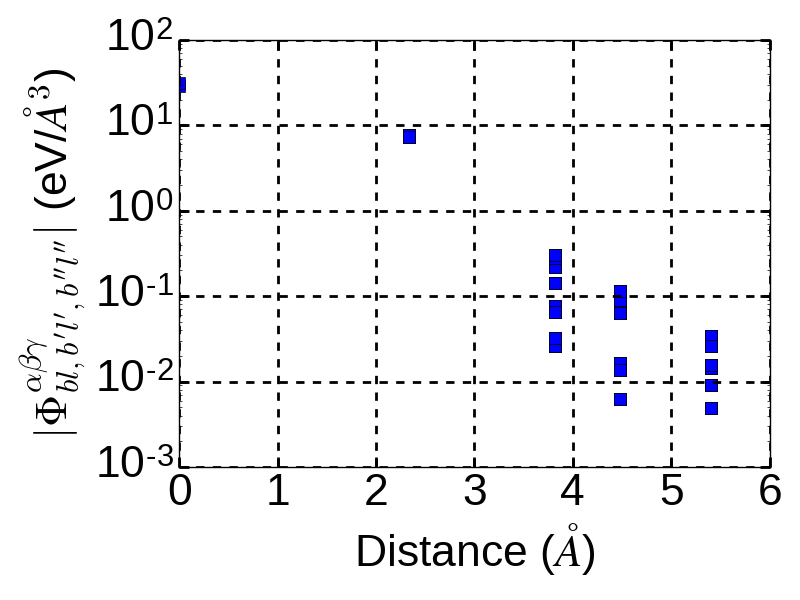}
    \caption{AlP}
    \end{subfigure}\qquad\qquad
    \begin{subfigure}[b]{0.4\textwidth}
    \includegraphics[width=60mm]{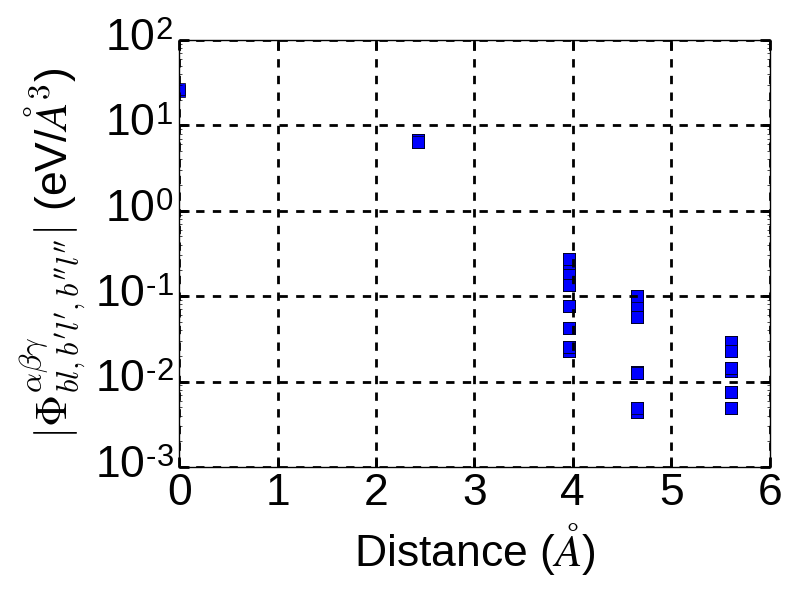}
    \caption{AlAs}
    \end{subfigure}\\
    \begin{subfigure}[b]{0.4\textwidth}
    \includegraphics[width=60mm]{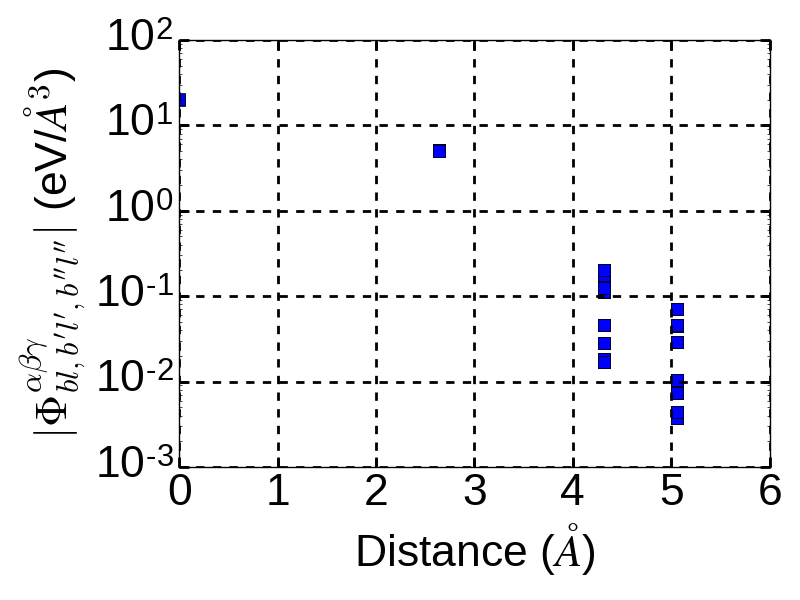}
    \caption{AlSb}
    \end{subfigure}
    \caption{Decay of real-space third-order force constants with atomic distance for AlP, AlAs and AlSb.}
    \label{Al_series_decay}
\end{figure}
\begin{figure}[H]
    \centering
    \begin{subfigure}[b]{0.4\textwidth}
    \includegraphics[width=60mm]{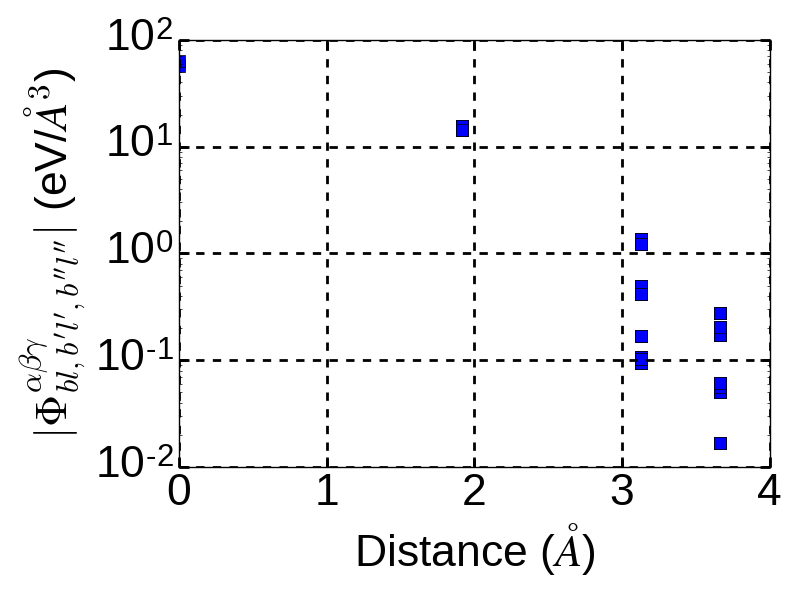}
    \caption{GaN}
    \end{subfigure}\qquad\qquad
    \begin{subfigure}[b]{0.4\textwidth}
    \includegraphics[width=60mm]{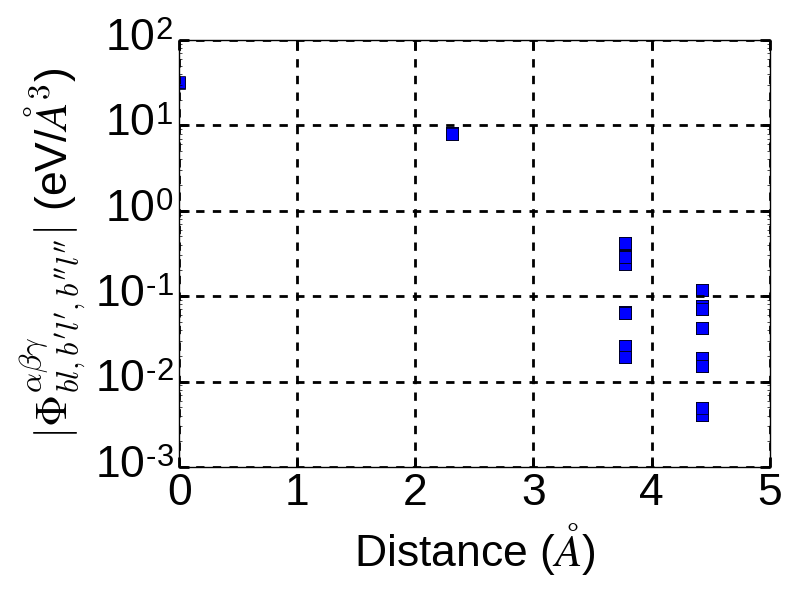}
    \caption{GaP}
    \end{subfigure}\\
    \begin{subfigure}[b]{0.4\textwidth}
    \includegraphics[width=60mm]{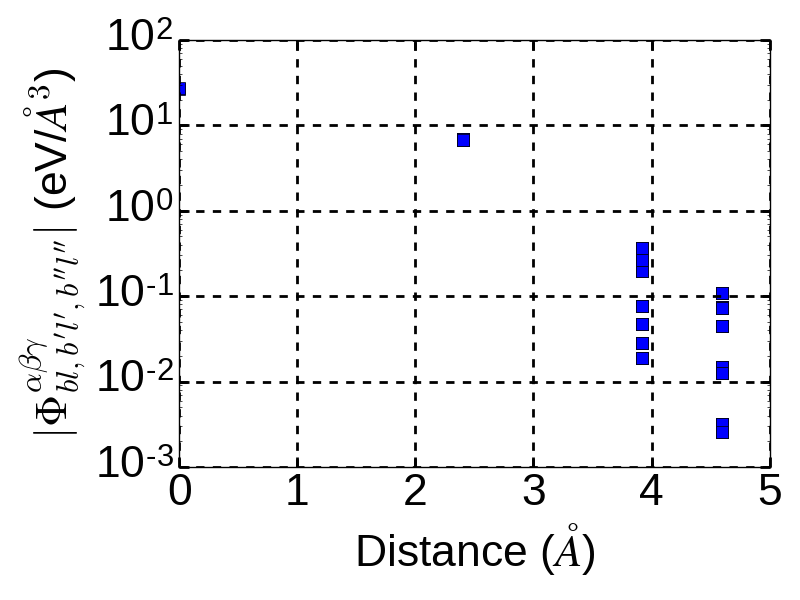}
    \caption{GaAs}
    \end{subfigure}
    \caption{Decay of real-space third-order force constants with atomic distance for GaN, GaP and GaAs.}
    \label{Ga_series_decay}
\end{figure}
\begin{figure}[H]
    \centering
    \begin{subfigure}[b]{0.4\textwidth}
    \includegraphics[width=60mm]{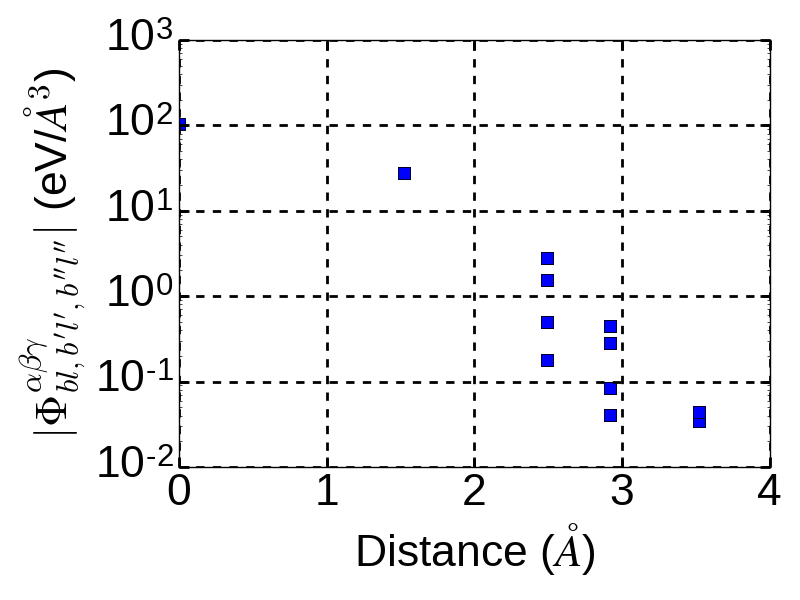}
    \caption{Diamond}
    \end{subfigure}\qquad\qquad
    \begin{subfigure}[b]{0.4\textwidth}
    \includegraphics[width=60mm]{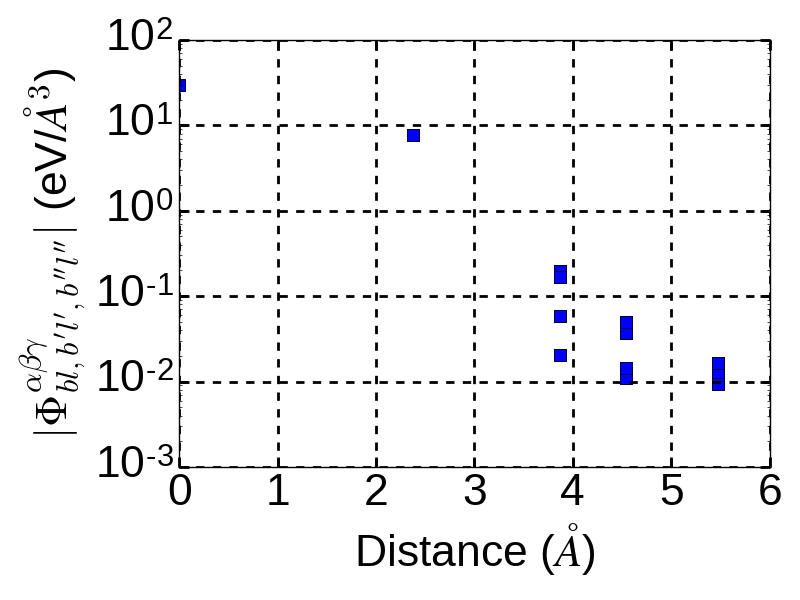}
    \caption{Si}
    \end{subfigure}
    \caption{Decay of real-space third-order force constants with atomic distance for diamond and Si.}
    \label{Si_diamond_decay}
\end{figure}
\subsection{Non-Equilibrium Phonon Distributions}
In this section, we present temperature maps of non-equilibrium phonon distributions in all III-V materials considered in this work (see Fig.~1 of main text for BN, BAs). In all materials, we observe that a small sub-set of long-wavelength optical phonons are nearly in equilibrium with electrons at short times while the temperature of remaining phonon modes is practically unchanged.

\subsubsection{Boron Phosphide (BP)}
 \begin{figure}[H]
	\begin{center}
	\includegraphics[width=0.90\textwidth]{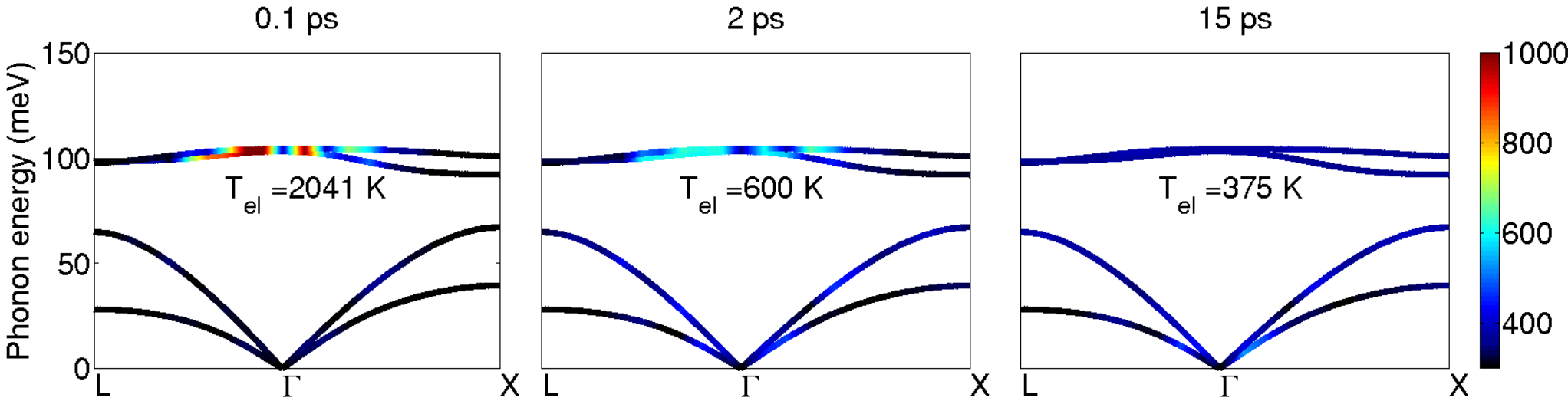}
	\end{center}
    \caption{Temperature map of phonon modes in the L-$\Gamma$-X direction of BP.}
    \label{BP_temp_map}
\end{figure}

\subsubsection{Boron Antimonide  (BSb)}
\begin{figure}[H]
	\begin{center}
	\includegraphics[width=0.90\textwidth]{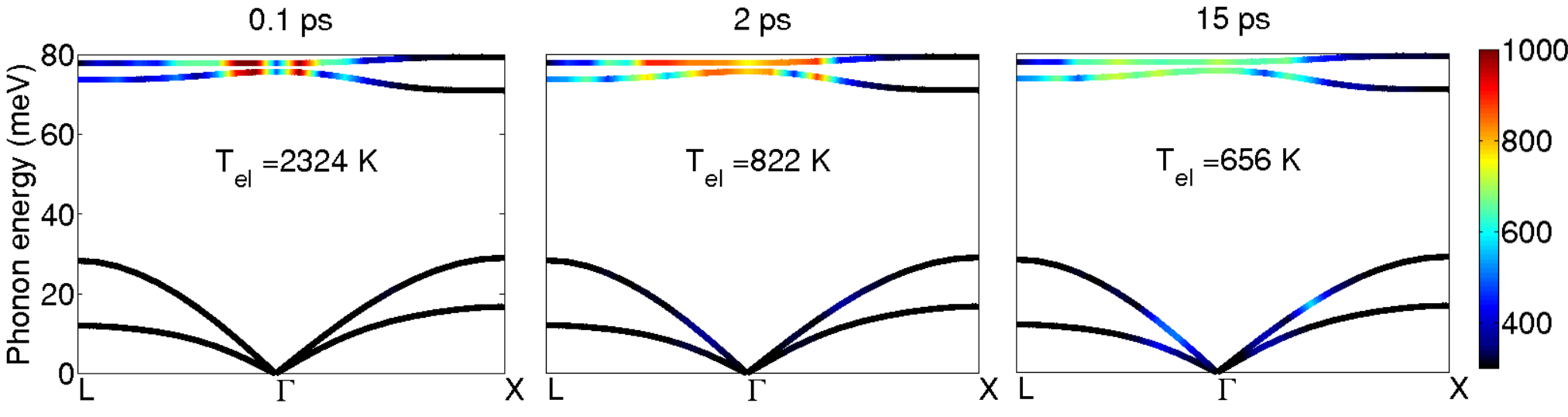}
	\end{center}
    \caption{Temperature map of phonon modes in the L-$\Gamma$-X direction of BSb.}
    \label{BSb_temp_map}
\end{figure}

\subsubsection{Aluminum Phosphide  (AlP)}

\begin{figure}[H]
	\begin{center}
	\includegraphics[width=0.90\textwidth]{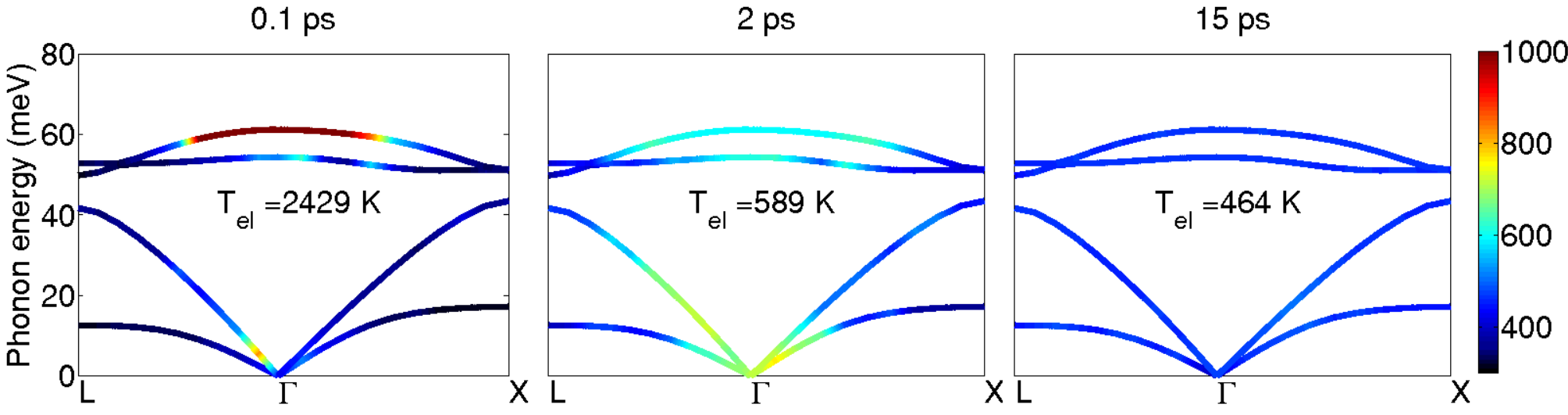}
	\end{center}
    \caption{Temperature map of phonon modes in the L-$\Gamma$-X direction of AlP.}
    \label{AlP_temp_map}
\end{figure}

\subsubsection{Aluminum Arsenide  (AlAs)}

\begin{figure}[H]
	\begin{center}
	\includegraphics[width=0.90\textwidth]{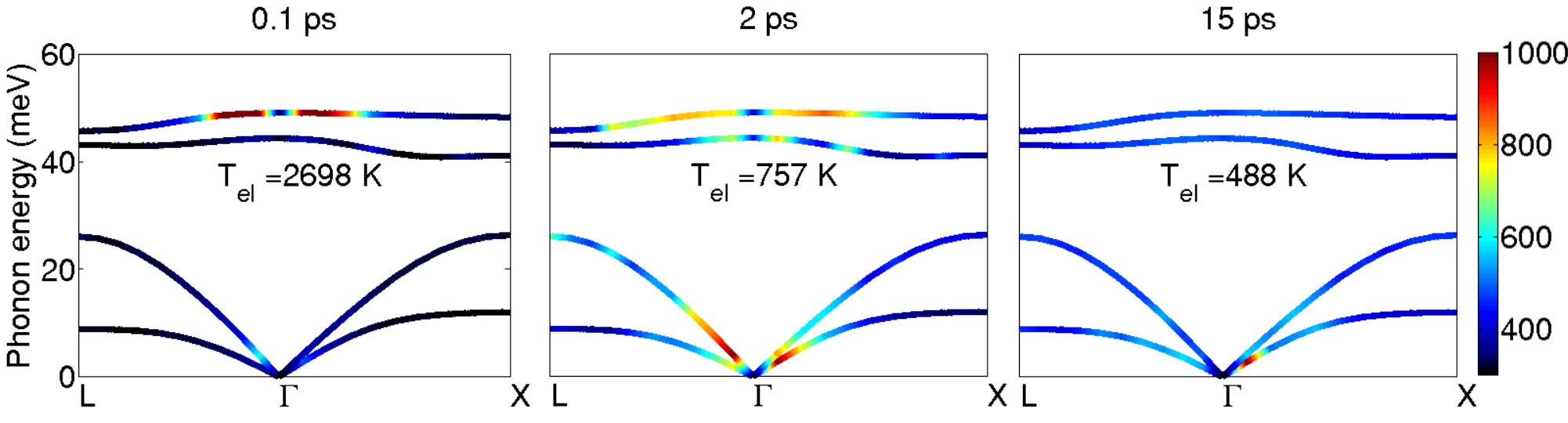}
	\end{center}
    \caption{Temperature map of phonon modes in the L-$\Gamma$-X direction of AlAs.}
    \label{AlAs_temp_map}
\end{figure}

\subsubsection{Aluminum Antimonide  (AlSb)}

\begin{figure}[H]
	\begin{center}
	\includegraphics[width=0.90\textwidth]{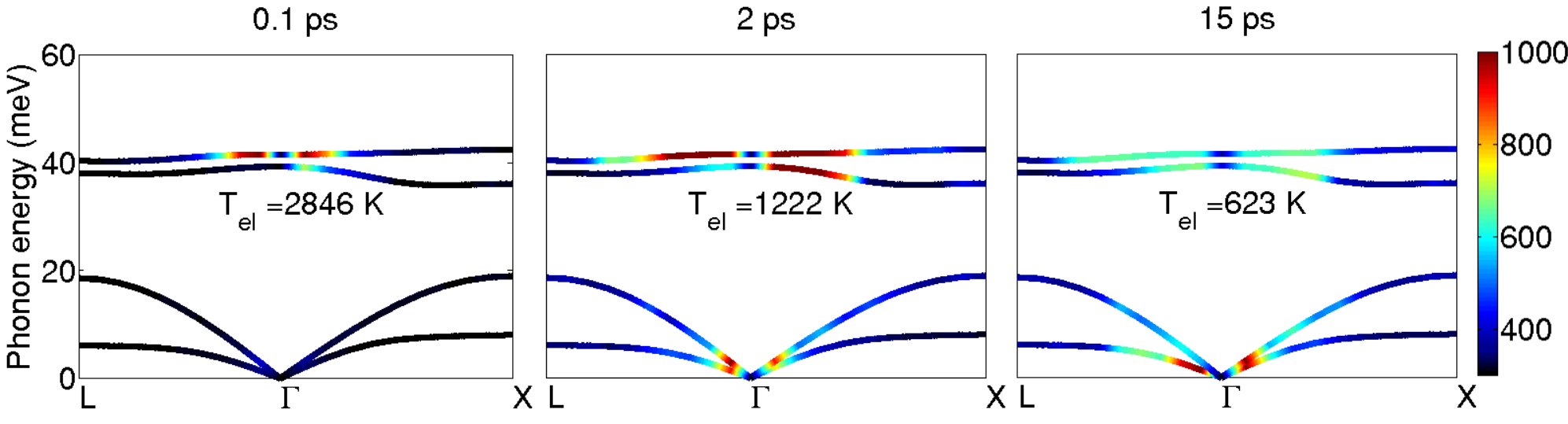}
	\end{center}
    \caption{Temperature map of phonon modes in the L-$\Gamma$-X direction of AlSb.}
    \label{AlSb_temp_map}
\end{figure}

\subsubsection{Gallium Nitride  (GaN)}

\begin{figure}[H]
	\begin{center}
	\includegraphics[width=0.90\textwidth]{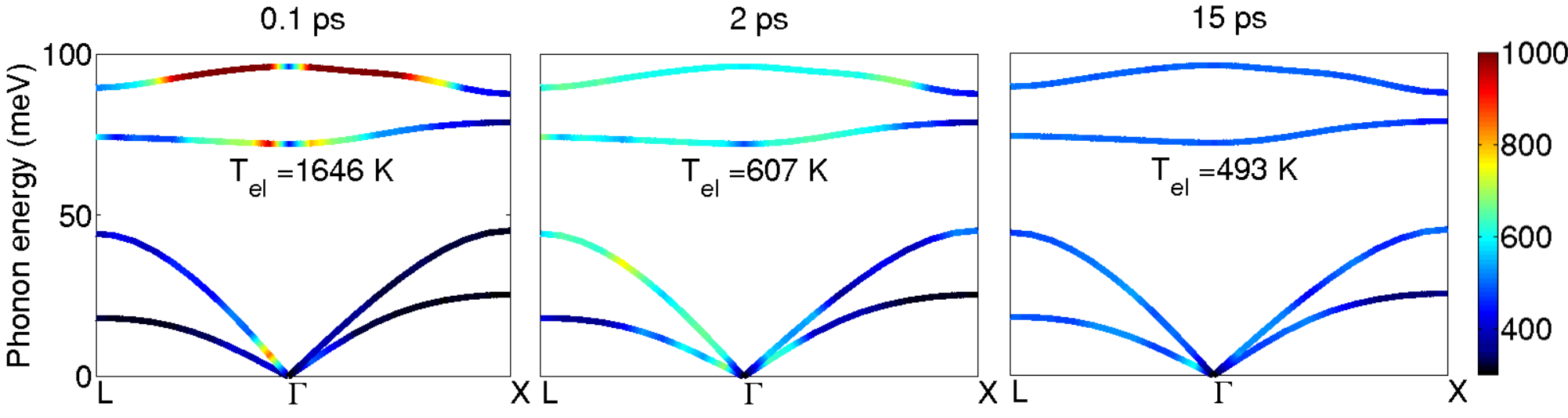}
	\end{center}
    \caption{Temperature map of phonon modes in the L-$\Gamma$-X direction of GaN.}
    \label{GaN_temp_map}
\end{figure}
\subsubsection{Gallium Phosphide  (GaP)}

\begin{figure}[H]
	\begin{center}
	\includegraphics[width=0.90\textwidth]{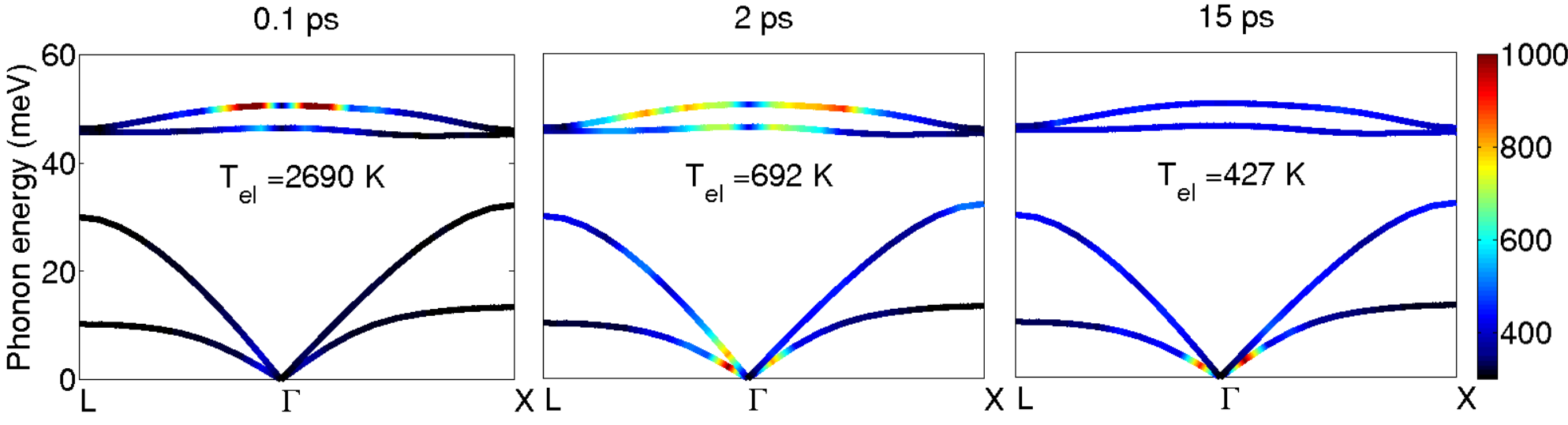}
	\end{center}
    \caption{Temperature map of phonon modes in the L-$\Gamma$-X direction of GaP.}
    \label{GaP_temp_map}
\end{figure}

\subsubsection{Gallium Arsenide  (GaAs)}

\begin{figure}[H]
	\begin{center}
	\includegraphics[width=0.90\textwidth]{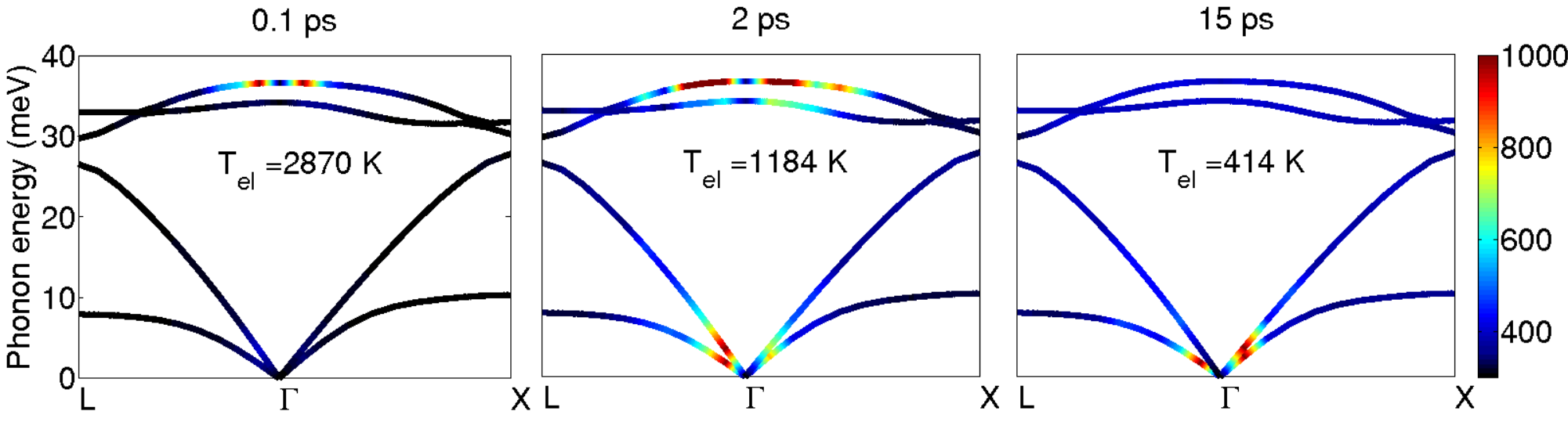}
	\end{center}
    \caption{Temperature map of phonon modes in the L-$\Gamma$-X direction of GaAs.}
    \label{GaAs_temp_map}
\end{figure}

\subsubsection{Diamond}

\begin{figure}[H]
	\begin{center}
	\includegraphics[width=0.90\textwidth]{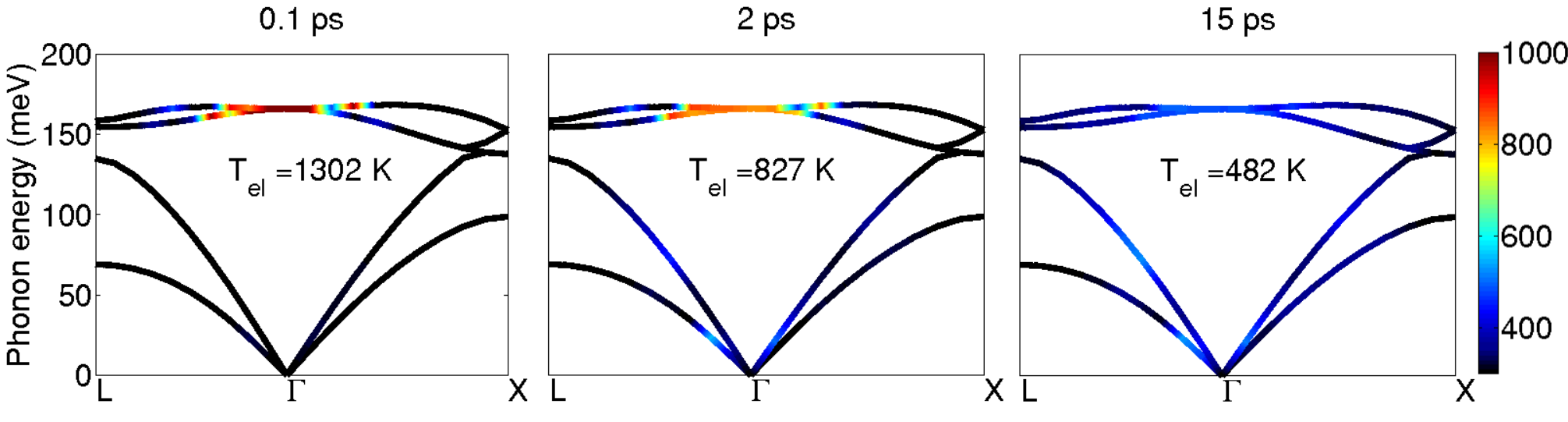}
	\end{center}
    \caption{Temperature map of phonon modes in the L-$\Gamma$-X direction of Diamond.}
    \label{diamond_temp_map}
\end{figure}

\subsubsection{Silicon}

\begin{figure}[H]
	\begin{center}
	\includegraphics[width=0.90\textwidth]{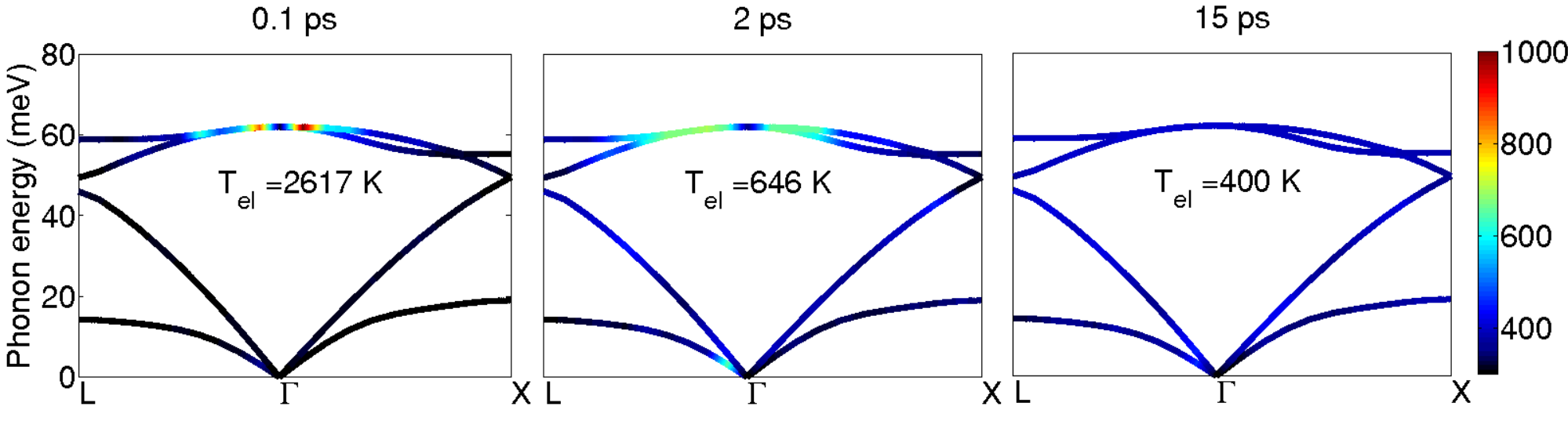}
	\end{center}
    \caption{Temperature map of phonon modes in the L-$\Gamma$-X direction of Si.}
    \label{Si_temp_map}
\end{figure}
\clearpage
\subsection{Polar vs. Non-Polar Semiconductors}
In this section, we discuss the similarities in electron-phonon thermalization dynamics between polar and non-polar materials materials. From the temperature map of phonon modes in polar materials such as BP, GaN (see Figs.~\ref{BP_temp_map},\ref{GaN_temp_map}) and non-polar materials such as diamond, Si (see Figs.~\ref{diamond_temp_map},\ref{Si_temp_map}), electrons transfer energy primarily to long-wavelength optical phonons at short time instants in both classes of semiconductors. The dominance of long-wavelength optical phonons in electron-phonon scattering for both polar and non-polar semiconductors is a consequence of the large electron-phonon scattering phase space associated with small-momentum optical phonons though the effect is further magnified by the Fr\"{o}hlich coupling in polar compounds. 

The phase space for electron-phonon scattering is an important factor that critically impacts the distribution of $\eta_{\textbf{q}\nu}$ in semiconductors: The large difference in the energy scales of electrons ($E_{el} \sim$ eV) and phonons ($E_{ph} \sim$ 100 meV) implies that intra-valley scattering is primarily dominated by small-momentum optical phonons. To illustrate the preference for small-momentum or long-wavelength phonon scattering, we consider a minimal model of a one-dimensional parabolic electronic bandstructure. At first order and neglecting Umklapp processes, conservation of energy and momentum in the electron-phonon scattering event involving states $\ket{k}$, $\ket{k+q}$ with energies $E_k = \hbar^2k^2/2m_{eff}$, $E_{k+q} = \hbar^2(k+q)^2/2m_{eff}$ and a phonon energy$<\hbar\omega_{q,max}$ ($\omega_{q,max}$ is the maximum energy of phonons in the material) can be written as:
\begin{equation}
\frac{\hbar^2(q^2+2kq)}{2m_{eff}}<\hbar\omega_{q,max}
\end{equation}
For simplicity, we consider scattering of an electron at the band minimum ($k = 0$) and a maximum optical phonon energy of 150 meV in diamond. Assuming a longitudinal effective mass $m_{eff} = 1.4 m_e$ in diamond \cite{nava1980electron}, we obtain $q<0.25\pi/a$ (for a transverse effective mass $m_{eff} = 0.36 m_e$, we obtain $q<0.12\pi/a$). The above analysis, albeit simplified, shows that conservation of energy and momentum in an intra-valley electron-phonon scattering event within a parabolic band dictates that only phonons with small momentum are allowed to participate. Inter-valley scattering could involve phonons with large momentum; however, in both BN and diamond, the valence band maximum occurs at  $\Gamma$ (see Figs.~\ref{band_structure}a,b) thus eliminating any inter-valley scattering processes with short-wavelength phonons. 
\begin{figure}[h]
    \centering
    \begin{subfigure}[b]{0.4\textwidth}
    \includegraphics[width=60mm]{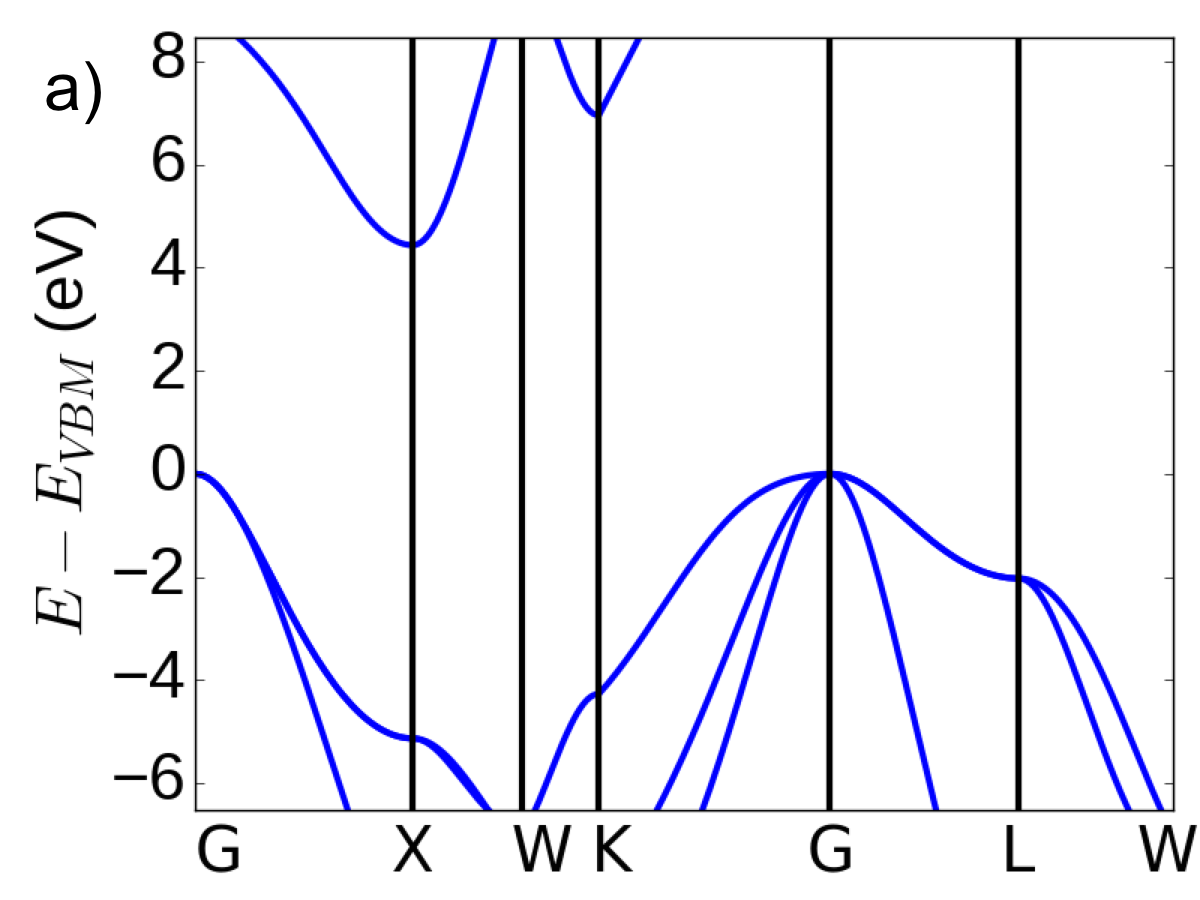}
    \end{subfigure}\quad
    \begin{subfigure}[b]{0.4\textwidth}
    \includegraphics[width=60mm]{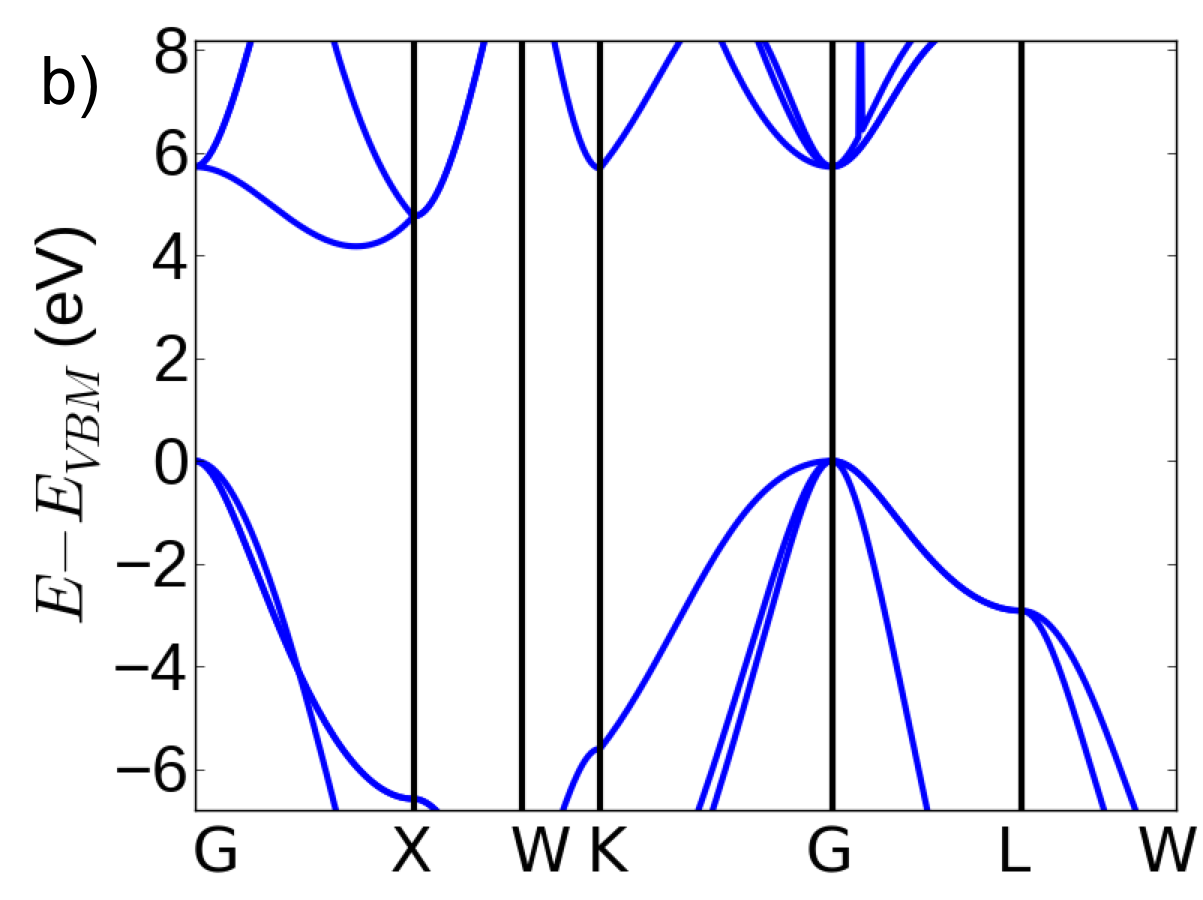}
    \end{subfigure}\quad
    \caption{Electron bandstructures (as described by DFT-LDA) of a) BN and b) diamond.}
    \label{band_structure}
\end{figure}

Beyond the scattering phase space considerations discussed above, the preference for small-momentum optical phonon scattering is further magnified by the larger magnitude of electron-phonon matrix elements for optical phonons relative to acoustic modes. As shown in Fig.~\ref{eph_matrix_element_plot}b, the magnitude of electron-phonon coupling matrix elements in diamond is about ten-fold larger for optical phonons near $\Gamma$ in comparison to acoustic modes (see also previous first-principles calculations of electron-phonon coupling in diamond in Ref.~\cite{PhysRevB.72.014306}). In a polar compound such as BN, the divergence of electron-phonon coupling matrix elements for long-wavelength LO phonons coupled via Fr\"{o}hlich interactions leads to a qualitatively similar (though quantitatively different) effect (Fig.~\ref{eph_matrix_element_plot}a).
\begin{figure}[h]
    \centering
    \begin{subfigure}[b]{0.4\textwidth}
    \includegraphics[width=60mm]{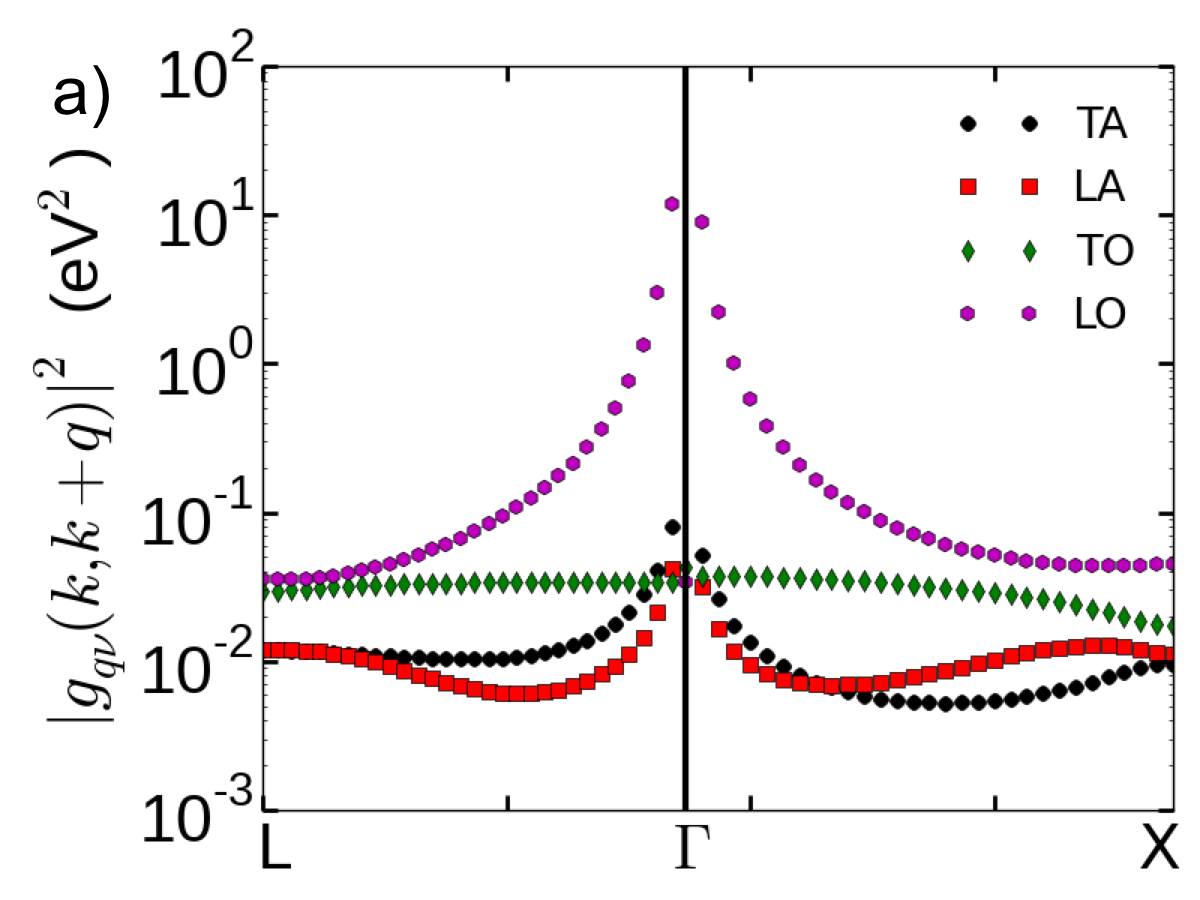}
    \end{subfigure}\qquad
    \begin{subfigure}[b]{0.4\textwidth}
    \includegraphics[width=60mm]{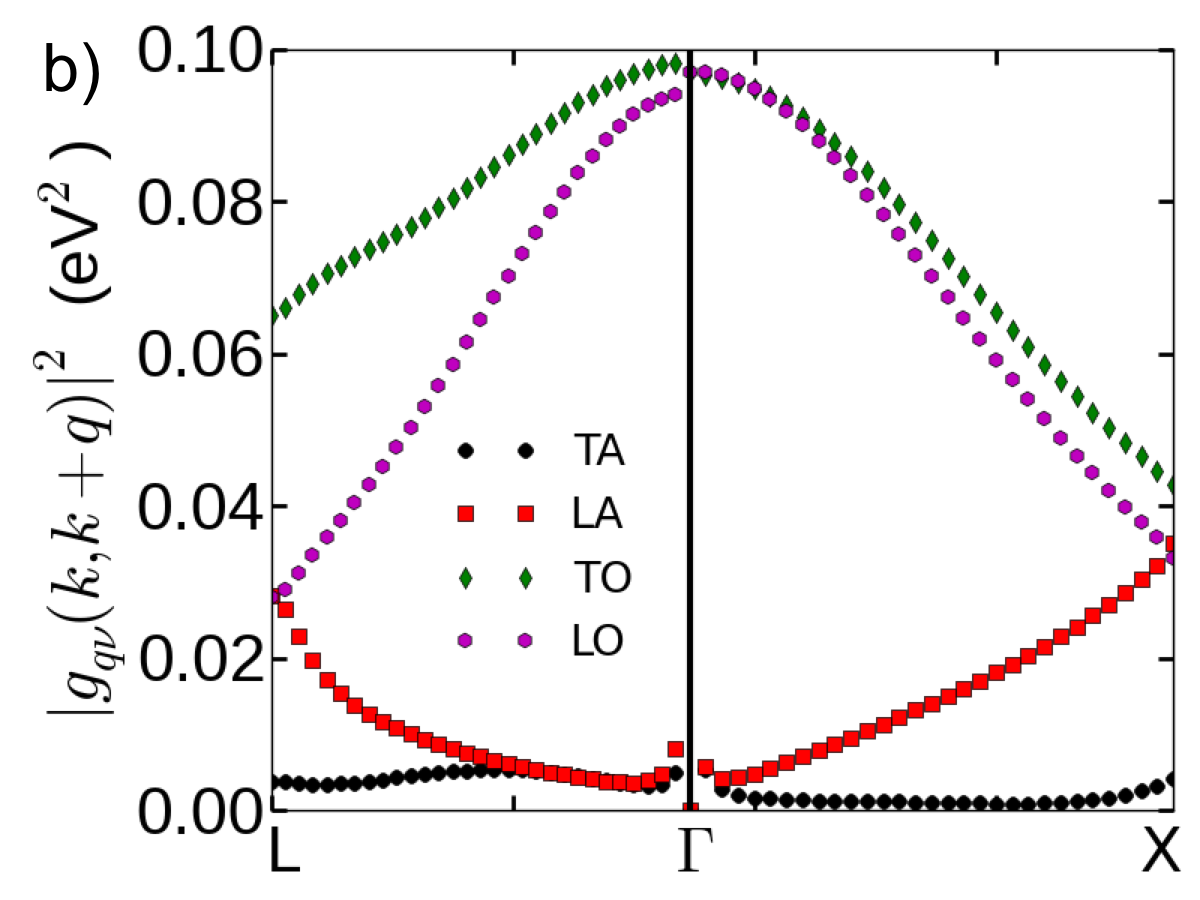}
    \end{subfigure}\qquad
    \caption{Squared-magnitude of electron-phonon coupling matrix elements as a function of phonon wavevector for an electronic state at the VBM of BN (a) and diamond (b). Since the VBM is triply degenerate for both BN and diamond, the above plots represent an average over all possible transitions within the three degenerate bands.}
    \label{eph_matrix_element_plot}
\end{figure}

In summary, long wavelength optical phonons are excited at short times in both polar and non-polar semiconductors due to the restriction of phase space to small momentum phonons and the larger electron-phonon matrix elements associated with optical phonon modes relative to acoustic phonons. 
\subsection{Grid and Smearing Dependence of Electronic Temperature Decay}
In this section, we verify that the results for electronic temperature decay obtained from the BTE simulation are independent of the $\textbf{k}$ and $\textbf{q}$ grids used in the calculation (see Fig.~\ref{grid_conv1}). We also verify the independence of results with respect to the Gaussian smearing used in energy conservation delta functions for electron-phonon (Fig.~\ref{grid_conv2}) and phonon-phonon (Fig.~\ref{grid_conv3}) scattering. 
\begin{figure}[h]
    \centering
    \begin{subfigure}[b]{0.4\textwidth}
    \includegraphics[width=60mm]{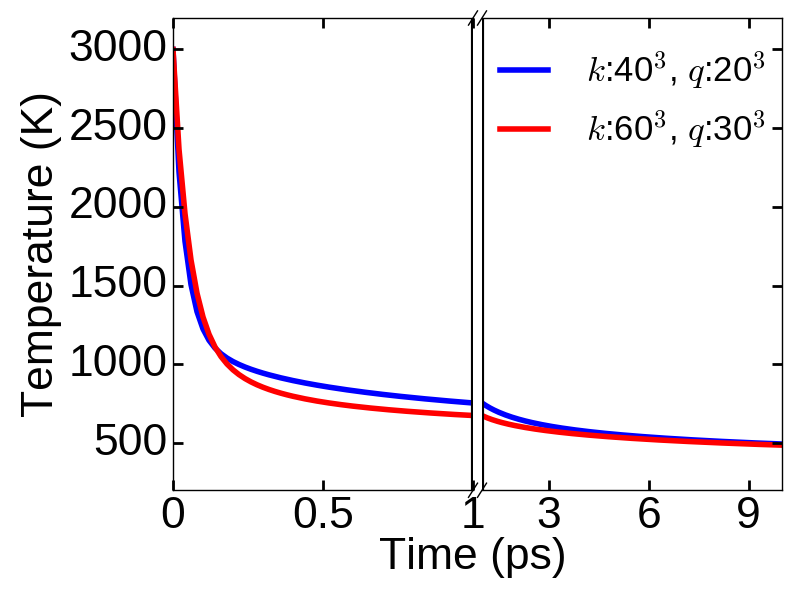}
    \caption{$\textbf{k}$, $\textbf{q}$ grid independence.}
    \label{grid_conv1}
    \end{subfigure}\qquad
    \begin{subfigure}[b]{0.4\textwidth}
    \includegraphics[width=60mm]{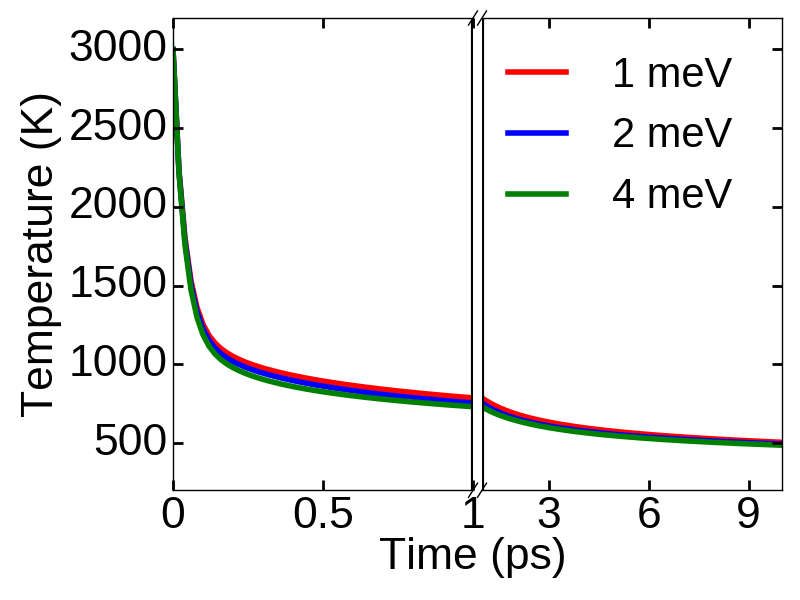}
    \caption{Electron-phonon smearing independence.}
    \label{grid_conv2}
    \end{subfigure}
     \begin{subfigure}[b]{0.4\textwidth}
    \includegraphics[width=60mm]{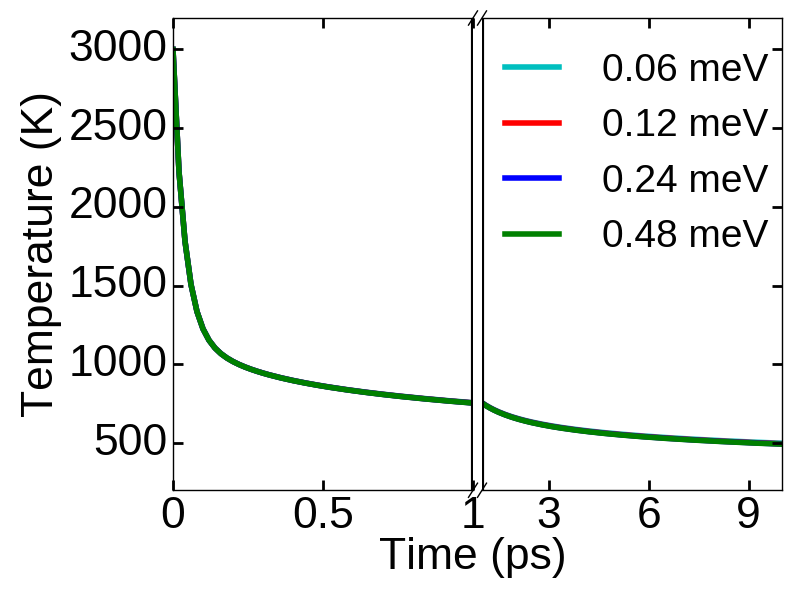}
    \caption{Phonon-phonon smearing independence.}
    \label{grid_conv3}
    \end{subfigure}
    \caption{Comparison of electronic temperature decay in BN for (a) different $\textbf{k}$ and $\textbf{q}$ grids (b) different values of Gaussian broadening used in energy-conserving delta functions for electron-phonon coupling (c) different values of Gaussian broadening used in energy-conserving delta functions for phonon-phonon coupling.}
\end{figure}
\clearpage

\section{2-Temperature, 3-Temperature, and Successive Thermalization Models}
In this section, we present a comparison between the electronic temperature decay obtained from numerical solution of the BTE with predictions from the 2T, 3T and successive thermalization (ST) models, as well as the mathematical expressions used to compute the parameters of these models from first-principles.

\subsection{2-Temperature model: Definitions}
The 2T model assigns one effective temperature for electrons ($T_{\textrm{el}}$) and phonons ($T_{\textrm{ph}}$), and assumes an effective coupling rate $G_{ep}(T_{\textrm{el}},T_{\textrm{ph}})$ that determines the rate of energy transfer between electrons and phonons:
\begin{equation}
C_{el}(T_{\textrm{el}})\frac{d T_{\textrm{el}}}{d t} = G_{ep}(T_{\textrm{el}},T_{\textrm{ph}})(T_{\textrm{ph}}-T_{\textrm{el}}) \qquad C_{ph}(T_{\textrm{ph}})\frac{d T_{\textrm{ph}}}{d t} = G_{ep}(T_{\textrm{el}},T_{\textrm{ph}})(T_{\textrm{el}}-T_{\textrm{ph}})
\end{equation}
All the parameters required in the 2T model $C_{el}(T_{\textrm{el}})$, $C_{ph}(T_{\textrm{ph}})$, $G_{ep}(T_{\textrm{el}},T_{\textrm{ph}})$ can be obtained from the first-principles calculations reported earlier. The electronic and the lattice heat capacities can be obtained as follows:
\begin{equation}
C_{el}(T_{\textrm{el}}) = \frac{2}{V}\sum\limits_{\textbf{k},m}(E_{\textbf{k},m}-E_f)\frac{\partial f_{FD}^o}{\partial T}\qquad
C_{ph}(T_{\textrm{ph}}) = \frac{1}{V}\sum\limits_{\textbf{q},\nu}\hbar\omega_{\textbf{q},\nu}\frac{\partial f_{BE}^o}{\partial T}
\end{equation}
The temperature-dependent electron-phonon coupling coefficient $G_{ep}(T_{\textrm{el}},T_{\textrm{ph}})$ can be obtained from the electron-phonon matrix elements computed from DFPT:
\begin{equation}
\begin{split}
G_{ep}(T_{\textrm{el}},T_{\textrm{ph}}) = \frac{4\pi}{\hbar V(T_{\textrm{el}}-T_{\textrm{ph}})}  & \sum\limits_{\textbf{k},\textbf{q},m,n,\nu}\hbar\omega_{\textbf{q}\nu}[f_{m\textbf{k}+\textbf{q}}(1-f_{n\textbf{k}})(n_{\textbf{q}\nu}+1)-\\&(1-f_{m\textbf{k}+\textbf{q}})f_{n\textbf{k}}n_{\textbf{q}\nu}]|g_{\textbf{q}\nu}(m\textbf{k}+\textbf{q},n\textbf{k})|^2 \delta(E_{m\textbf{k}+\textbf{q}}-E_{n\textbf{k}}-\hbar\omega_{\textbf{q}\nu})
 \end{split}
 \label{eqn_Gep}
\end{equation}
where the equilibrium electron ($f_{n\textbf{k}}$) and phonon ($n_{\textbf{q}\nu}$) occupation functions are evaluated at temperatures $T_{\textrm{el}}$ and $T_{\textrm{ph}}$ respectively. 

\subsection{3-Temperature model: Definitions}

To account for the selective coupling of electrons with certain phonon branches, Waldecker et al.~\cite{waldecker2016electron} proposed a three-temperature (3T) model where the phonon branches are sub-divided into two categories depending on the coupling strength. Following this approach, we assume a separate temperature $T_{\textrm{ph,o}}$ for the optical phonon branches and a temperature $T_{\textrm{ph,a}}$ for the acoustic phonon branches. The equations for the temperature evolution of electrons, optical phonons, and acoustic phonons are given below:
\begin{equation}
\begin{split}
C_{el}\frac{d T_{\textrm{el}}}{d t} &= G_{ep,o}(T_{\textrm{ph,o}}-T_{\textrm{el}})+G_{ep,a}(T_{\textrm{ph,a}}-T_{\textrm{el}}) \\
C_{ph,o}\frac{dT_{\textrm{ph,o}}}{d t} &= G_{ep,o}(T_{\textrm{el}}-T_{\textrm{ph,o}})+G_{pp}(T_{\textrm{ph,a}}-T_{\textrm{ph,o}}) \\
C_{ph,a}\frac{d T_{\textrm{ph,a}}}{d t} &= G_{ep,a}(T_{\textrm{el}}-T_{\textrm{ph,a}})+G_{pp}(T_{\textrm{ph,o}}-T_{\textrm{ph,a}})
\end{split}
\end{equation}
where $C_{ph,o}$, $C_{ph,a}$ denote the heat capacities of optical and acoustic phonons respectively (temperature dependencies omitted for simplicity of notation). $G_{ep,o}$, $G_{ep,a}$ denote the electron-phonon coupling constant for optical and acoustic branches and are obtained from Eq.~(\ref{eqn_Gep}) where the sum over phonon modes runs over optical branches for $G_{ep,o}$ and over acoustic branches for $G_{ep,a}$. $G_{pp}$ denotes the phonon-phonon coupling constant between optical and acoustic phonon modes and is computed from first-principles phonon-phonon matrix elements. 

The 2T and 3T models are found to significantly under-predict the equilibration time between electrons and phonons for almost all the semiconductors considered here (GaAs is a notable exception). 

\subsection{Successive Thermalization model: Definitions}

We perform constrained simulations of electron cooling in which the phonons are partitioned into multiple subspaces sorted (on a logarithmic grid) by their interaction strength $\eta_{\textbf{q}\nu}$ approximated by:  
\begin{equation}
\begin{split}
\eta_{\textbf{q}\nu} &= \hbar\omega_{\textbf{q}\nu}\Big\{\sum\limits_{\textbf{k},m,n} |g_{\textbf{q}\nu}(m\textbf{k}+\textbf{q},n\textbf{k})|^2 \delta(E_{m\textbf{k}+\textbf{q}}-E_{n\textbf{k}}-\hbar\omega_{\textbf{q}\nu}) \\ &
+ \sum\limits_{\textbf{q}'\nu',\textbf{q}''\nu''}^{ \eta_{\textbf{q}'\nu'} > \eta_{\textbf{q}\nu}} |\Psi_{\textbf{q}\textbf{q}'\textbf{q}\pm\textbf{q}'+\textbf{G}}^{\nu\nu'\nu''}|^2\delta(\hbar(\omega_{\textbf{q}\nu}\pm\omega_{\textbf{q}'\nu'}-\omega_{\textbf{q}_1''\nu''})) \Big\},
\end{split}
\label{Eq:Eta}
\end{equation}
where the terms on the right approximate the scattering due to EPI and PPI for each mode $\vert\textbf{q}\nu\rangle$ (the PPI term only includes modes with a larger interaction strength and is computed self-consistently). At time $t=0$,  only the modes belonging to the subspace with the largest $\eta_{\textbf{q}\nu}$ are allowed to interact with electrons until thermalization. Subsequent to thermalization of the first subspace, the next subspace is introduced in the simulation along with the now-thermalized system of electrons and the first subspace. This constrained, ``successive thermalization'' (ST) process is continued until all modes are included. The effective coupling coefficient between the interacting systems at each thermalization step is chosen to be proportional to the sum of all interaction strengths $\eta_{\textbf{q}\nu}$ of modes belonging to the subspace undergoing thermalization. We note that the proportionality constant is chosen to be the \textit{same for every subspace} and is \textit{independent of temperature}. 

For the ST approach, we present results corresponding to two different cutoffs ($c=0.1,0.5$) in choosing a subspace, \textit{i.e.}, all phonon modes with $\eta_{\textbf{q}\nu} > c\eta_{\textbf{q}\nu,max} $ are chosen to belong to a subspace after every thermalization step. While the specific choice of the cutoff parameter $c$ is found to not significantly alter the timescales of electronic temperature decay, the exact agreement between the ST approach and the full-BTE results can depend on the details of the cutoff parameter for materials with narrow distributions of phonon interaction strength. 

As the central assumption of the ST simulation is to neglect the interaction between the ``non-active" subspaces and the rest of the system, we expect the ST simulation to recover fully the BTE results in the limit of infinitely broad phonon coupling strength distributions (i.e. when the ``active" subspace interacts infinitely faster than the inactive ones). As can now be seen in Figs.~\ref{BN_predictions}-\ref{Si_predictions}, the compounds BN, BAs, BSb, diamond have the broadest distributions of phonon-scattering times (see Table \ref{std_dev}), and correspondingly, the ST model shows the best agreement with the full BTE simulations for these compounds. Compounds such as AlAs, AlSb, GaAs (see Table \ref{std_dev}) have the narrowest distribution and show the largest deviations between the ST and BTE predictions. However, in contrast to the 2T model, the successive thermalization approach captures the slow timescales of electron-phonon thermalization for all 12 semiconductors considered in this work and confirms our hypothesis that electronic cooling is limited by thermalization within the phonon-subsystem. 

\begin{table}[h]
\centering
\caption{Standard deviation of $log_{10}{\eta_{\textbf{q}\nu}}$ for all 12 semiconductors considered in this manuscript.}
\label{std_dev}
\begin{tabular}{|l|l|} \hline
Compound                 & Standard deviation in $log_{10}{\eta_{\textbf{q}\nu}}$ \\ \hline
BN & 0.94                                                   \\ \hline
BP & 0.63                             \\ \hline
BAs & 1.21                                                   \\ \hline
BSb & 1.05                                                   \\ \hline
AlP & 0.31                                                   \\ \hline
AlAs & 0.31                                                   \\ \hline
AlSb  & 0.33                                                   \\ \hline
GaN  & 0.58                                                   \\ \hline
GaP  & 0.4                                                    \\ \hline
GaAs & 0.31                                                   \\ \hline
Diamond  & 0.83                                                   \\ \hline
Si  & 0.49   \\     \hline
\end{tabular}
\end{table}

\subsection{Predictions of the different models}
\subsubsection{Boron Nitride (BN)}

\begin{figure}[h]
    \centering
   \begin{subfigure}[b]{0.4\textwidth}
    \includegraphics[width=60mm]{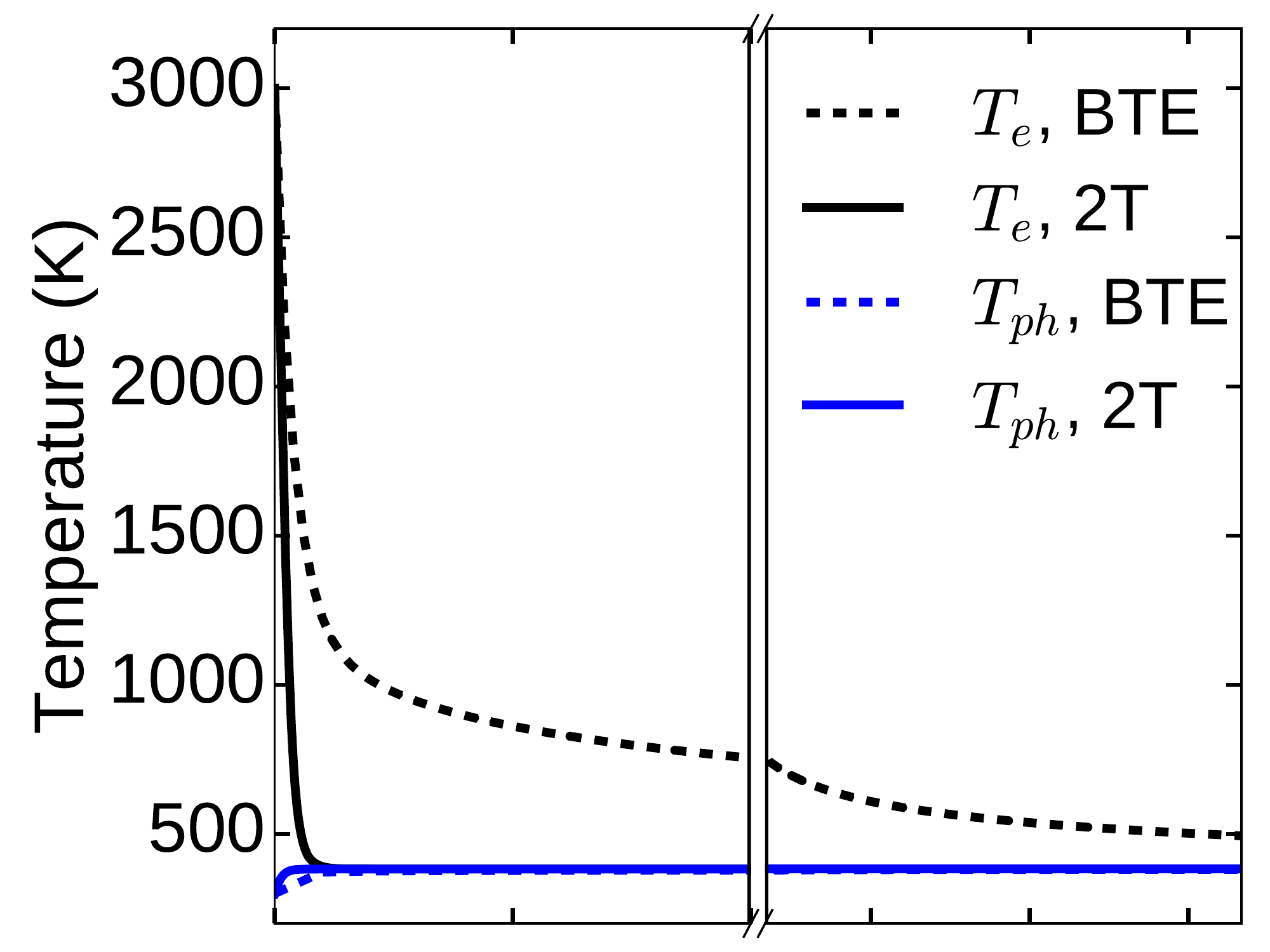}
    \caption{2T-BN}
    \end{subfigure}\qquad\qquad
    \begin{subfigure}[b]{0.4\textwidth}
    \includegraphics[width=60mm]{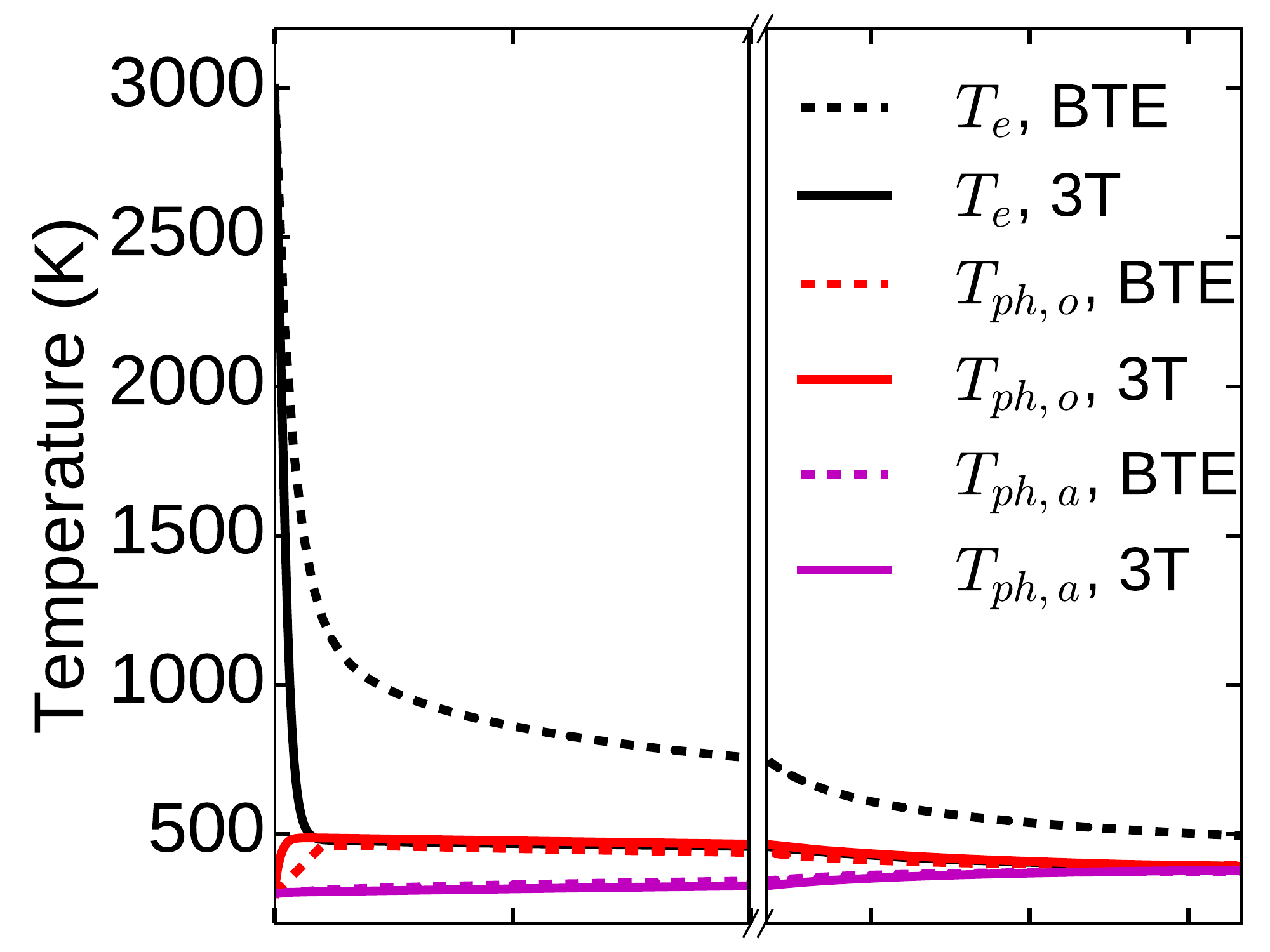}
    \caption{3T-BN}
    \end{subfigure}\\
    \begin{subfigure}[b]{0.4\textwidth}
    \includegraphics[width=60mm]{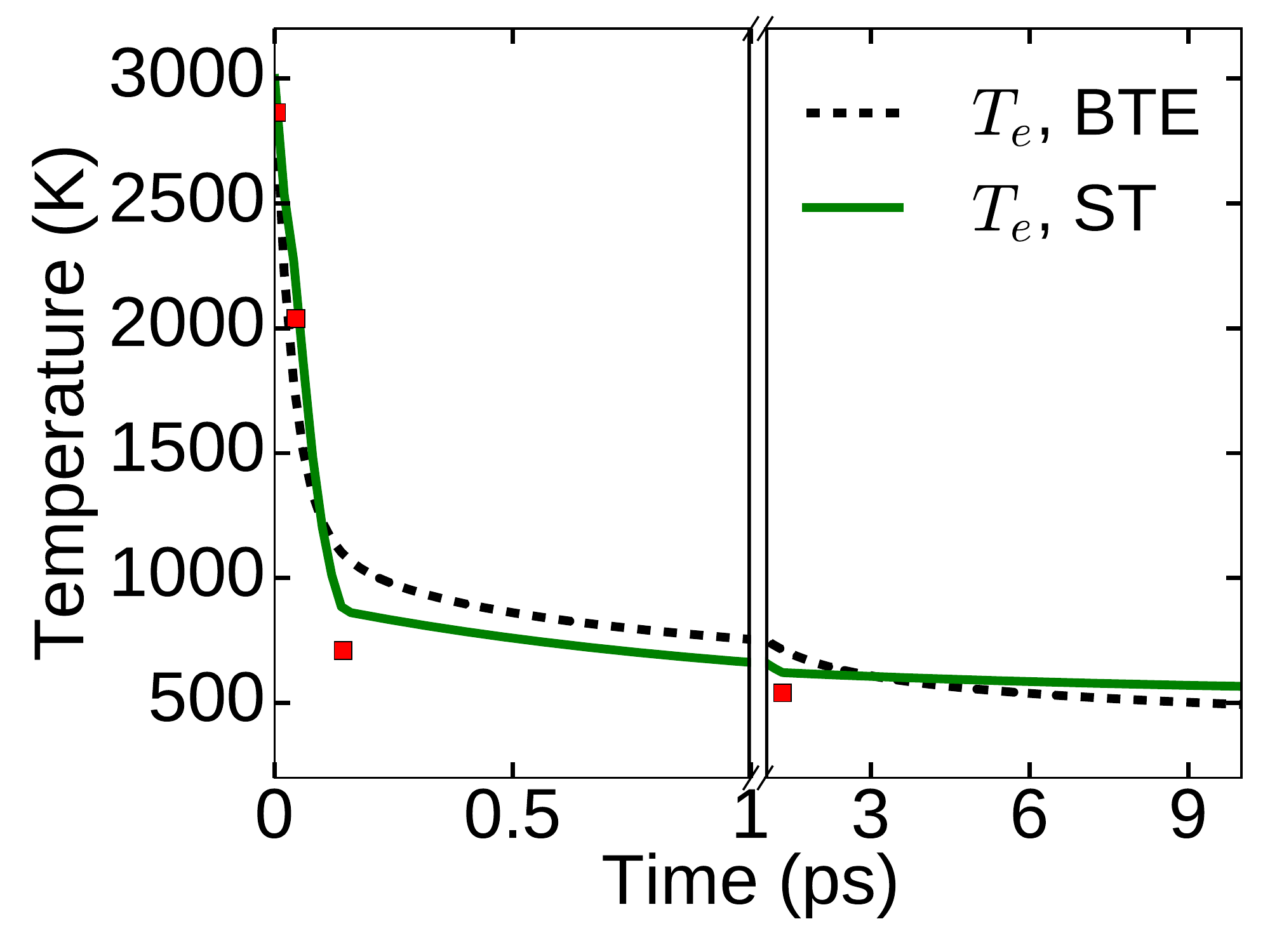}
    \caption{ST-BN ($c=0.1$)}
    \end{subfigure}\qquad\qquad
    \begin{subfigure}[b]{0.4\textwidth}
    \includegraphics[width=60mm]{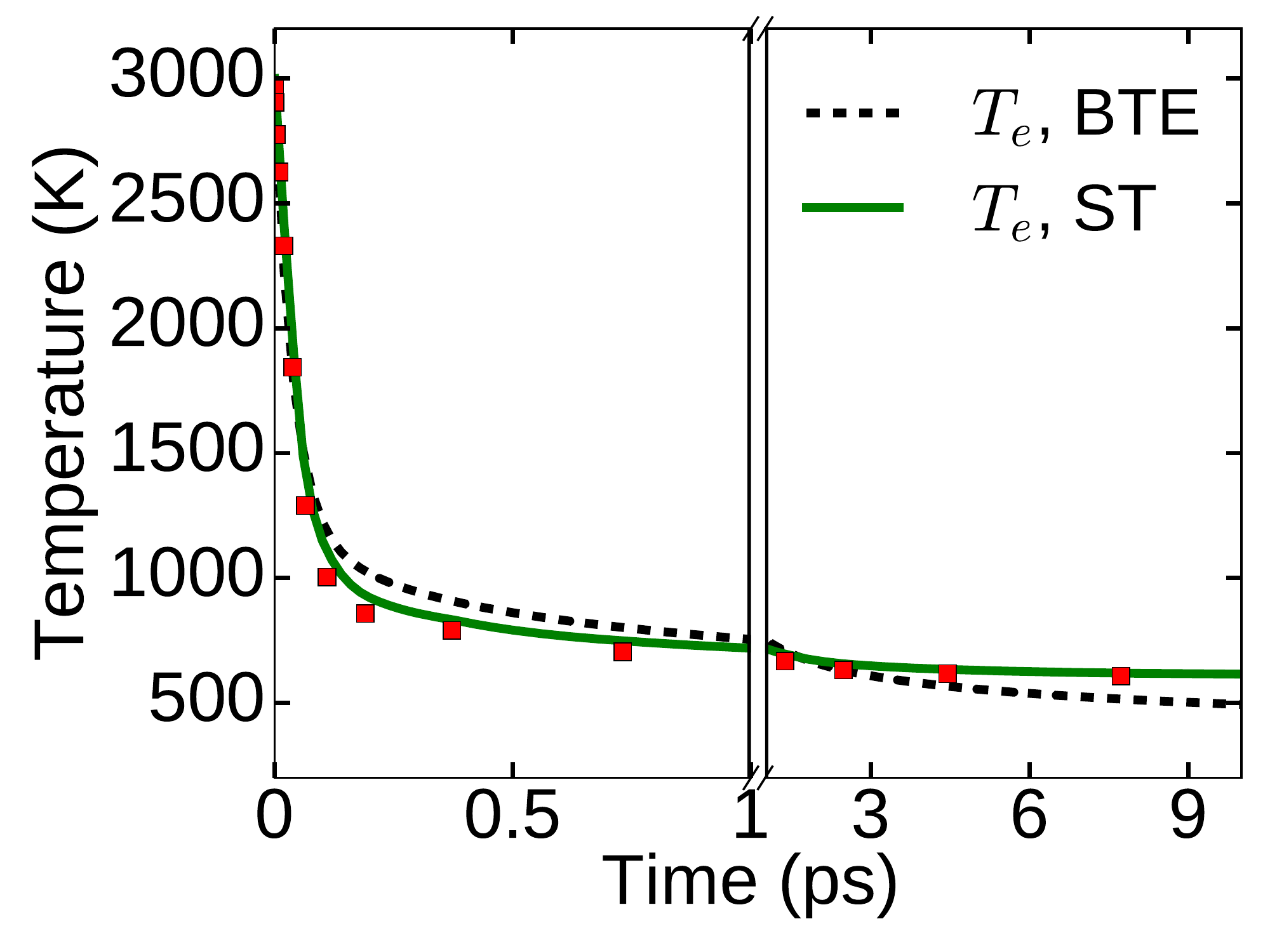}
    \caption{ST-BN ($c=0.5$)}
    \end{subfigure}\\
    \begin{centering}
    \begin{subfigure}[b]{0.4\textwidth}
    \includegraphics[width=60mm]{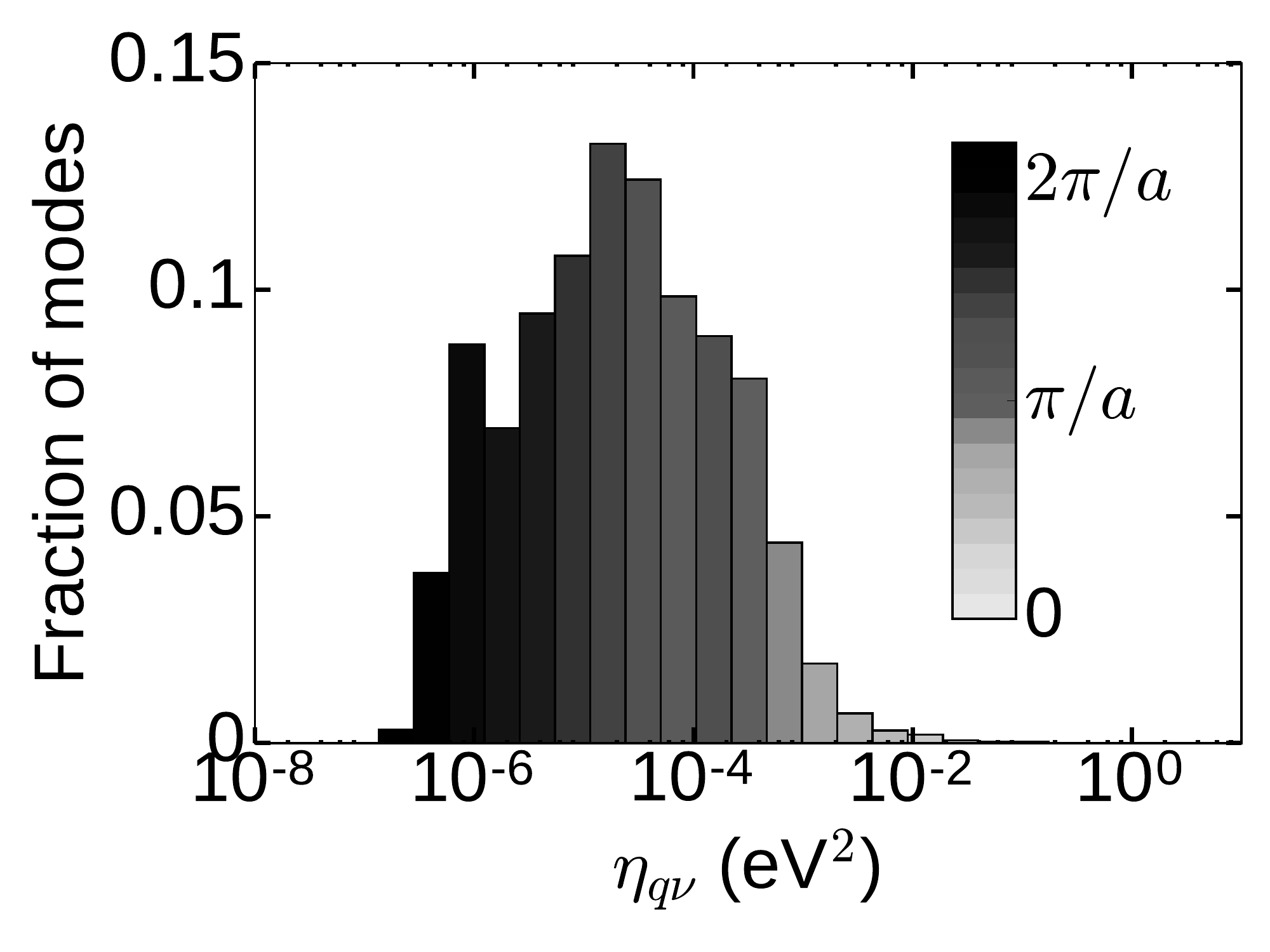}
    \caption{$\eta_{\textbf{q}\nu}$ - BN}
    \end{subfigure}
    \end{centering}
    \caption{Comparison between the electronic temperature decay obtained from a full-BTE solution and the 2T (a), 3T (b), and successive thermalization (c,d) models for BN. e) The distribution of phonon interaction strength $\eta_{\textbf{q}\nu}$ color-coded according to the average wavevector magnitude of phonons in each subset.}
    \label{BN_predictions}
\end{figure}

\newpage
\subsubsection{Boron Phosphide (BP)}

\begin{figure}[h]
    \centering
   \begin{subfigure}[b]{0.4\textwidth}
    \includegraphics[width=60mm]{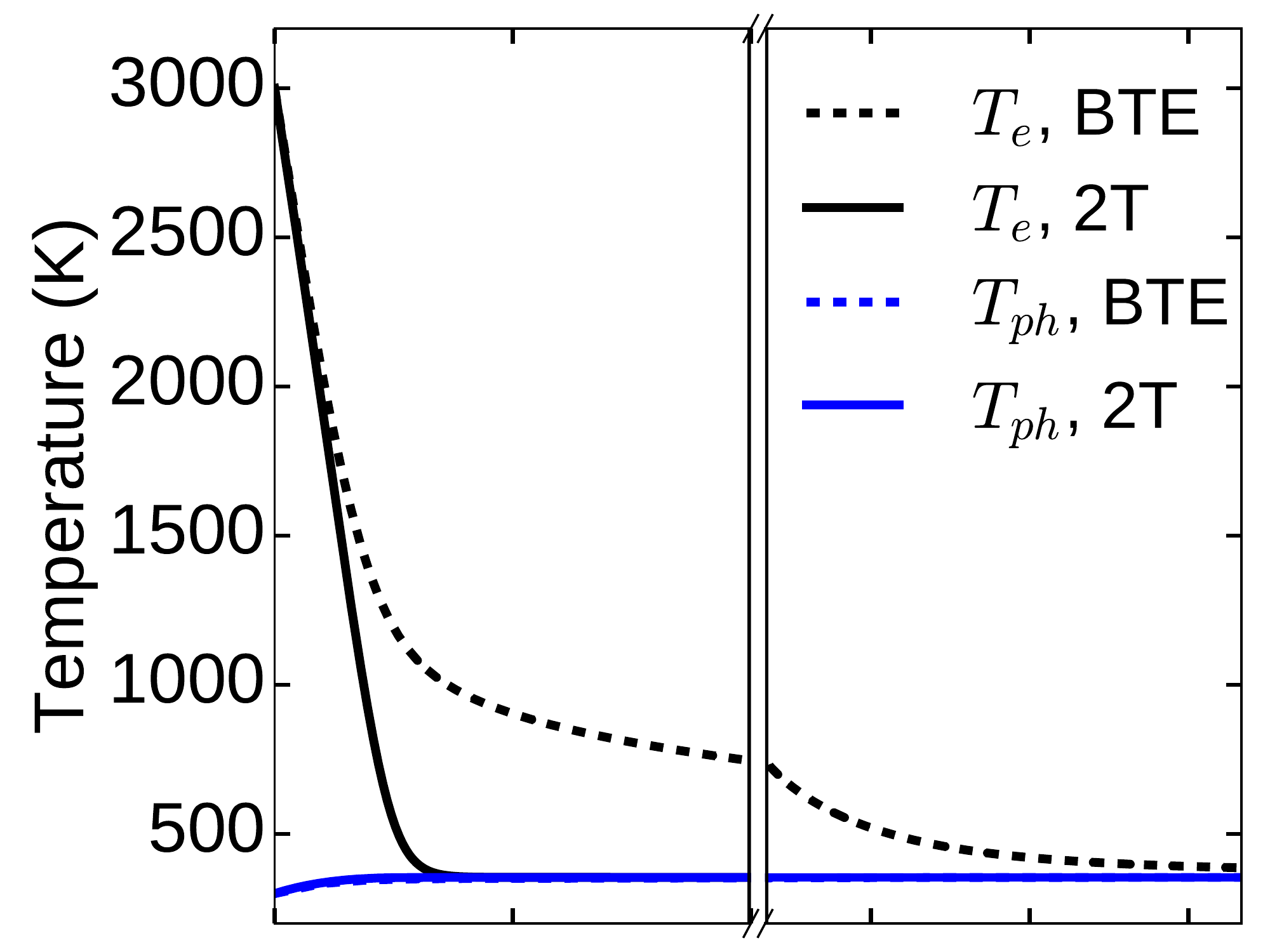}
    \caption{2T-BP}
    \end{subfigure}\qquad\qquad
    \begin{subfigure}[b]{0.4\textwidth}
    \includegraphics[width=60mm]{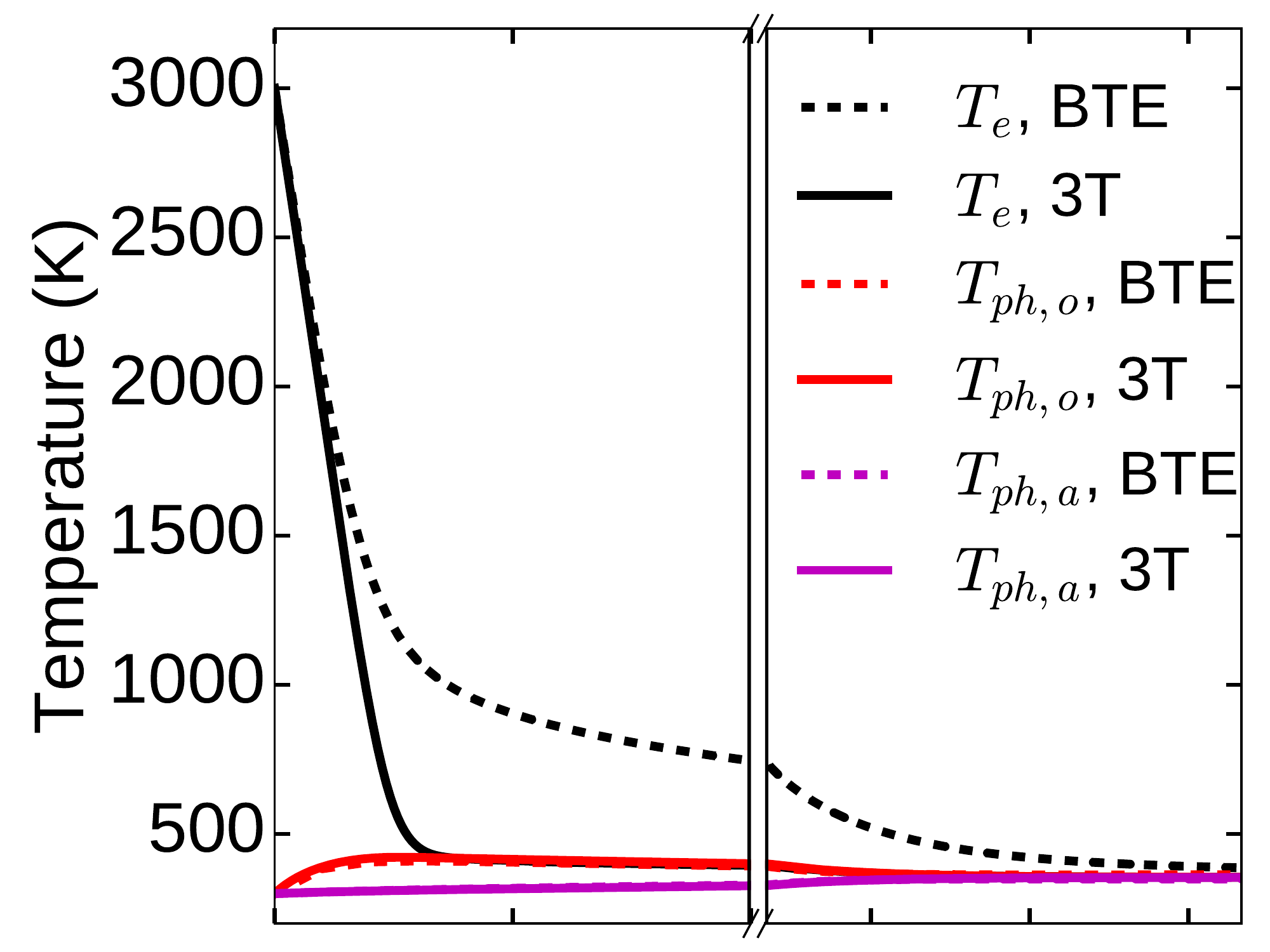}
    \caption{3T-BP}
    \end{subfigure}\\
    \begin{subfigure}[b]{0.4\textwidth}
    \includegraphics[width=60mm]{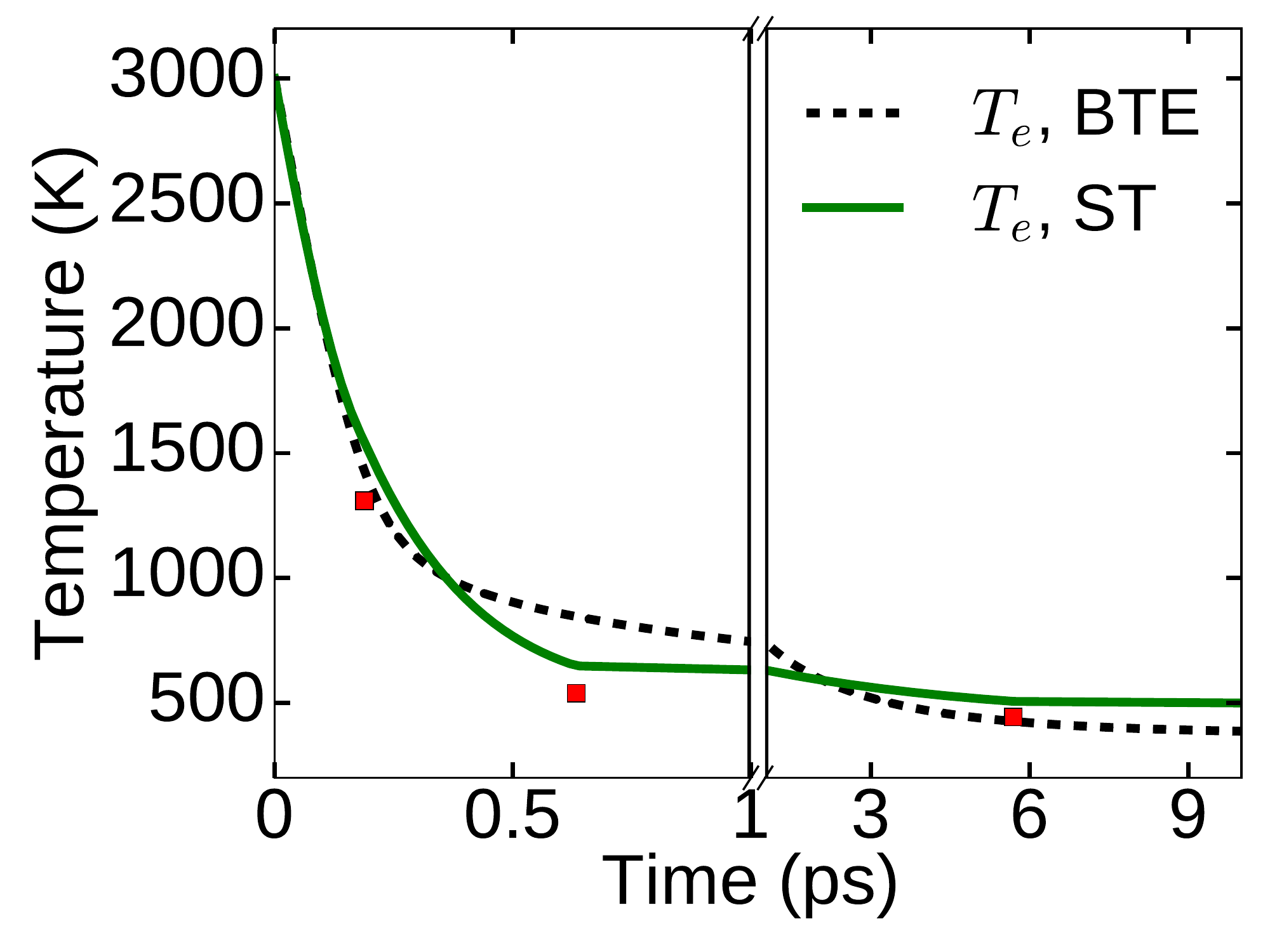}
    \caption{ST-BP ($c=0.1$)}
    \end{subfigure}\qquad\qquad
    \begin{subfigure}[b]{0.4\textwidth}
    \includegraphics[width=60mm]{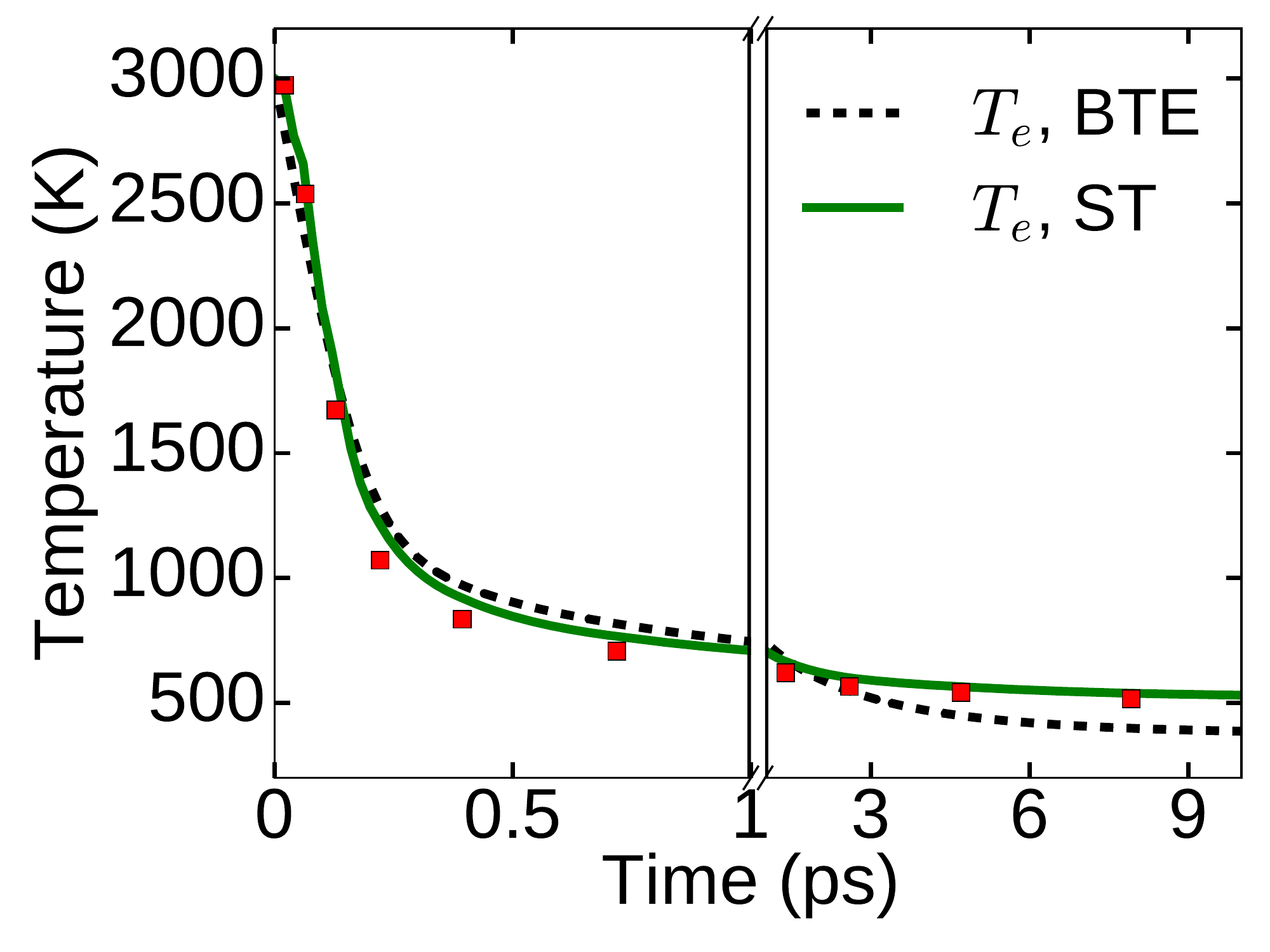}
    \caption{ST-BP ($c=0.5$)}
    \end{subfigure}\\
    \begin{centering}
    \begin{subfigure}[b]{0.4\textwidth}
    \includegraphics[width=60mm]{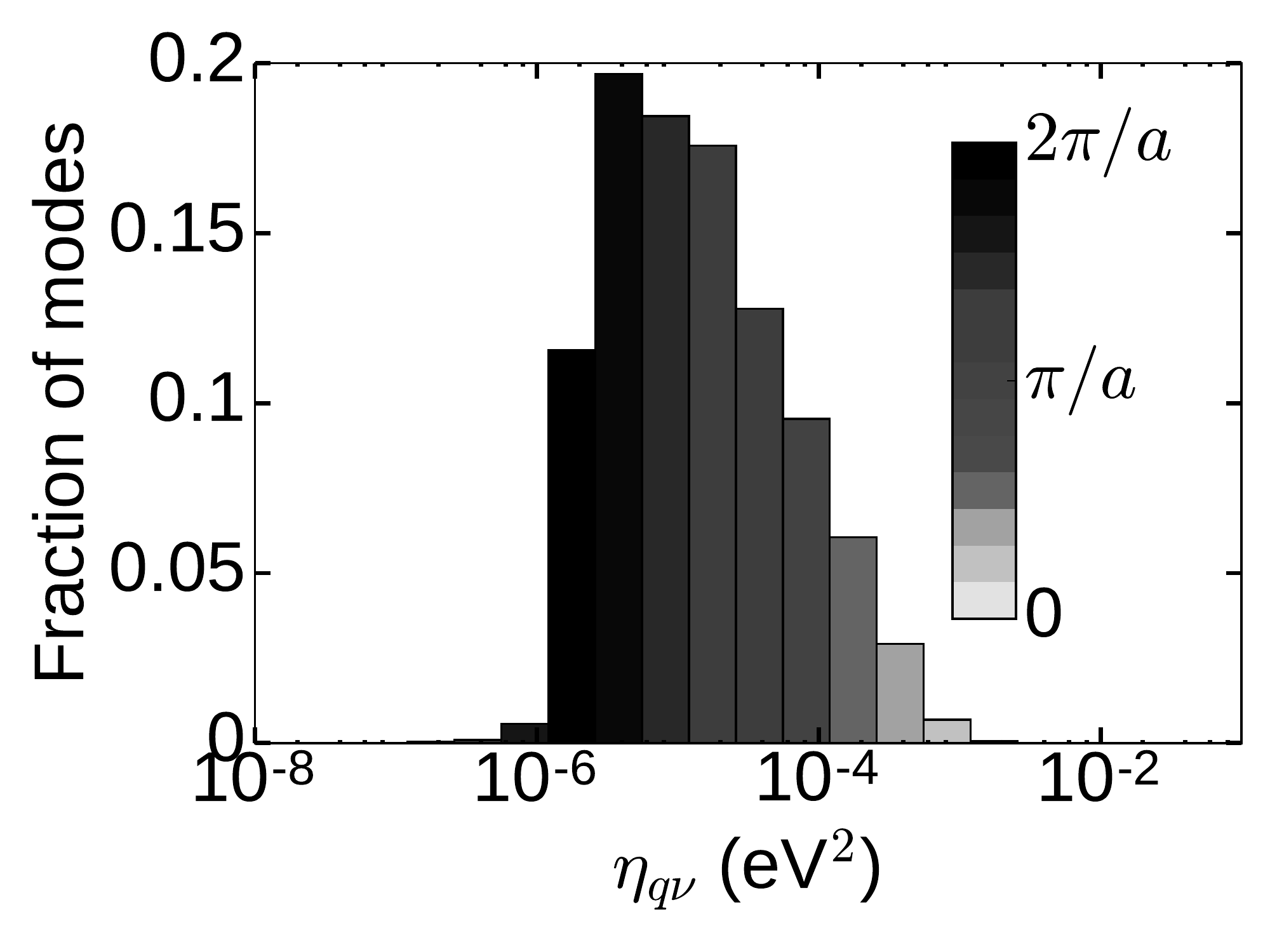}
    \caption{$\eta_{\textbf{q}\nu}$ - BP}
    \end{subfigure}
    \end{centering}
    \caption{Comparison between the electronic temperature decay obtained from a full-BTE solution and the 2T (a), 3T (b), and successive thermalization (c,d) models for BP. e) The distribution of phonon interaction strength $\eta_{\textbf{q}\nu}$ color-coded according to the average wavevector magnitude of phonons in each subset.}
\end{figure}

\newpage
\subsubsection{Boron Arsenide (BAs)}

\begin{figure}[h]
    \centering
   \begin{subfigure}[b]{0.4\textwidth}
    \includegraphics[width=60mm]{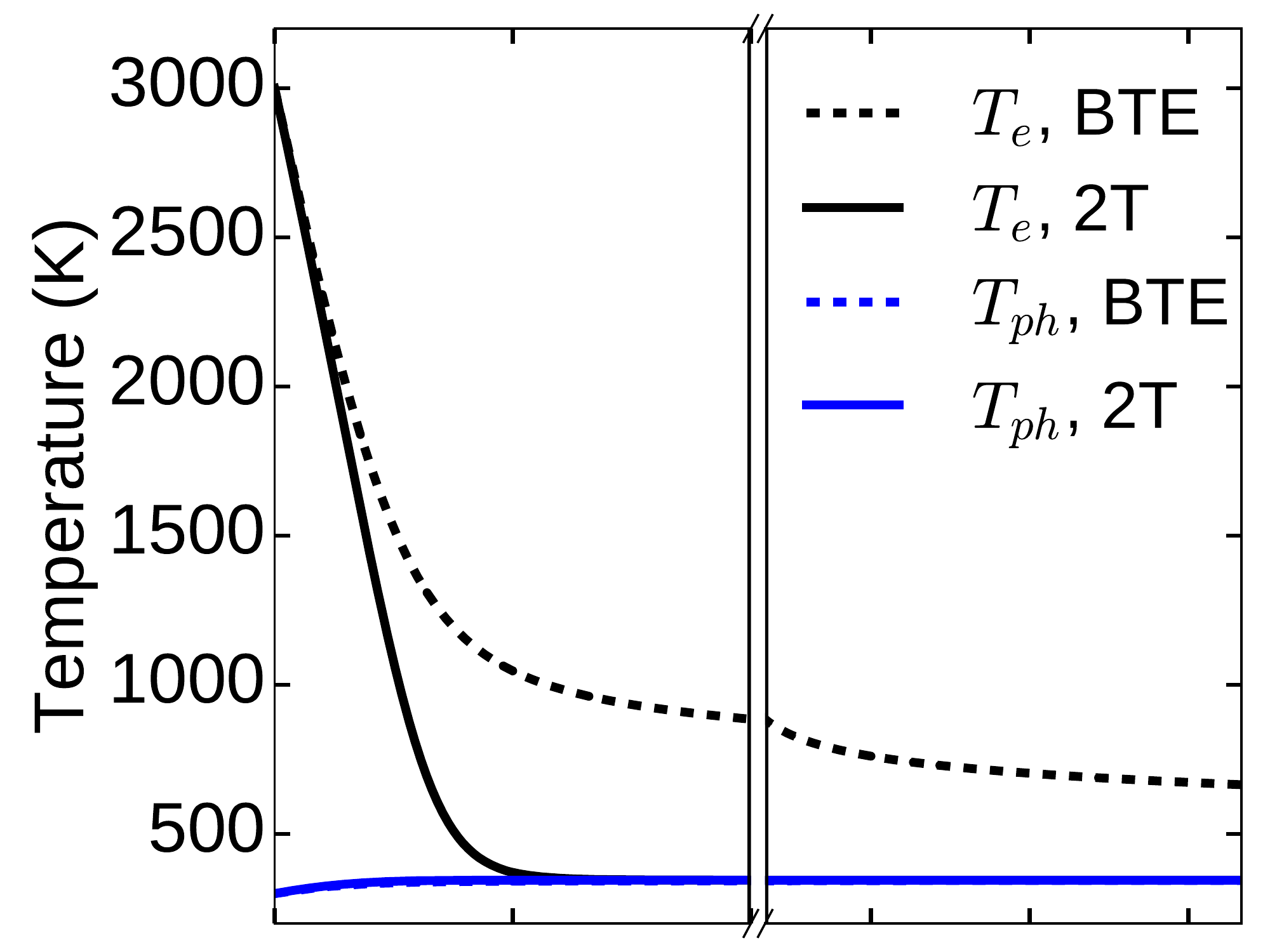}
    \caption{2T-BAs}
    \end{subfigure}\qquad\qquad
    \begin{subfigure}[b]{0.4\textwidth}
    \includegraphics[width=60mm]{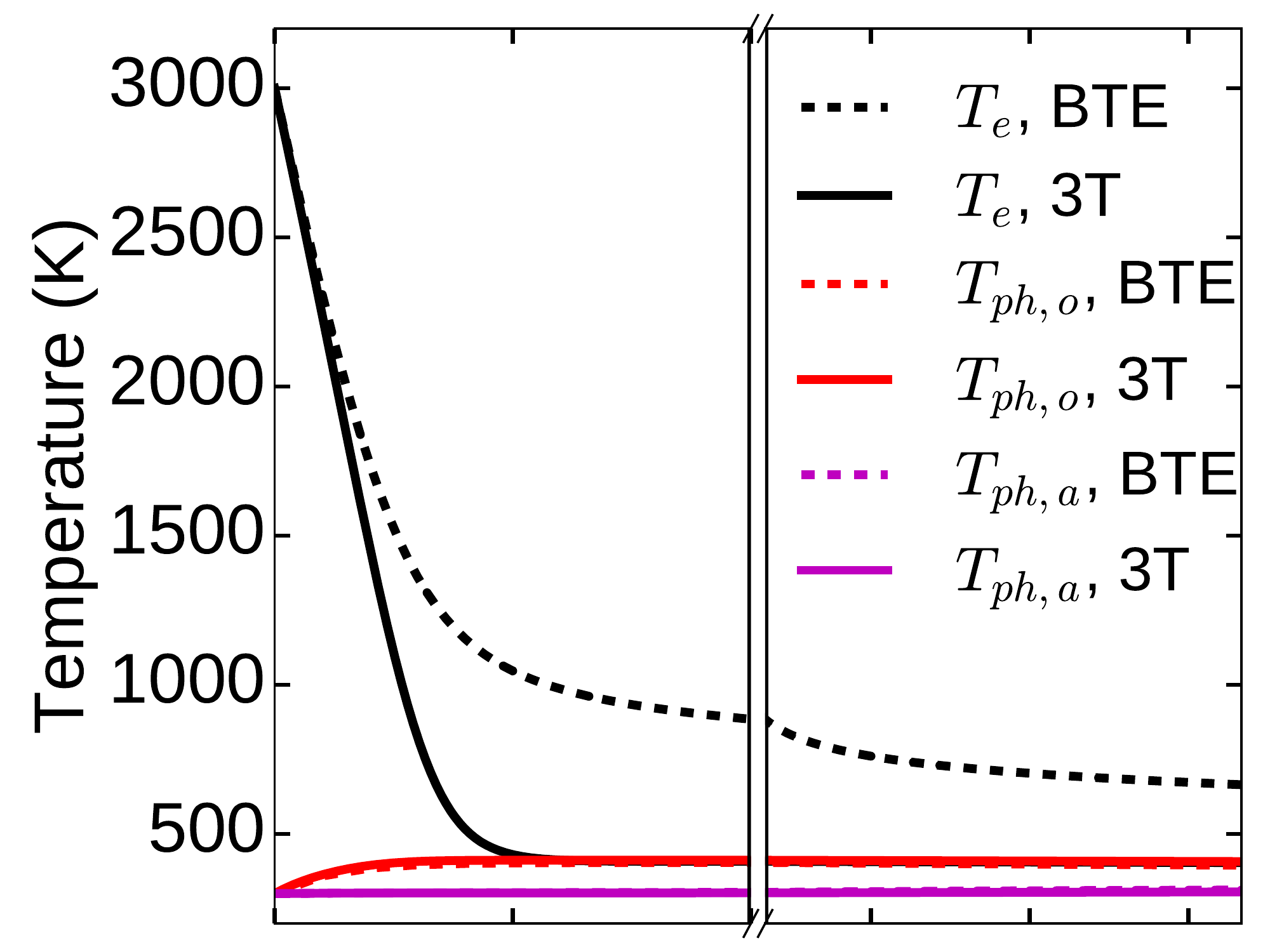}
    \caption{3T-BAs}
    \end{subfigure}\\
    \begin{subfigure}[b]{0.4\textwidth}
    \includegraphics[width=60mm]{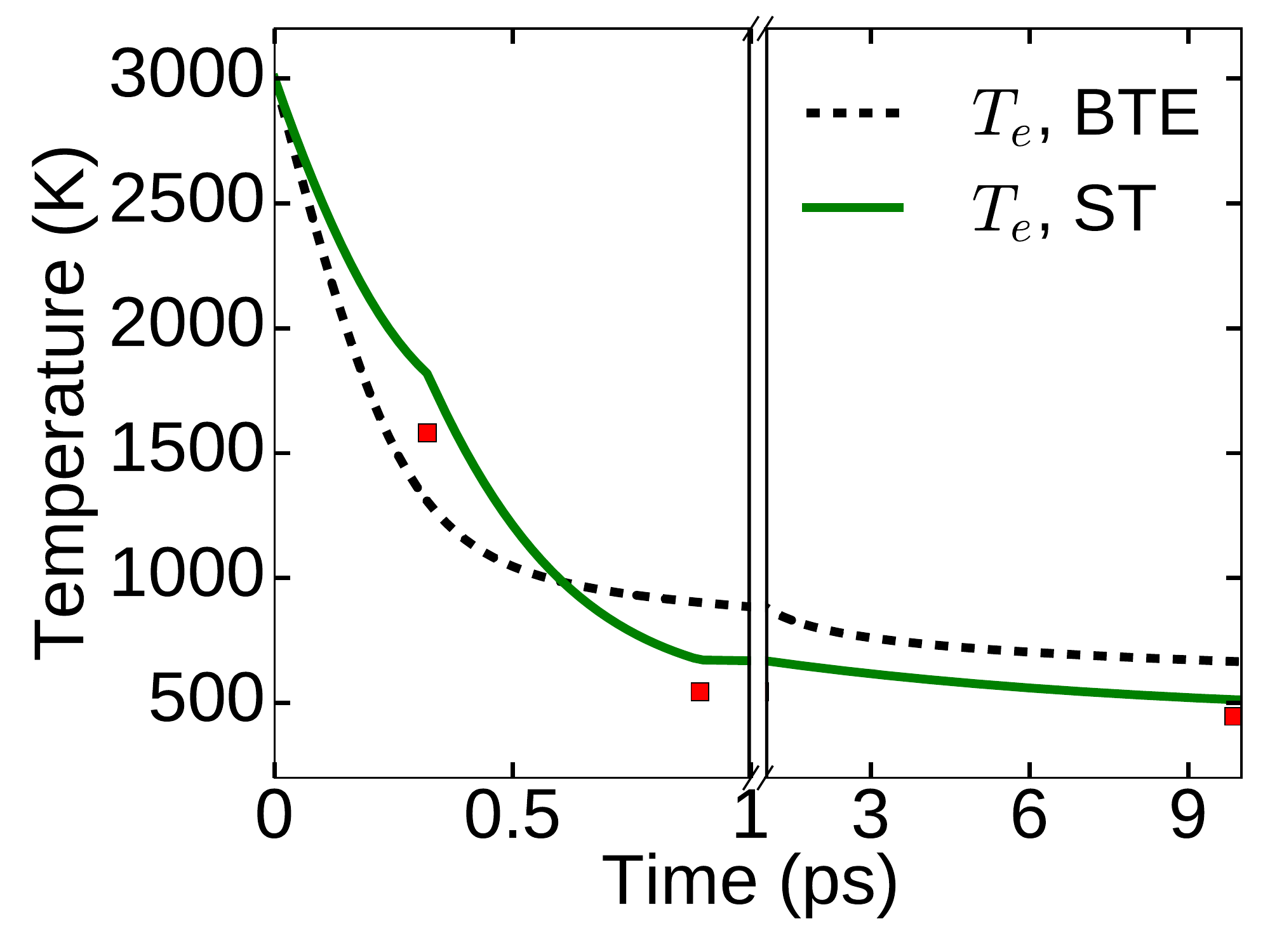}
    \caption{ST-BAs ($c=0.1$)}
    \end{subfigure}\qquad\qquad
    \begin{subfigure}[b]{0.4\textwidth}
    \includegraphics[width=60mm]{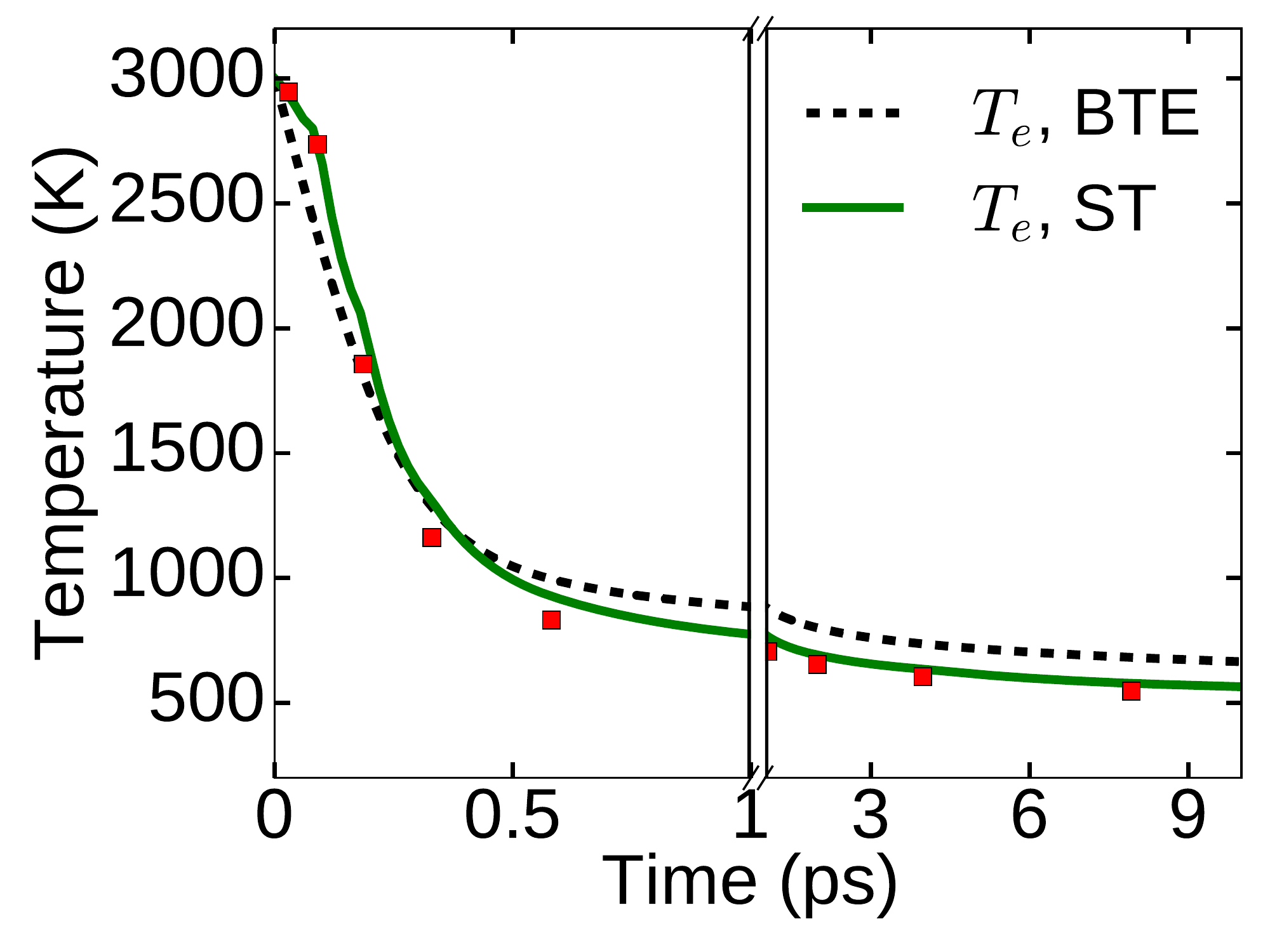}
    \caption{ST-BAs ($c=0.5$)}
    \end{subfigure}\\
    \begin{centering}
    \begin{subfigure}[b]{0.4\textwidth}
    \includegraphics[width=60mm]{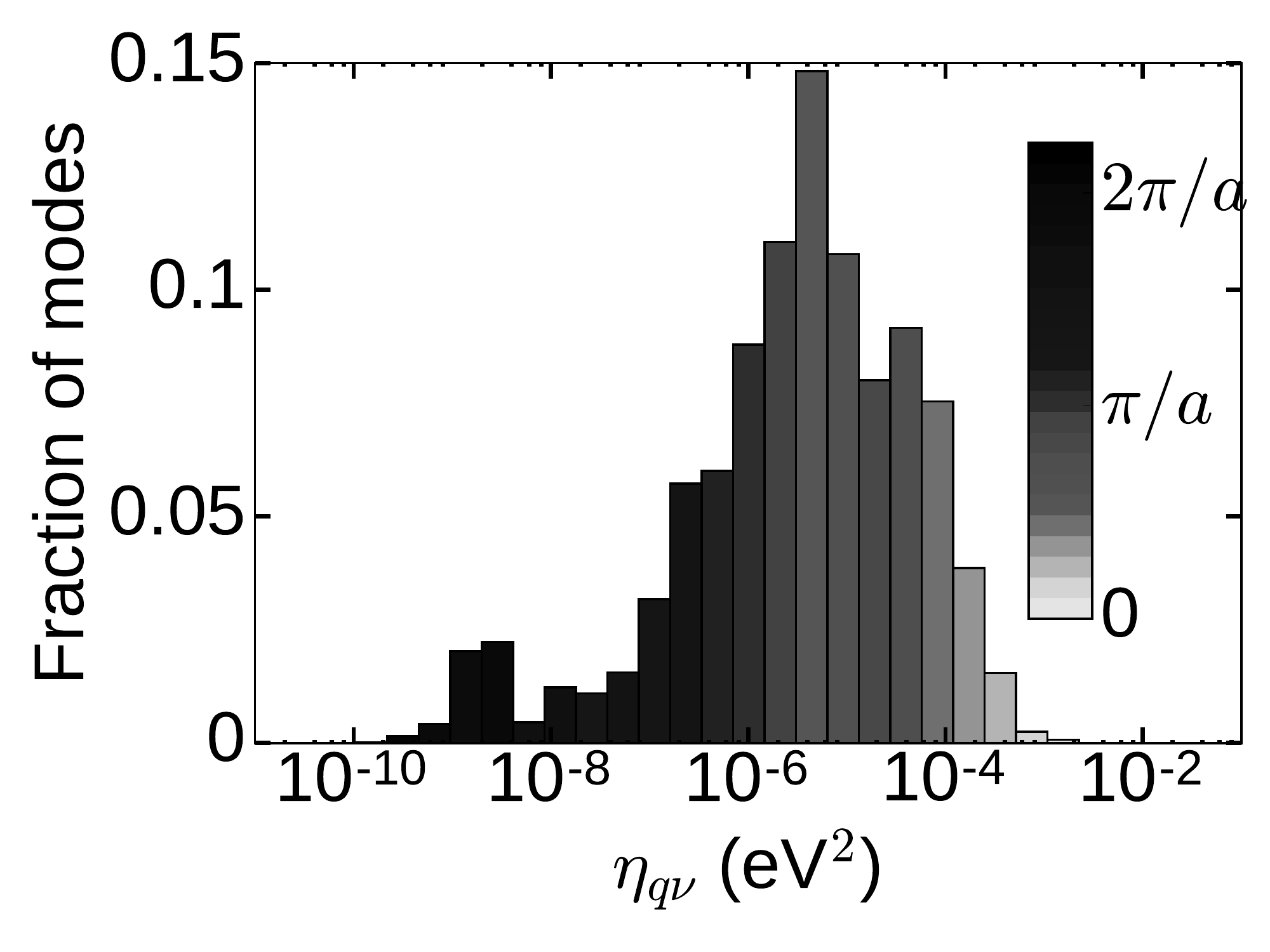}
    \caption{$\eta_{\textbf{q}\nu}$ - BAs}
    \end{subfigure}
    \end{centering}
    \caption{Comparison between the electronic temperature decay obtained from a full-BTE solution and the 2T (a), 3T (b), and successive thermalization (c,d) models for BAs. e) The distribution of phonon interaction strength $\eta_{\textbf{q}\nu}$ color-coded according to the average wavevector magnitude of phonons in each subset.}
\end{figure}

\newpage
\subsubsection{Boron Antimonide (BSb)}

\begin{figure}[h]
    \centering
   \begin{subfigure}[b]{0.4\textwidth}
    \includegraphics[width=60mm]{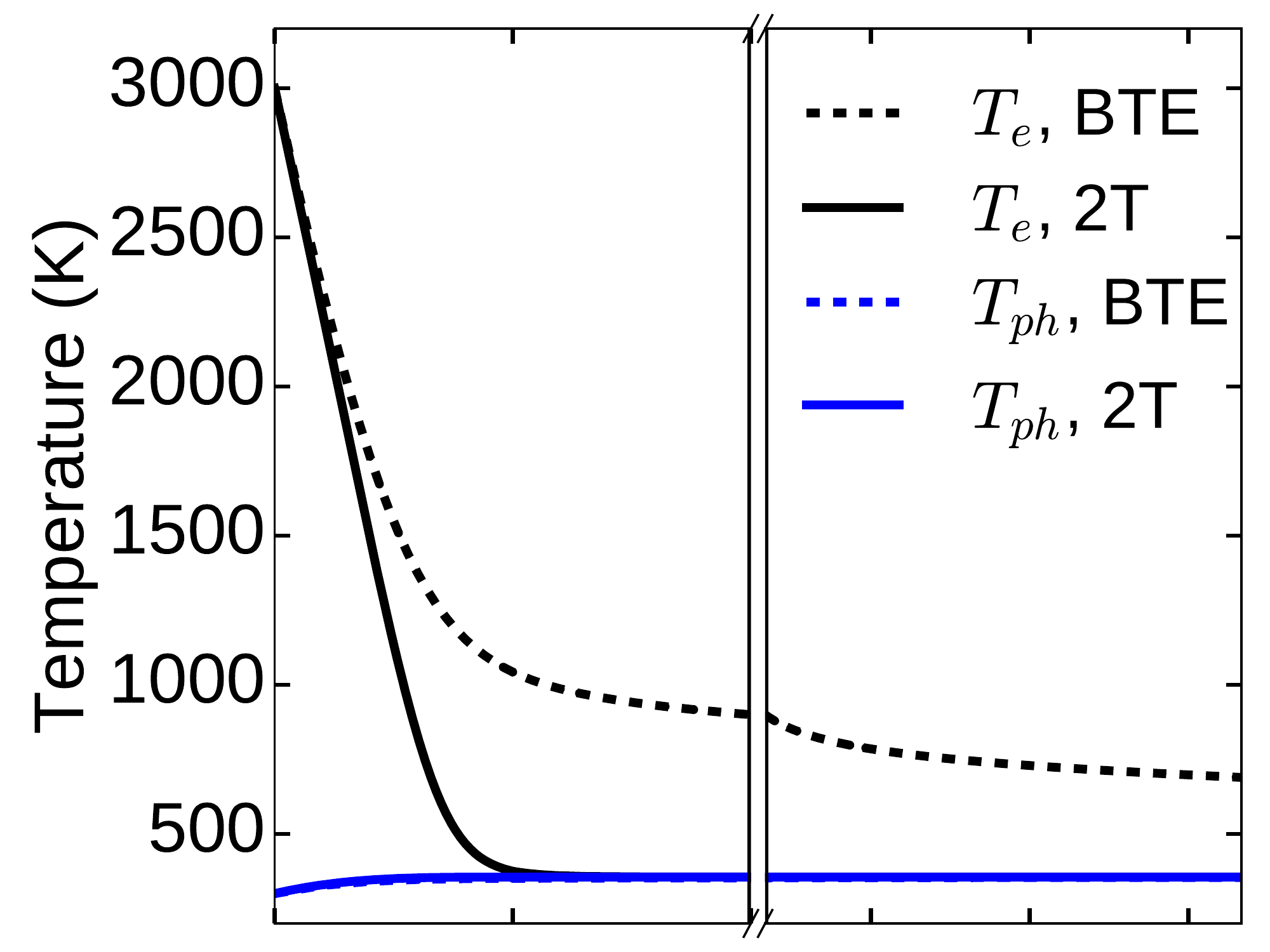}
    \caption{2T-BSb}
    \end{subfigure}\qquad\qquad
    \begin{subfigure}[b]{0.4\textwidth}
    \includegraphics[width=60mm]{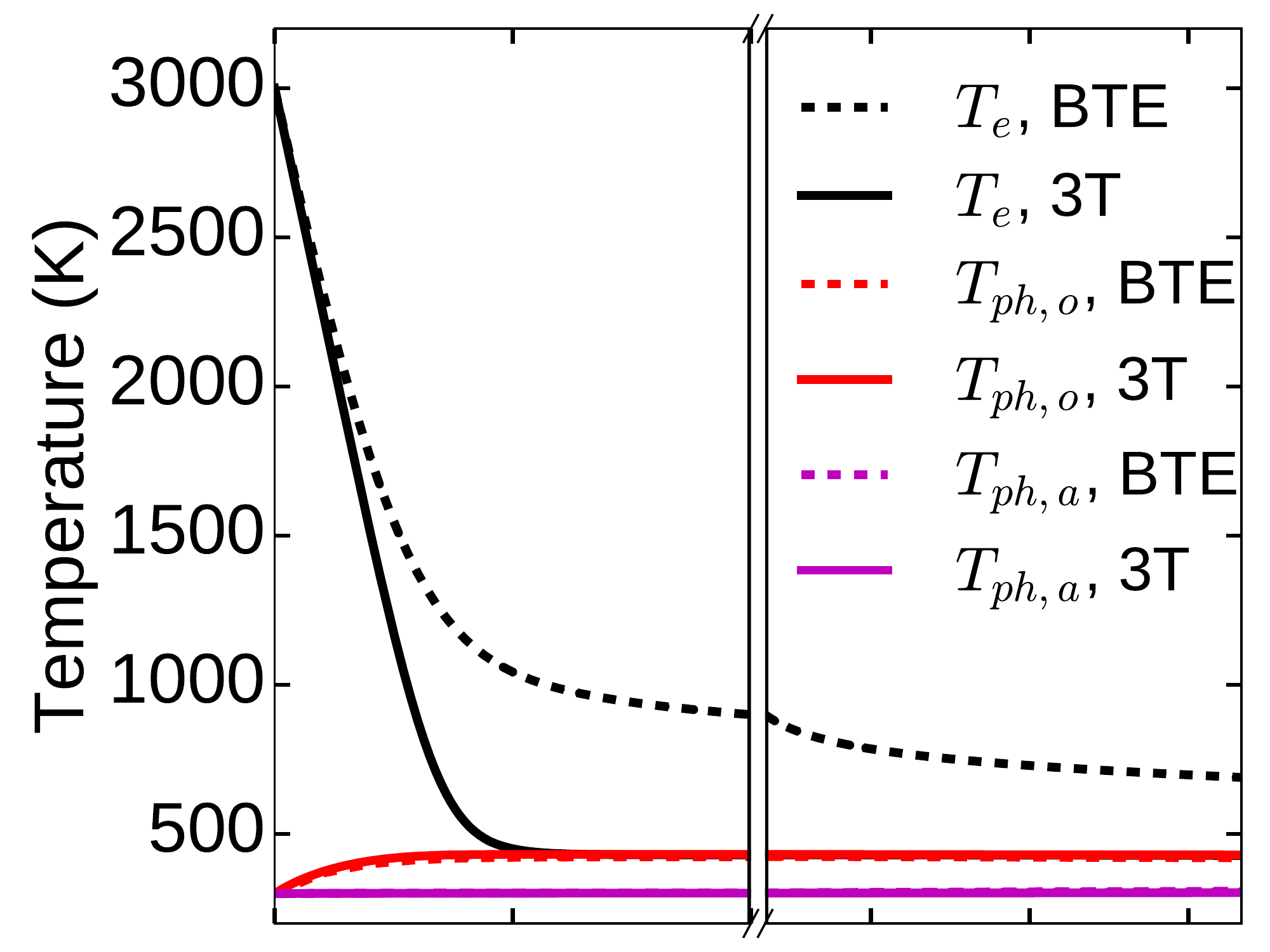}
    \caption{3T-BSb}
    \end{subfigure}\\
    \begin{subfigure}[b]{0.4\textwidth}
    \includegraphics[width=60mm]{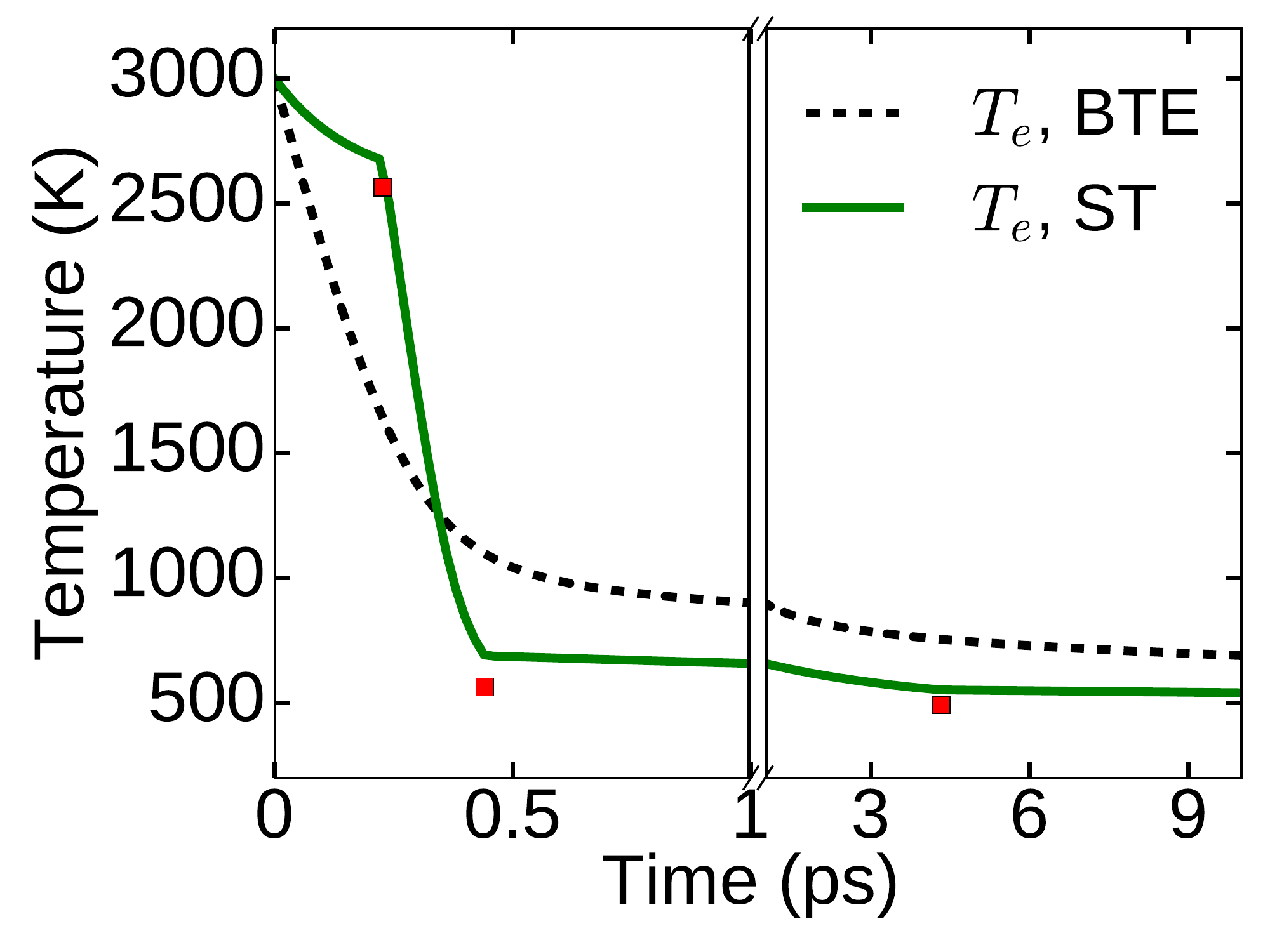}
    \caption{ST-BSb ($c=0.1$)}
    \end{subfigure}\qquad\qquad
    \begin{subfigure}[b]{0.4\textwidth}
    \includegraphics[width=60mm]{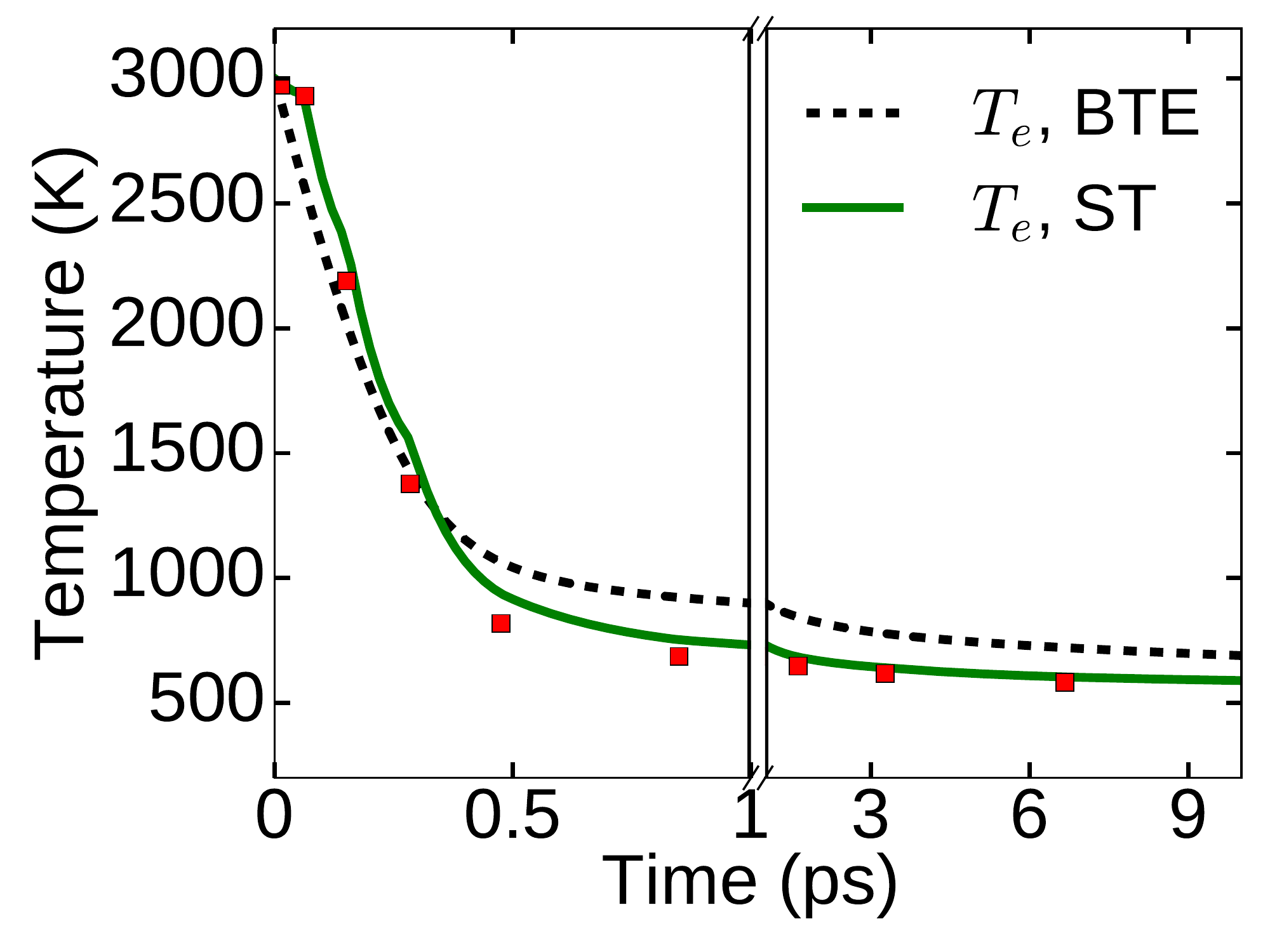}
    \caption{ST-BSb ($c=0.5$)}
    \end{subfigure}\\
    \begin{centering}
    \begin{subfigure}[b]{0.4\textwidth}
    \includegraphics[width=60mm]{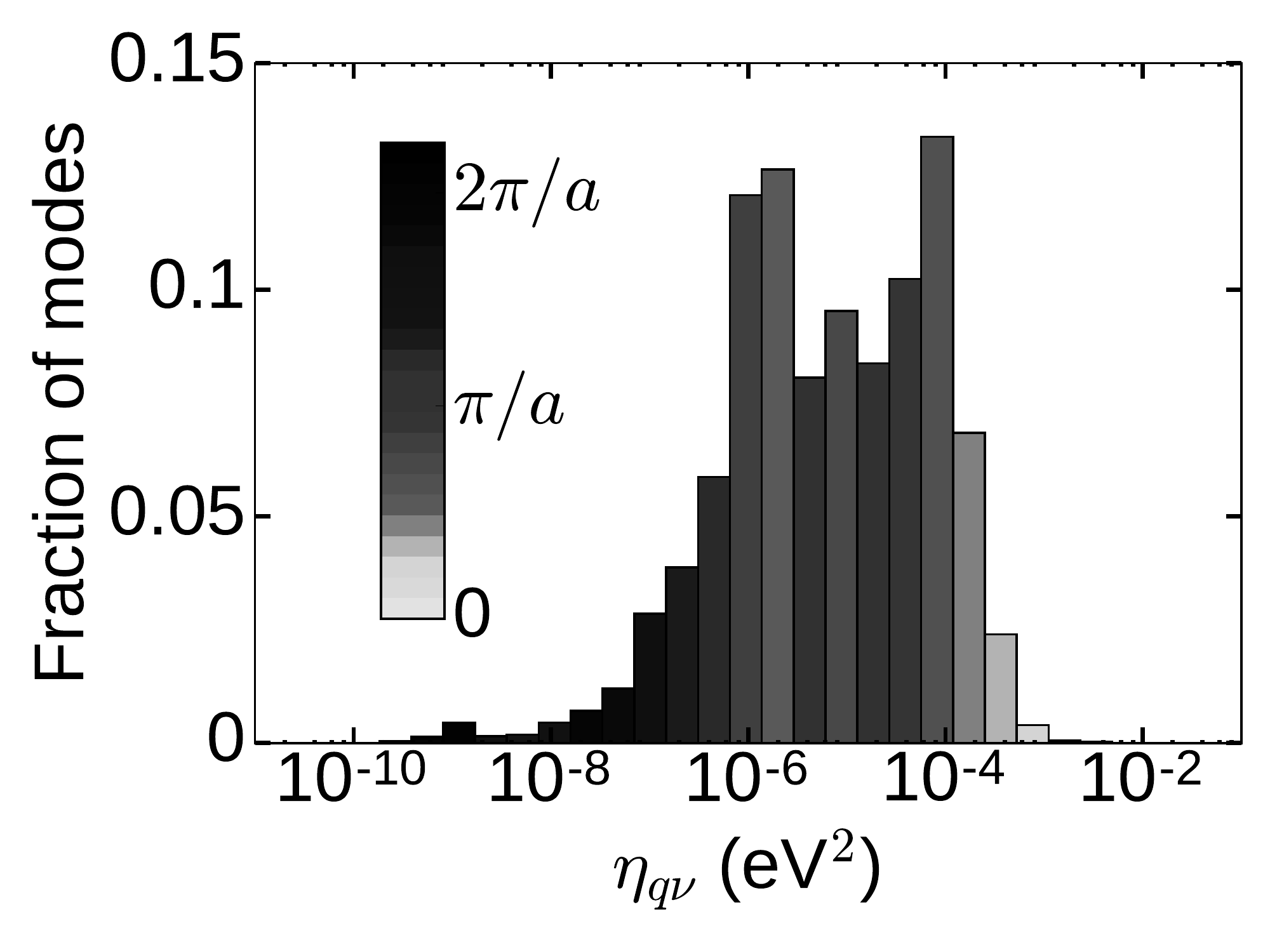}
    \caption{$\eta_{\textbf{q}\nu}$ - BSb}
    \end{subfigure}
    \end{centering}
    \caption{Comparison between the electronic temperature decay obtained from a full-BTE solution and the 2T (a), 3T (b), and successive thermalization (c,d) models for BSb. e) The distribution of phonon interaction strength $\eta_{\textbf{q}\nu}$ color-coded according to the average wavevector magnitude of phonons in each subset.}
\end{figure}

\newpage

\subsubsection{Aluminum Phosphide (AlP)}

\begin{figure}[h]
    \centering
   \begin{subfigure}[b]{0.4\textwidth}
    \includegraphics[width=60mm]{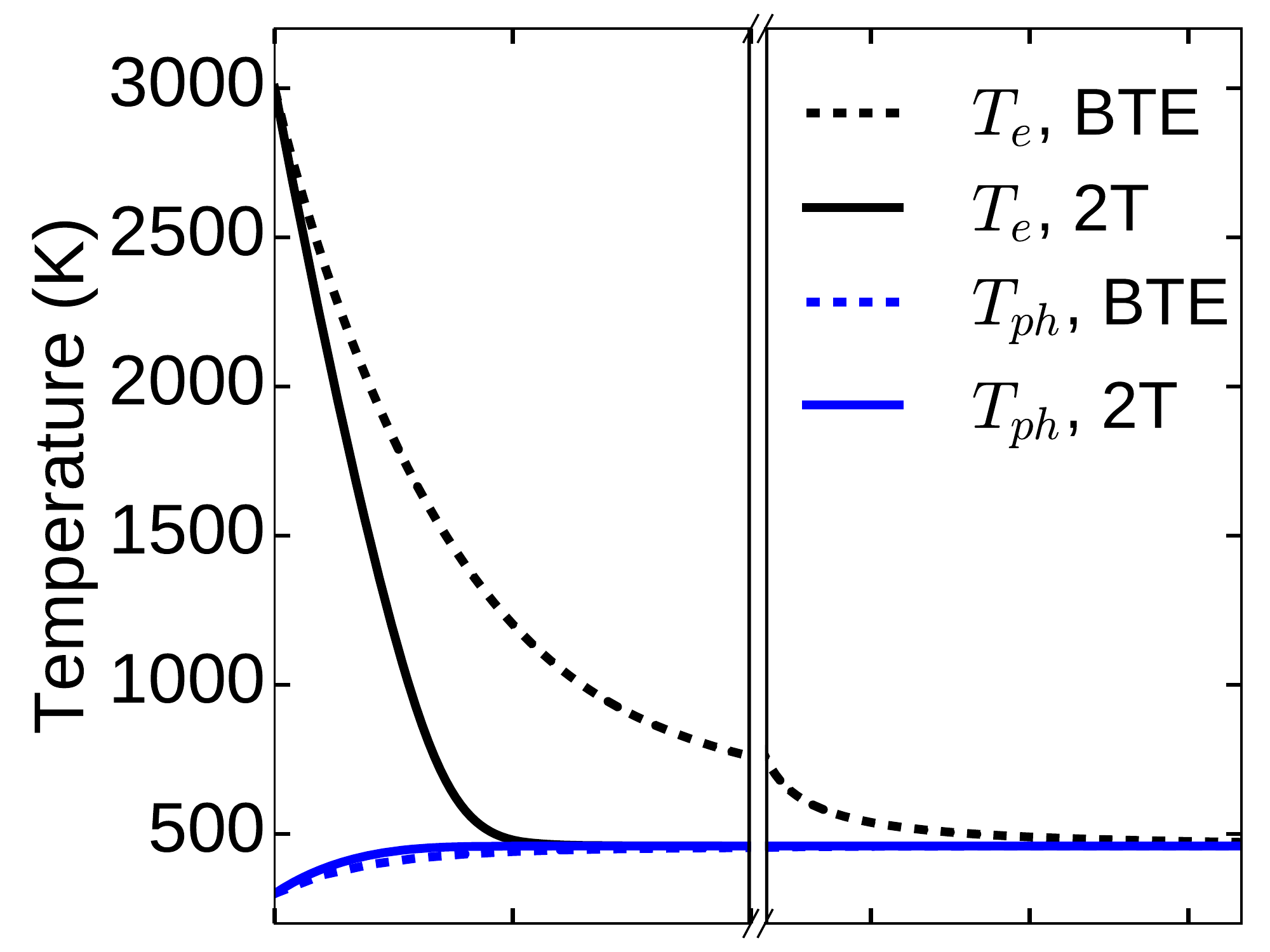}
    \caption{2T-AlP}
    \end{subfigure}\qquad\qquad
    \begin{subfigure}[b]{0.4\textwidth}
    \includegraphics[width=60mm]{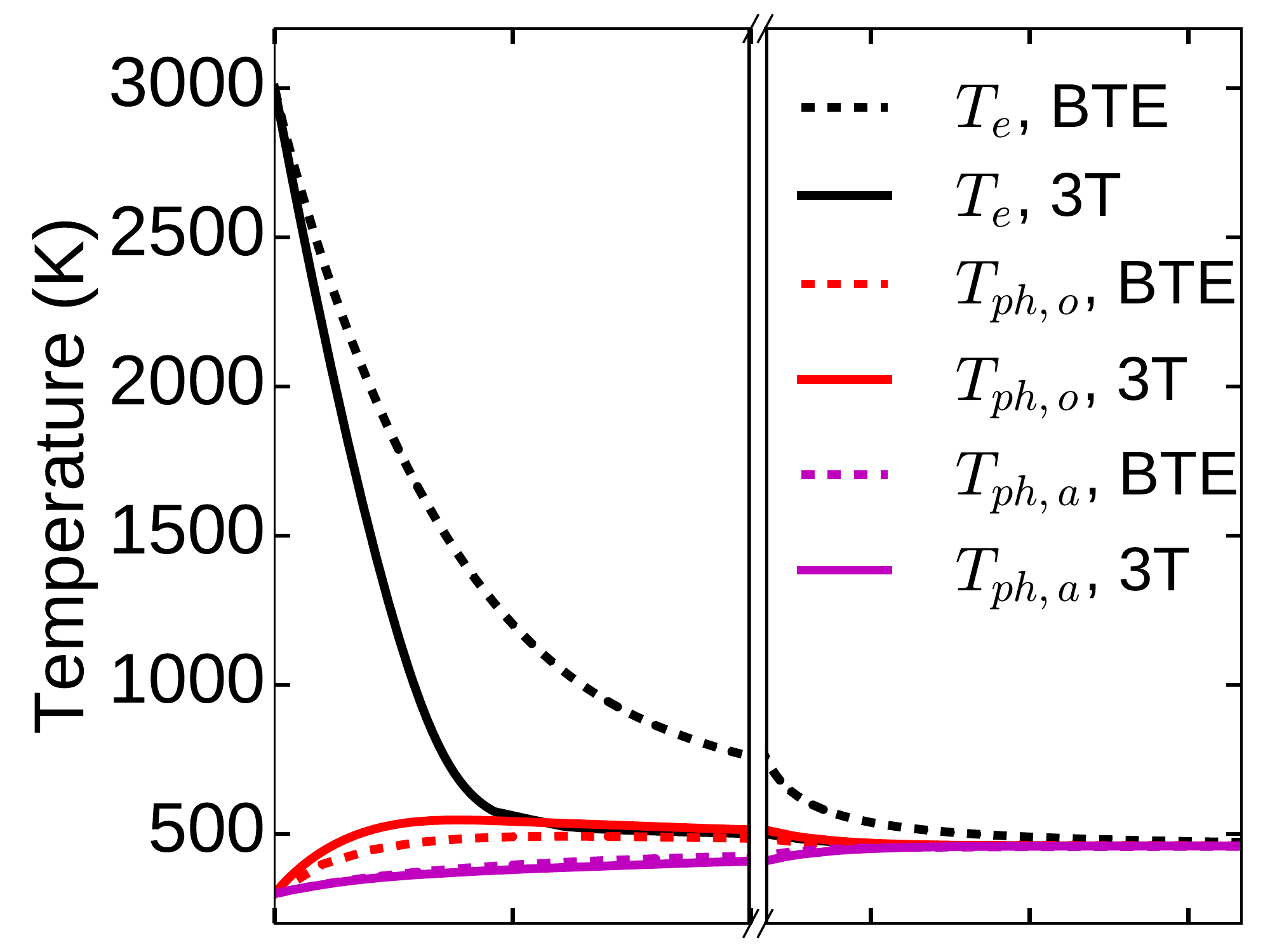}
    \caption{3T-AlP}
    \end{subfigure}\\
    \begin{subfigure}[b]{0.4\textwidth}
    \includegraphics[width=60mm]{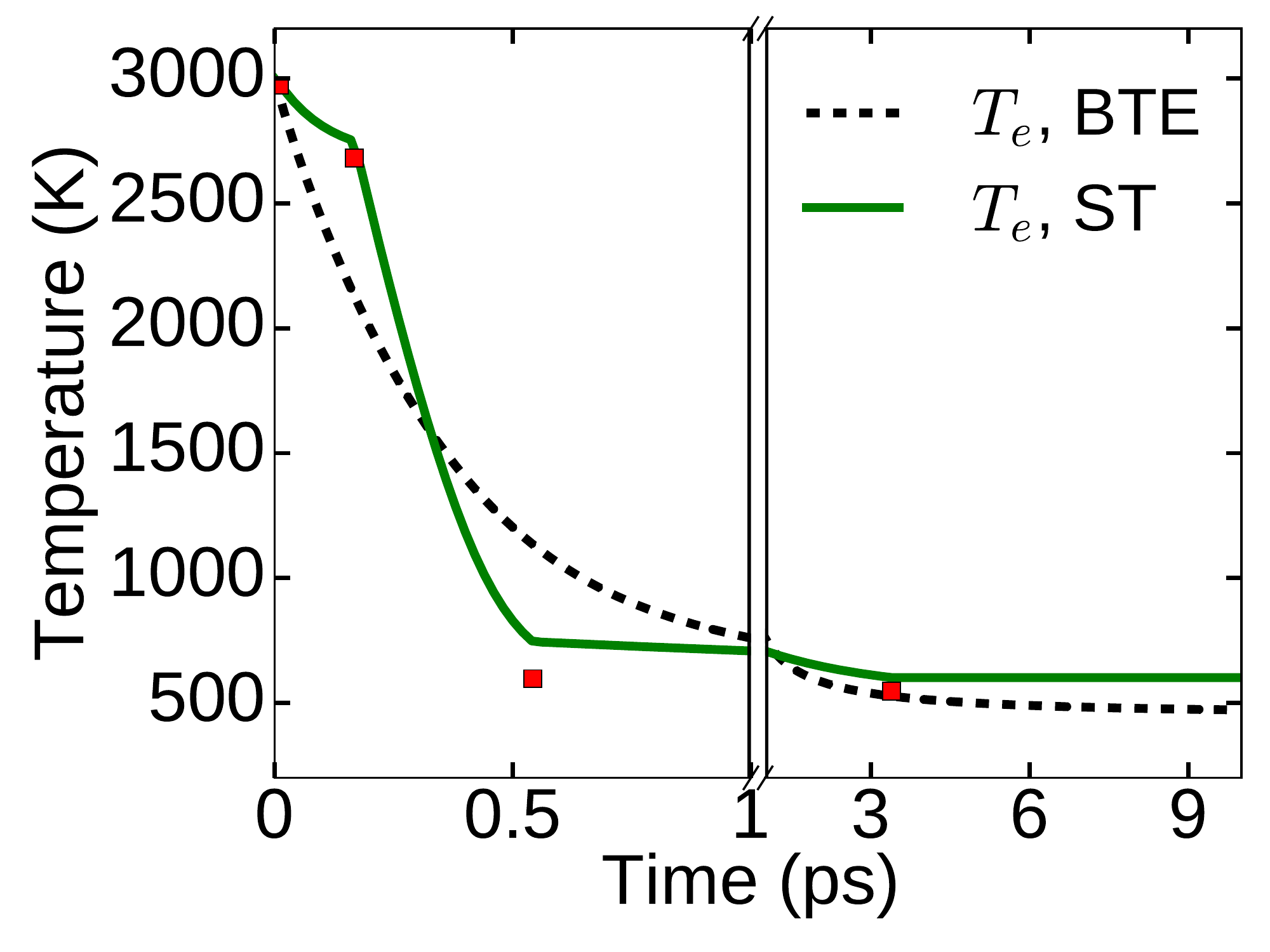}
    \caption{ST-AlP ($c=0.1$)}
    \end{subfigure}\qquad\qquad
    \begin{subfigure}[b]{0.4\textwidth}
    \includegraphics[width=60mm]{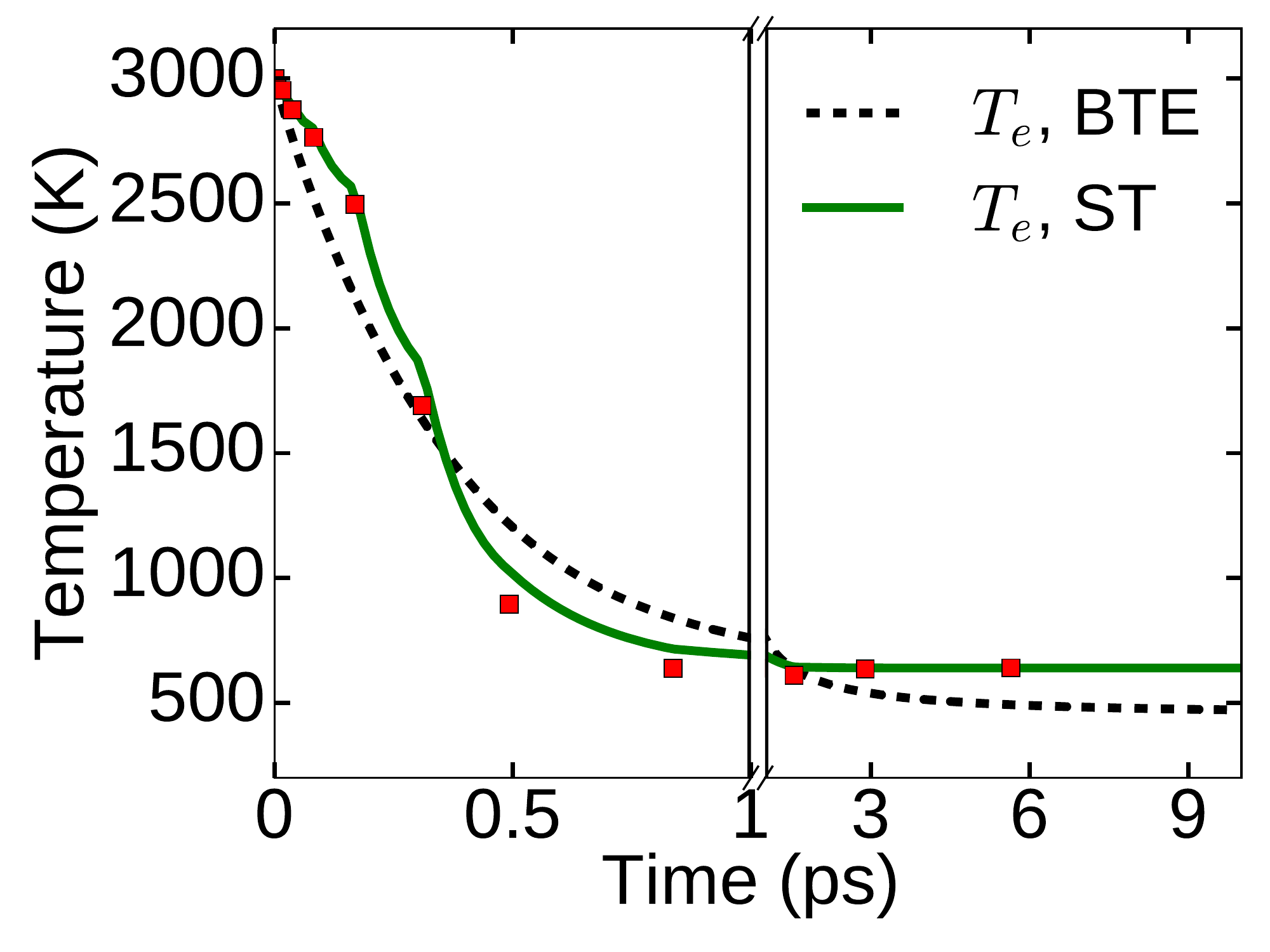}
    \caption{ST-AlP ($c=0.5$)}
    \end{subfigure}\\
    \begin{centering}
    \begin{subfigure}[b]{0.4\textwidth}
    \includegraphics[width=60mm]{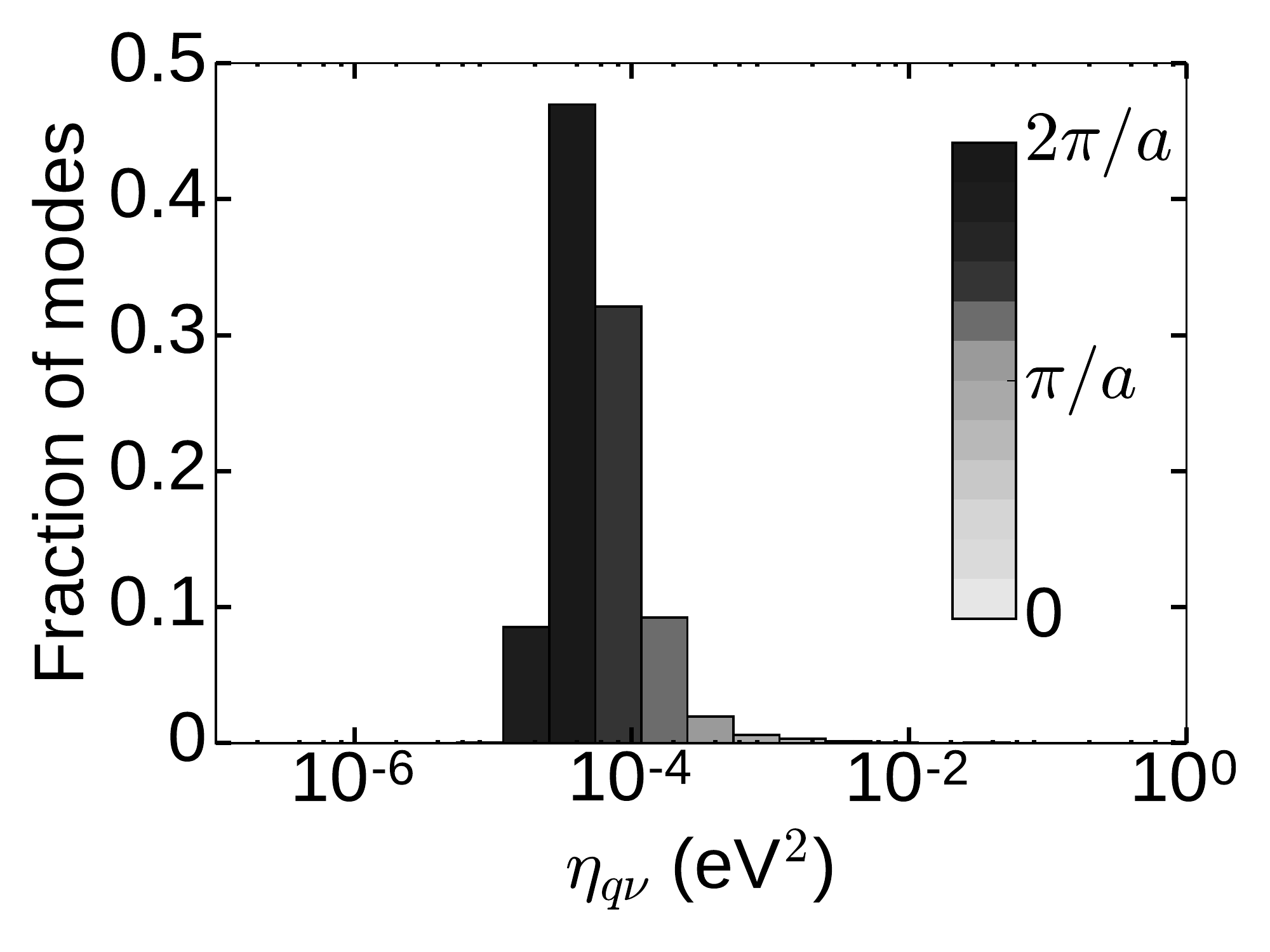}
    \caption{$\eta_{\textbf{q}\nu}$ - AlP}
    \end{subfigure}
    \end{centering}
    \caption{Comparison between the electronic temperature decay obtained from a full-BTE solution and the 2T (a), 3T (b), and successive thermalization (c,d) models for AlP. e) The distribution of phonon interaction strength $\eta_{\textbf{q}\nu}$ color-coded according to the average wavevector magnitude of phonons in each subset.}
\end{figure}

\newpage
\subsubsection{Aluminum Arsenide (AlAs)}

\begin{figure}[h]
    \centering
   \begin{subfigure}[b]{0.4\textwidth}
    \includegraphics[width=60mm]{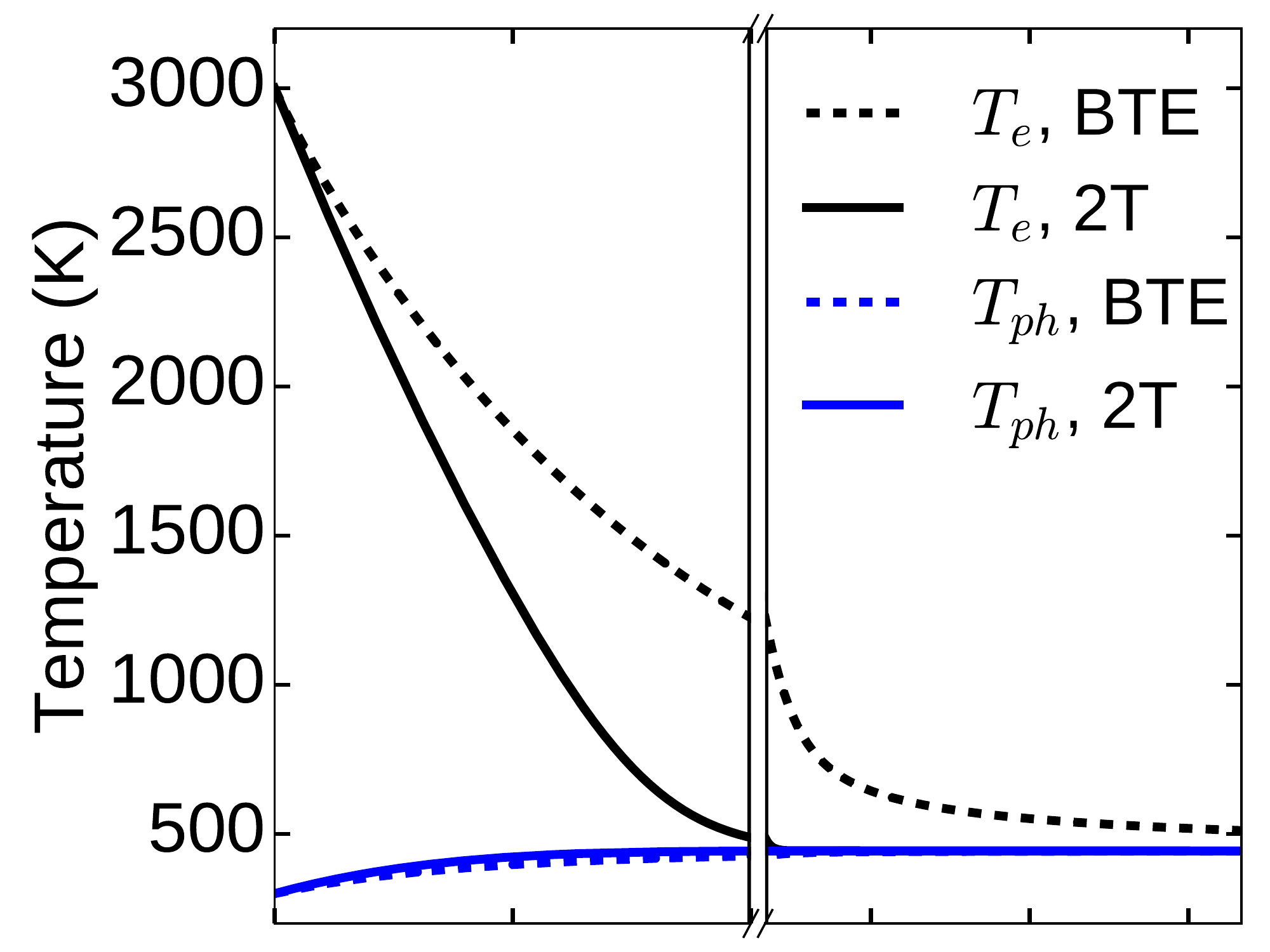}
    \caption{2T-AlAs}
    \end{subfigure}\qquad\qquad
    \begin{subfigure}[b]{0.4\textwidth}
    \includegraphics[width=60mm]{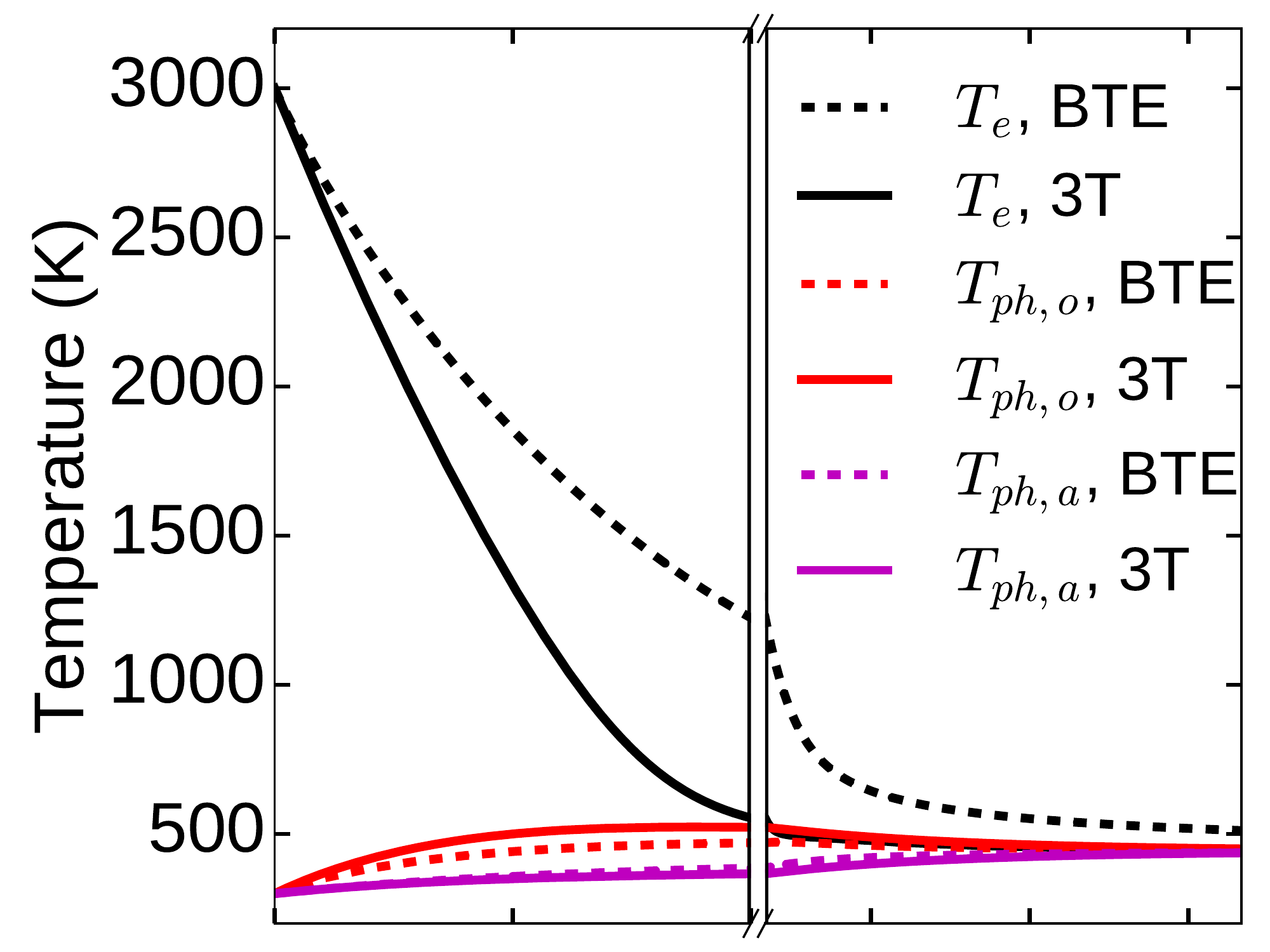}
    \caption{3T-AlAs}
    \end{subfigure}\\
    \begin{subfigure}[b]{0.4\textwidth}
    \includegraphics[width=60mm]{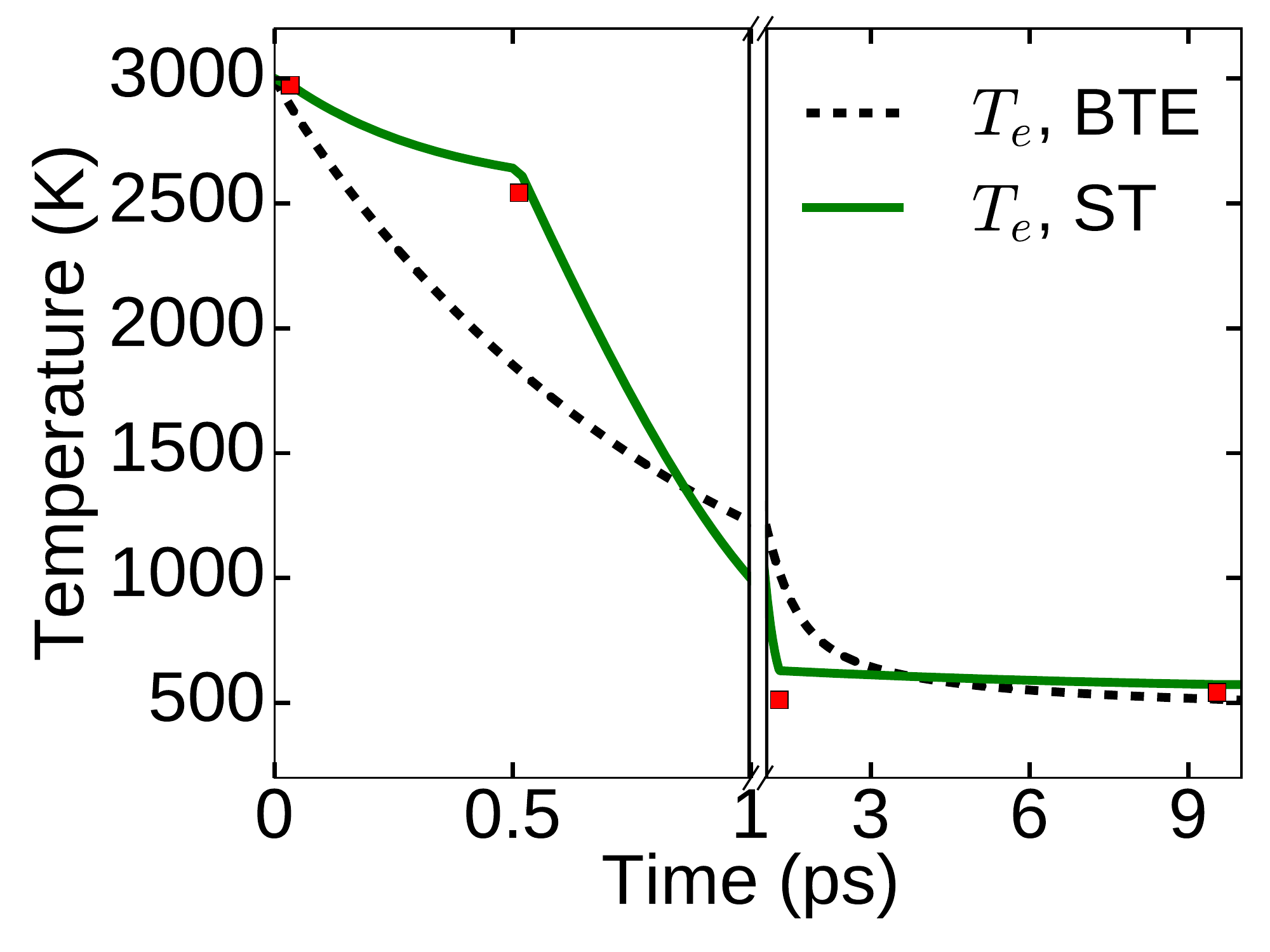}
    \caption{ST-AlAs ($c=0.1$)}
    \end{subfigure}\qquad\qquad
    \begin{subfigure}[b]{0.4\textwidth}
    \includegraphics[width=60mm]{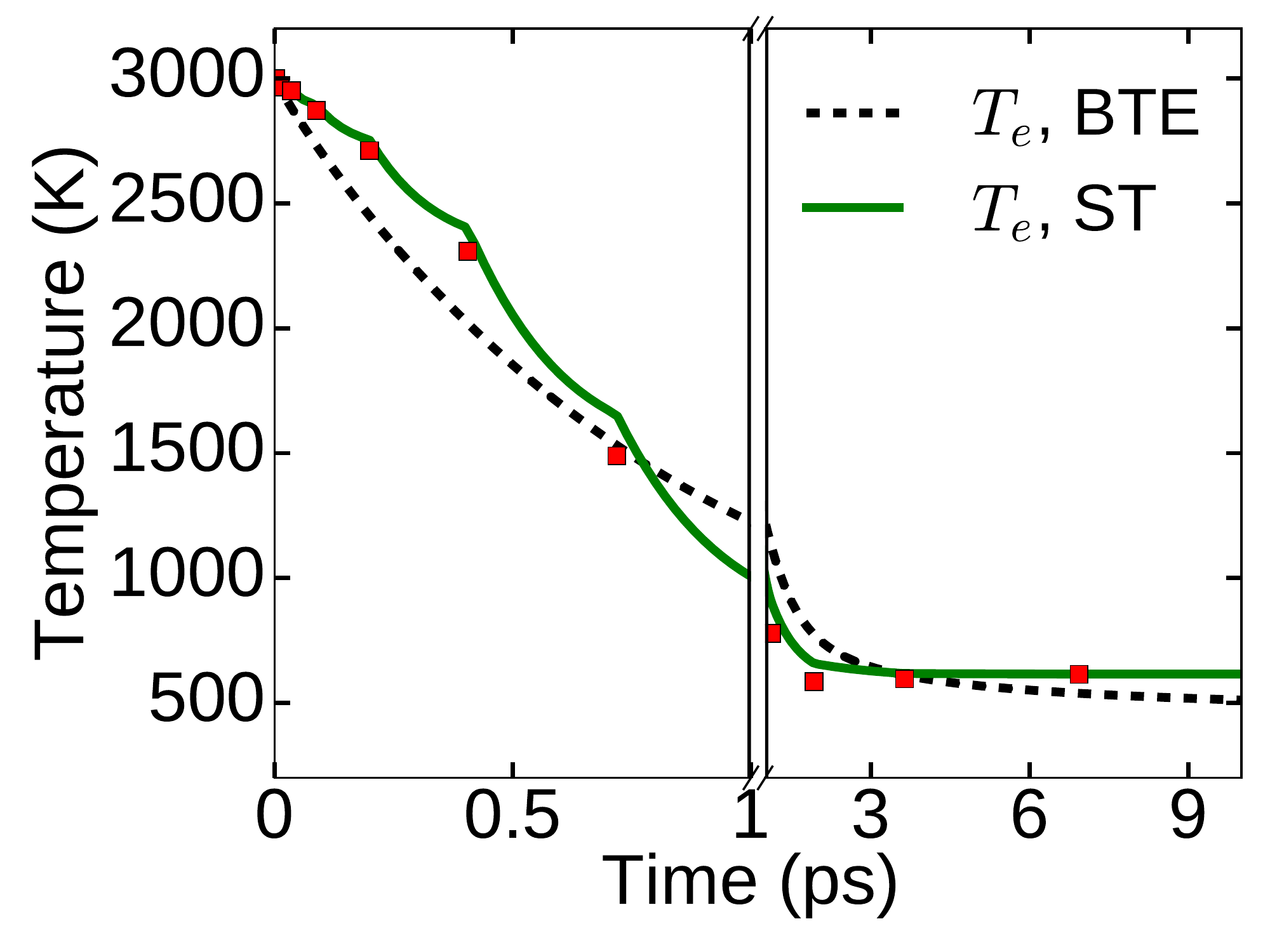}
    \caption{ST-AlAs ($c=0.5$)}
    \end{subfigure}\\
    \begin{centering}
    \begin{subfigure}[b]{0.4\textwidth}
    \includegraphics[width=60mm]{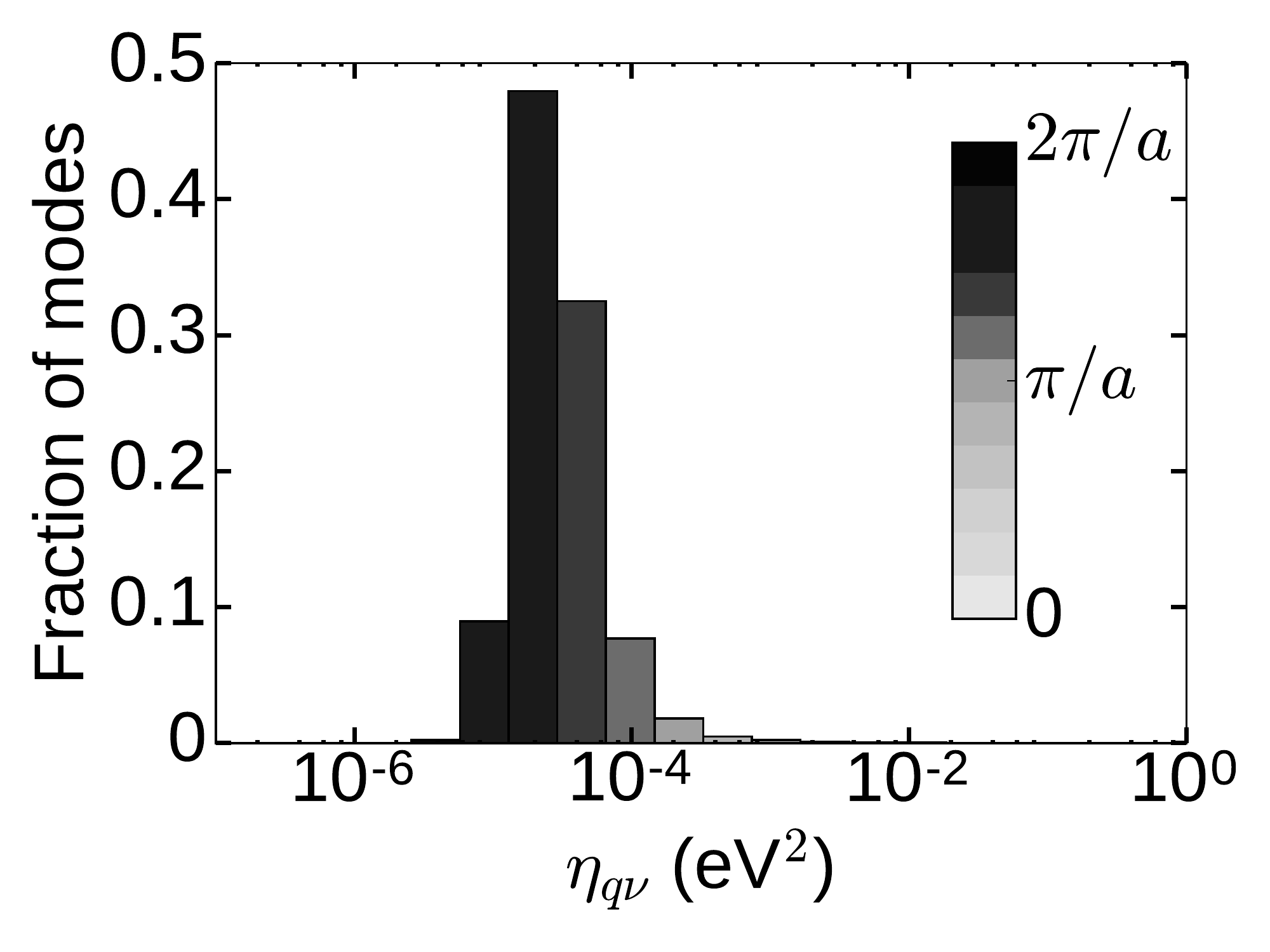}
    \caption{$\eta_{\textbf{q}\nu}$ - AlAs}
    \end{subfigure}
    \end{centering}
    \caption{Comparison between the electronic temperature decay obtained from a full-BTE solution and the 2T (a), 3T (b), and successive thermalization (c,d) models for AlAs. e) The distribution of phonon interaction strength $\eta_{\textbf{q}\nu}$ color-coded according to the average wavevector magnitude of phonons in each subset.}
\end{figure}

\newpage
\subsubsection{Aluminum Antimonide (AlSb)}

\begin{figure}[h]
    \centering
   \begin{subfigure}[b]{0.4\textwidth}
    \includegraphics[width=60mm]{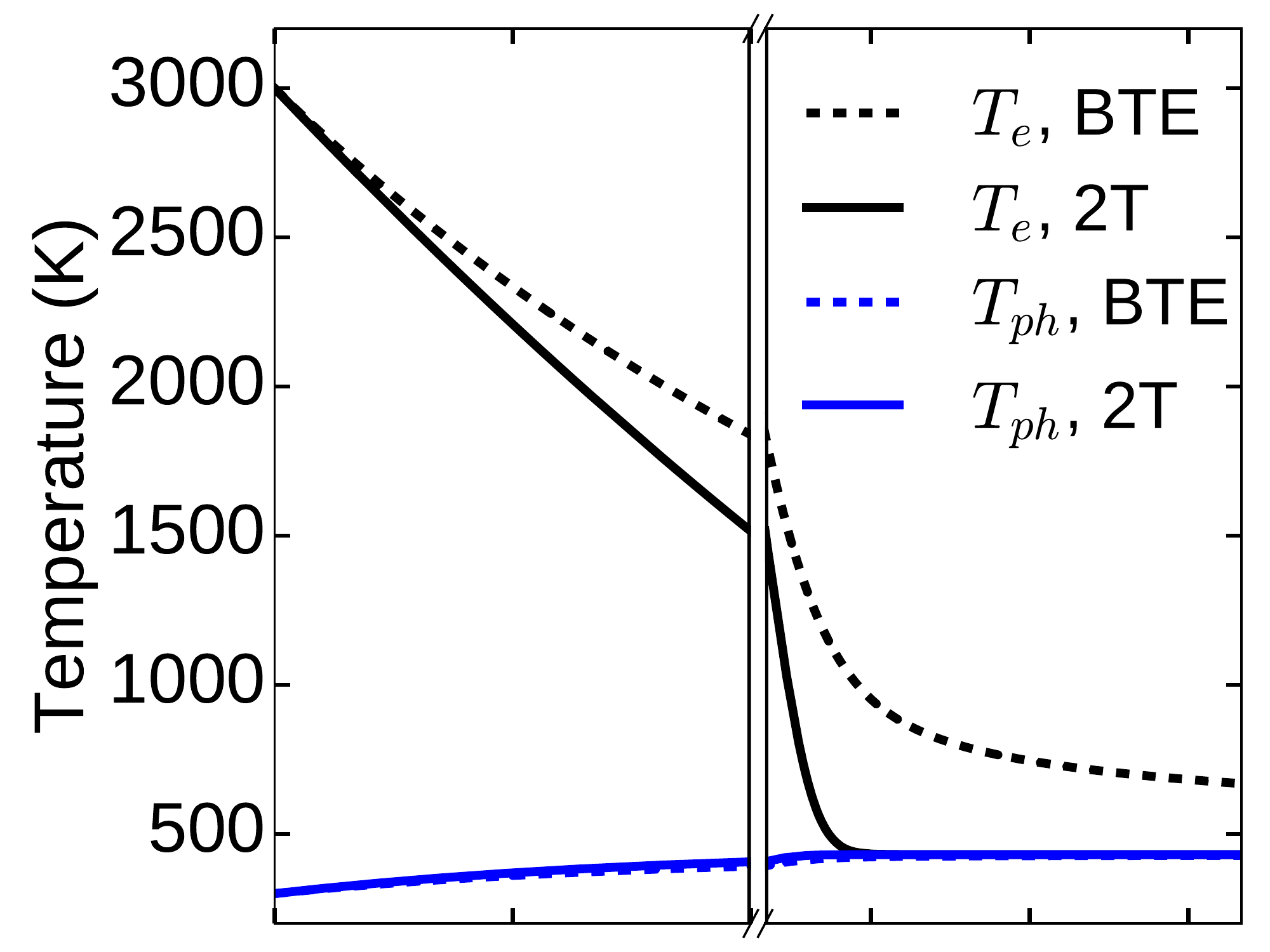}
    \caption{2T-AlSb}
    \end{subfigure}\qquad\qquad
    \begin{subfigure}[b]{0.4\textwidth}
    \includegraphics[width=60mm]{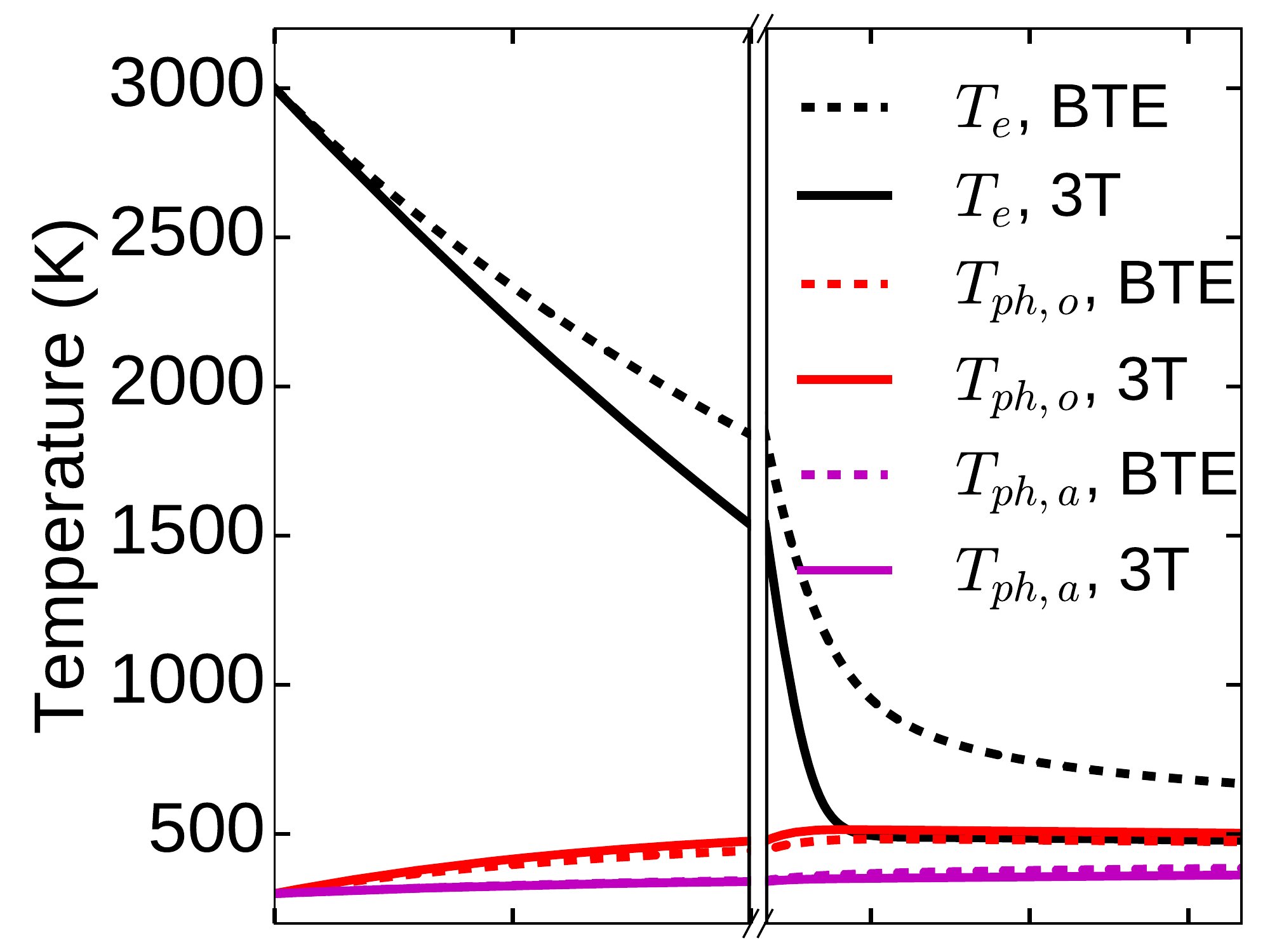}
    \caption{3T-AlSb}
    \end{subfigure}\\
    \begin{subfigure}[b]{0.4\textwidth}
    \includegraphics[width=60mm]{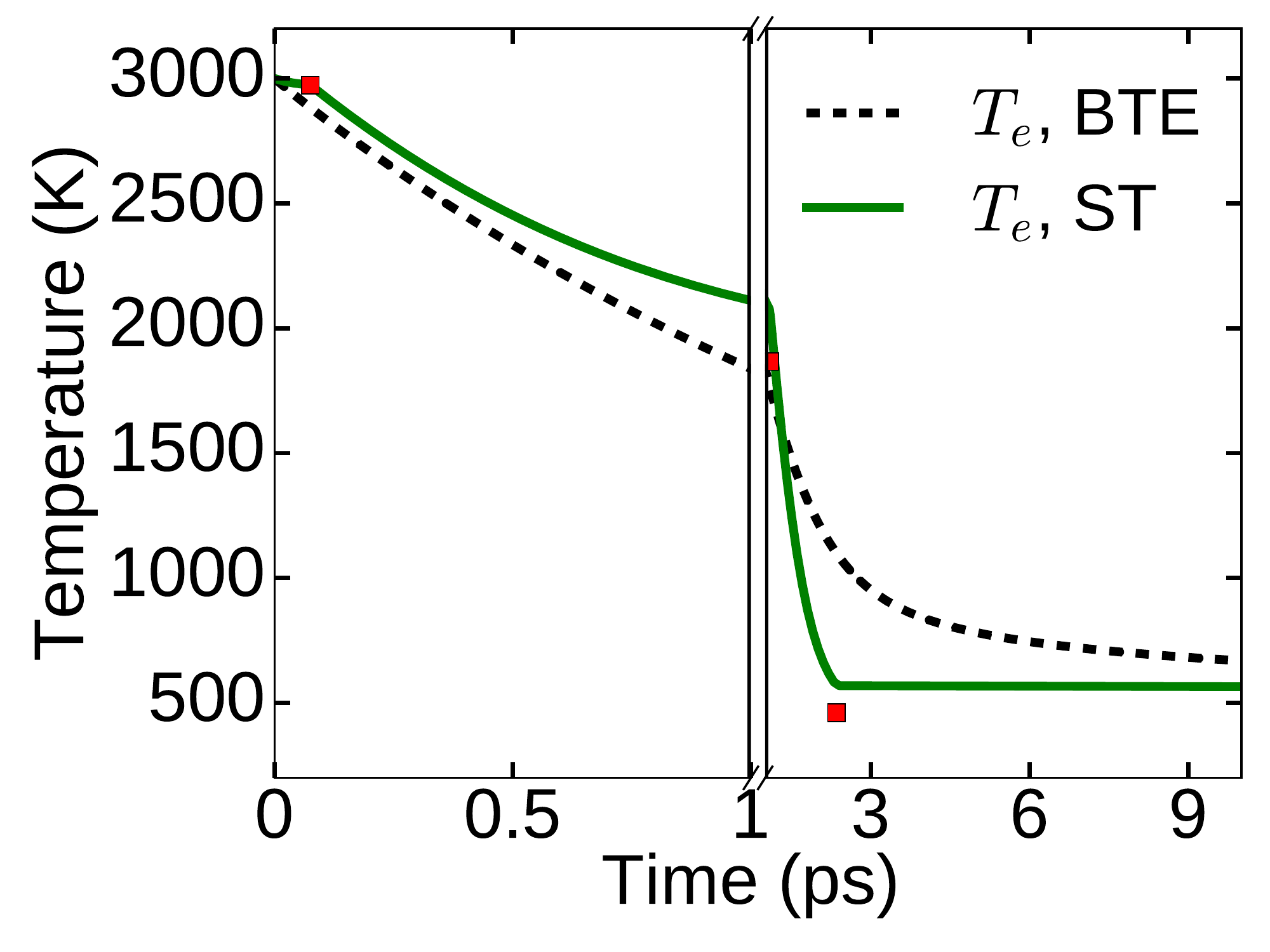}
    \caption{ST-AlSb ($c=0.1$)}
    \end{subfigure}\qquad\qquad
    \begin{subfigure}[b]{0.4\textwidth}
    \includegraphics[width=60mm]{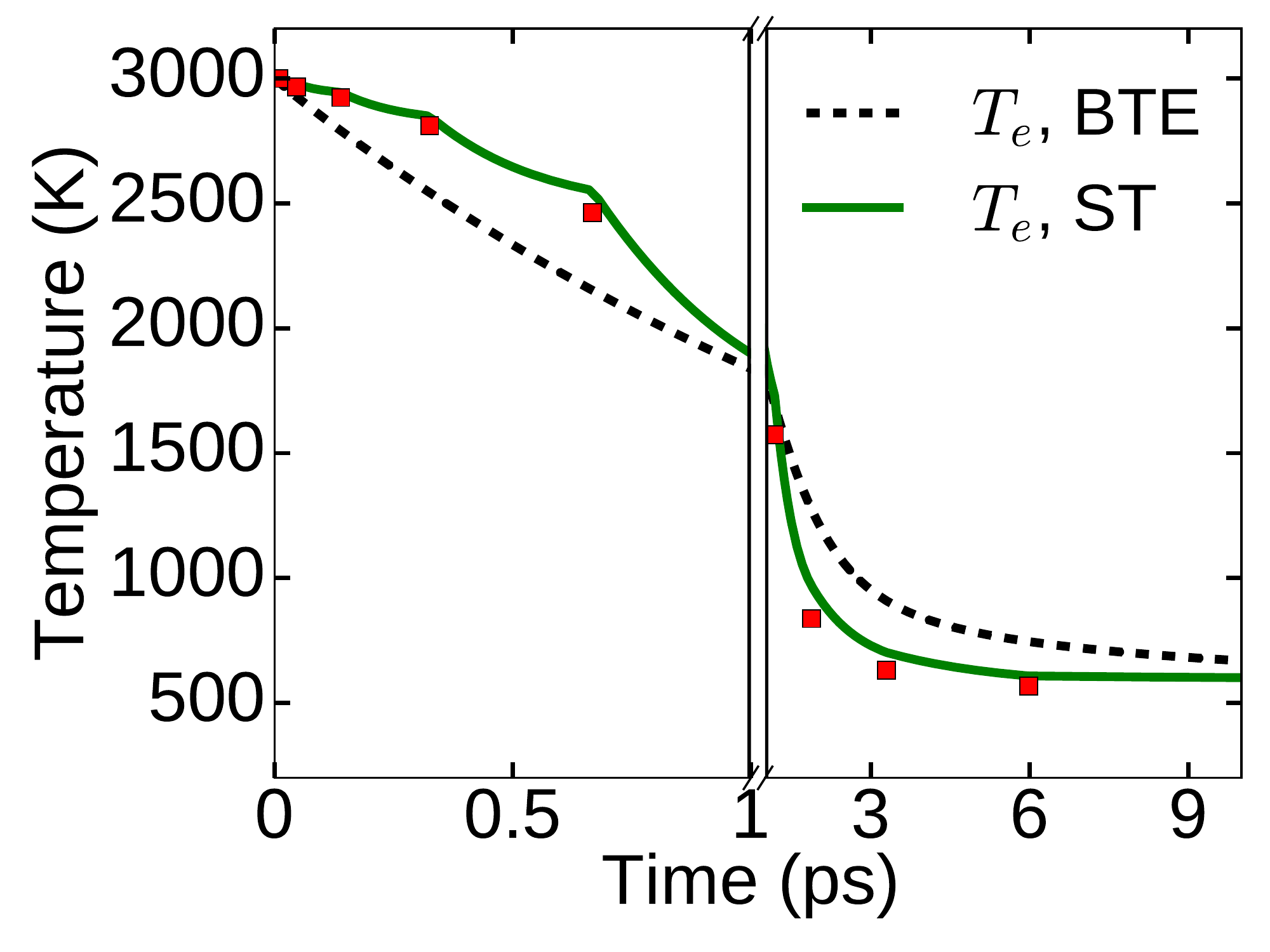}
    \caption{ST-AlSb ($c=0.5$)}
    \end{subfigure}\\
    \begin{centering}
    \begin{subfigure}[b]{0.4\textwidth}
    \includegraphics[width=60mm]{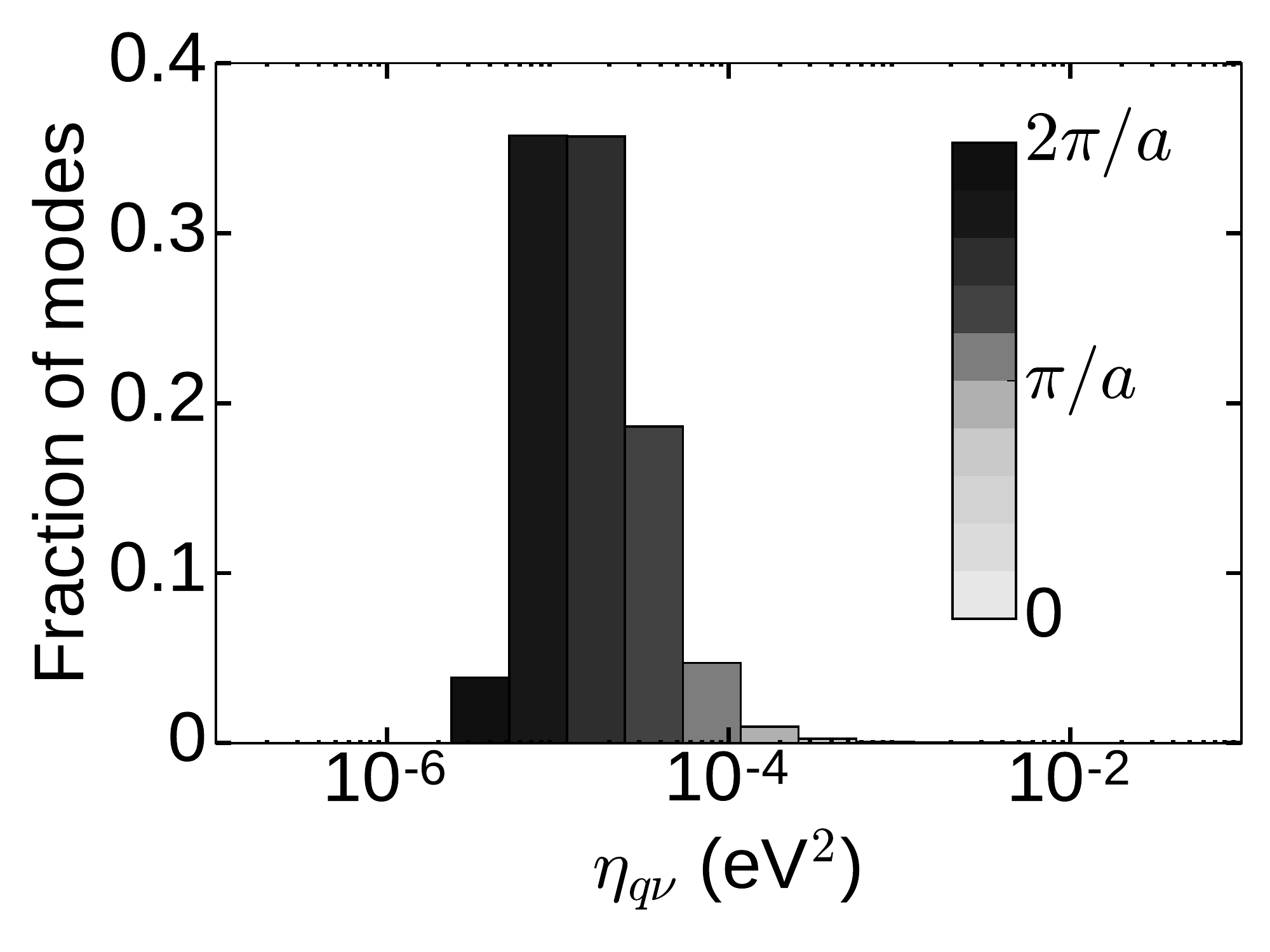}
    \caption{$\eta_{\textbf{q}\nu}$ - AlSb}
    \end{subfigure}
    \end{centering}
    \caption{Comparison between the electronic temperature decay obtained from a full-BTE solution and the 2T (a), 3T (b), and successive thermalization (c,d) models for AlSb. e) The distribution of phonon interaction strength $\eta_{\textbf{q}\nu}$ color-coded according to the average wavevector magnitude of phonons in each subset.}
\end{figure}

\newpage

\subsubsection{Gallium Nitride (GaN)}

\begin{figure}[h]
    \centering
   \begin{subfigure}[b]{0.4\textwidth}
    \includegraphics[width=60mm]{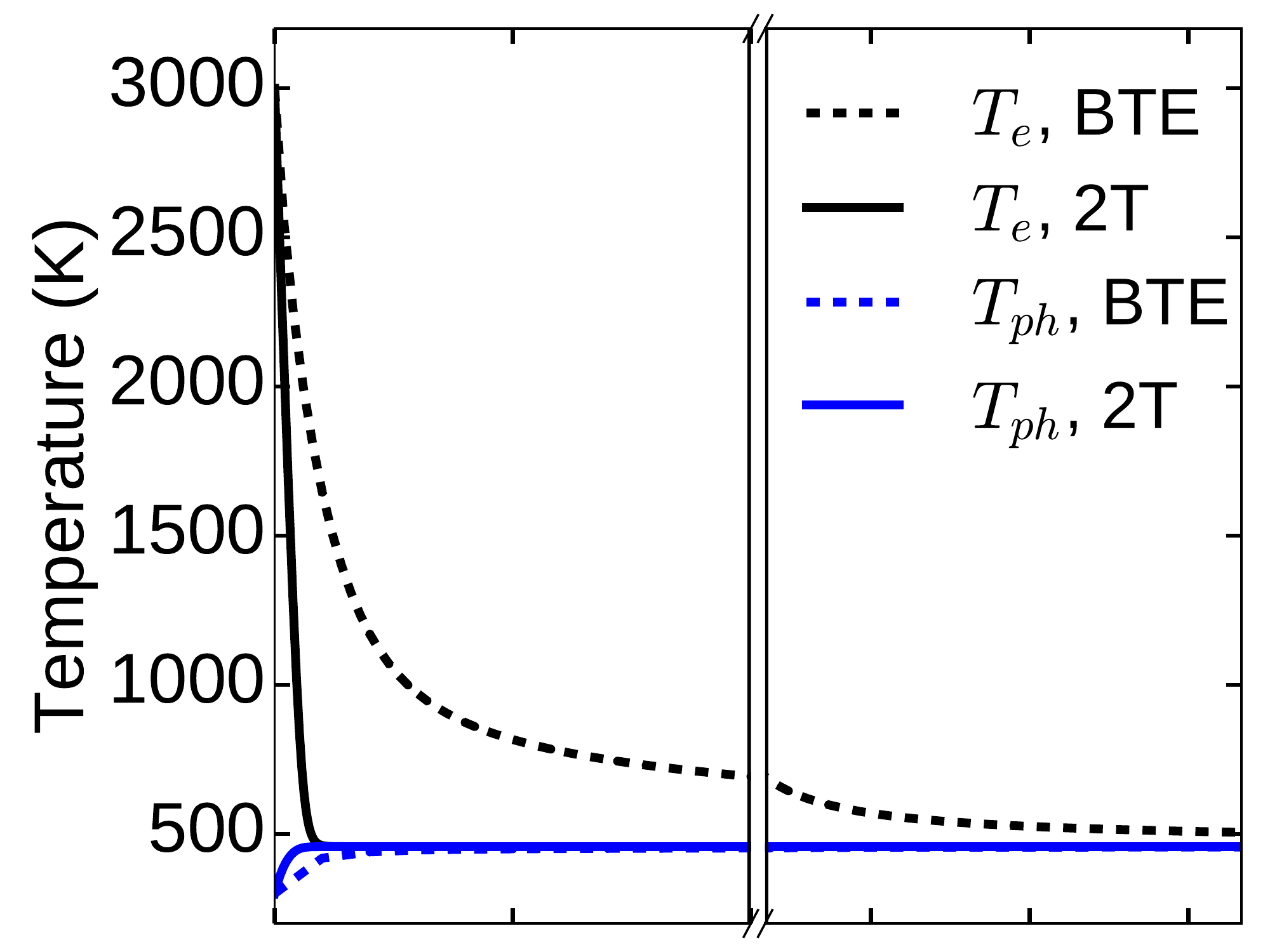}
    \caption{2T-GaN}
    \end{subfigure}\qquad\qquad
    \begin{subfigure}[b]{0.4\textwidth}
    \includegraphics[width=60mm]{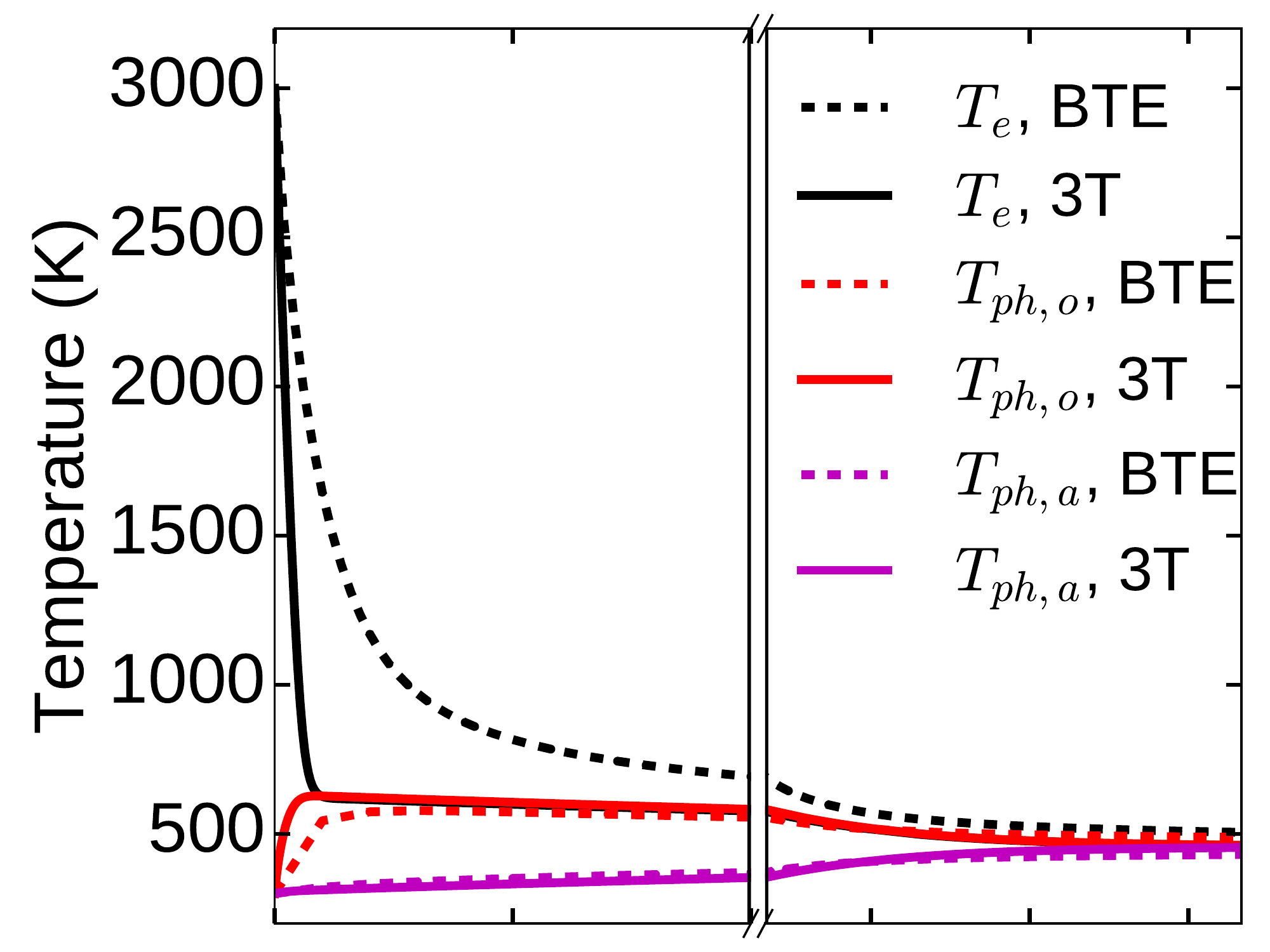}
    \caption{3T-GaN}
    \end{subfigure}\\
    \begin{subfigure}[b]{0.4\textwidth}
    \includegraphics[width=60mm]{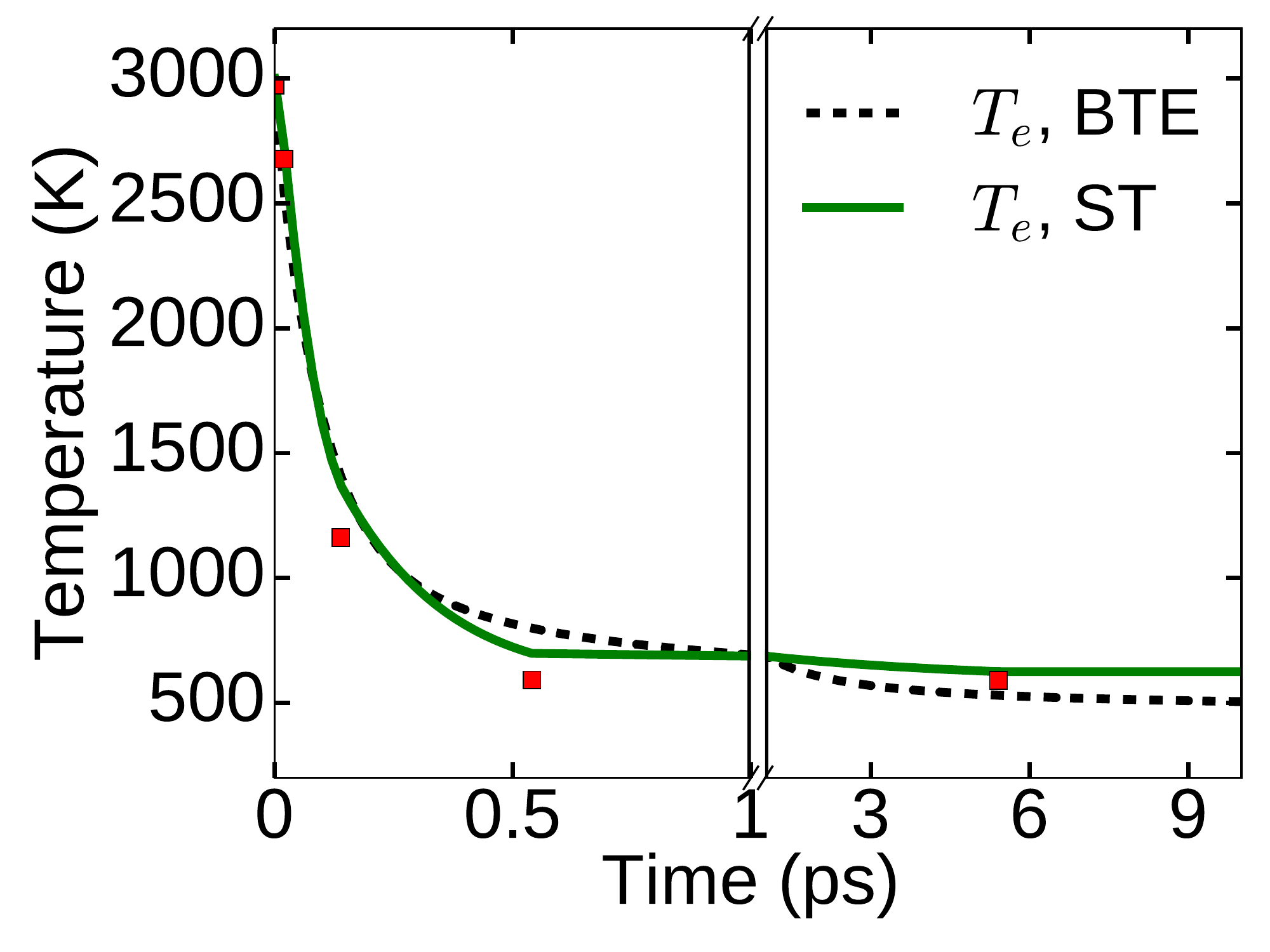}
    \caption{ST-GaN ($c=0.1$)}
    \end{subfigure}\qquad\qquad
    \begin{subfigure}[b]{0.4\textwidth}
    \includegraphics[width=60mm]{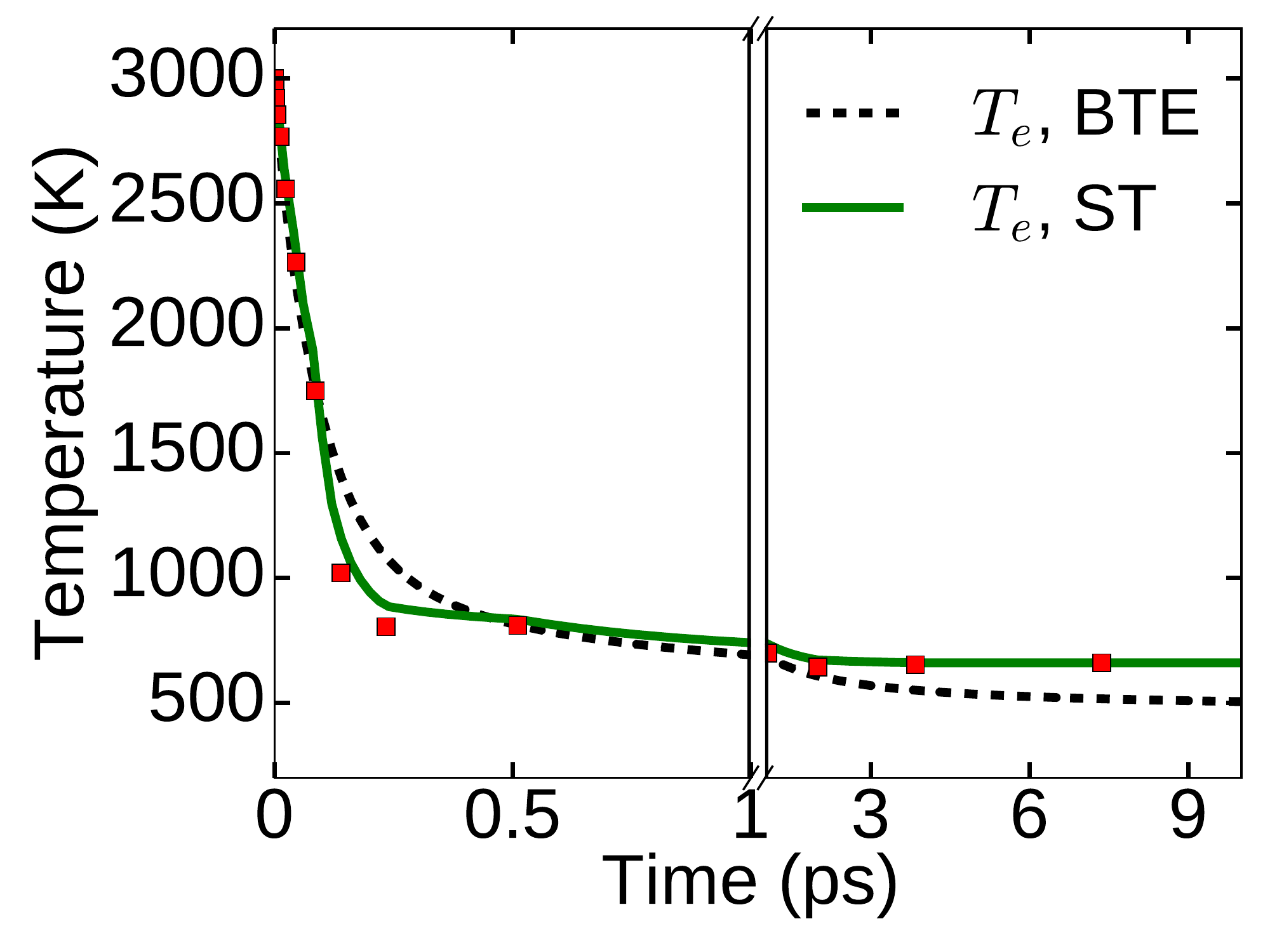}
    \caption{ST-GaN ($c=0.5$)}
    \end{subfigure}\\
    \begin{centering}
    \begin{subfigure}[b]{0.4\textwidth}
    \includegraphics[width=60mm]{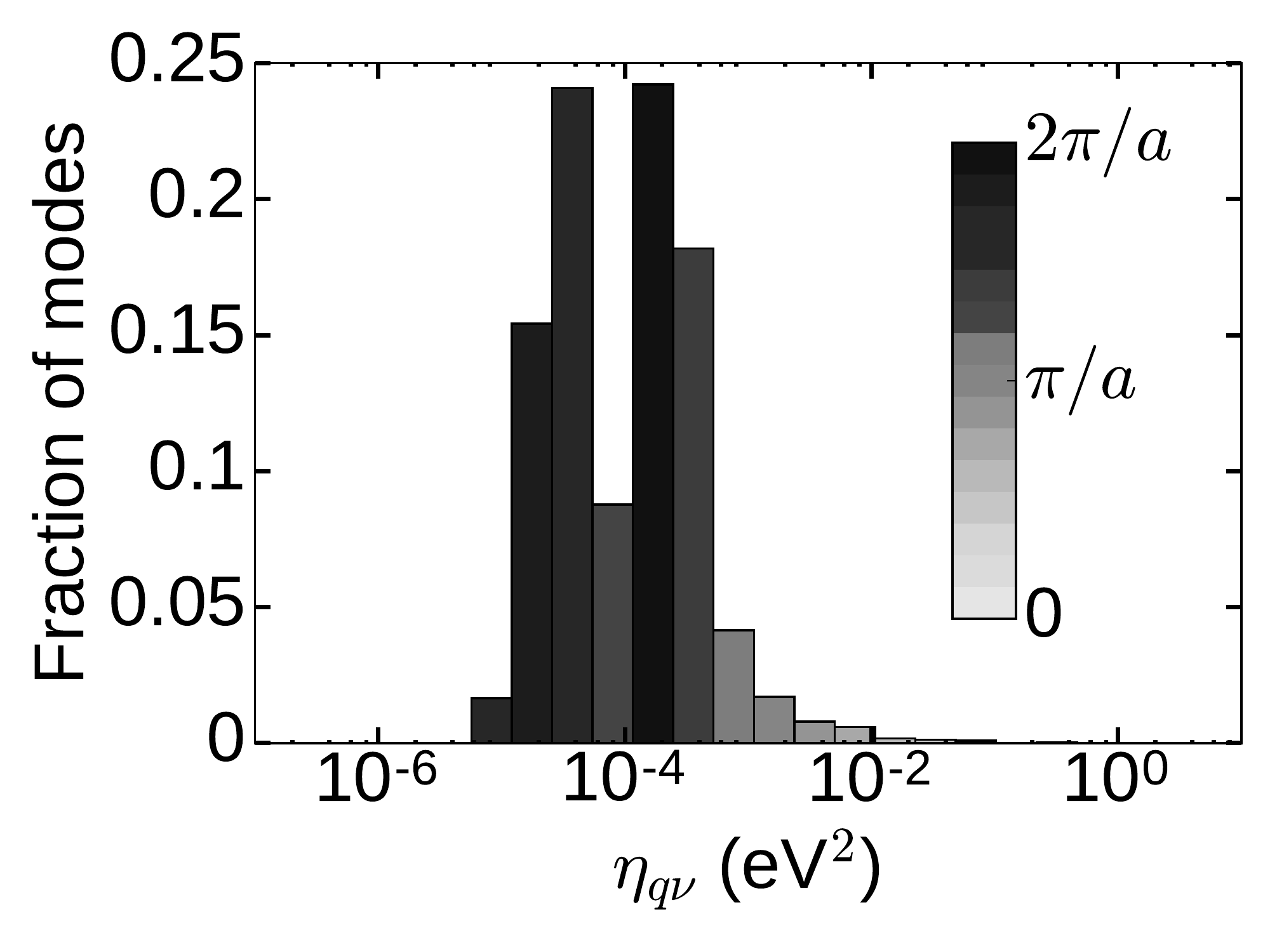}
    \caption{$\eta_{\textbf{q}\nu}$ - GaN}
    \end{subfigure}
    \end{centering}
    \caption{Comparison between the electronic temperature decay obtained from a full-BTE solution and the 2T (a), 3T (b), and successive thermalization (c,d) models for GaN. e) The distribution of phonon interaction strength $\eta_{\textbf{q}\nu}$ color-coded according to the average wavevector magnitude of phonons in each subset.}
\end{figure}

\newpage

\subsubsection{Gallium Phosphide (GaP)}

\begin{figure}[h]
    \centering
   \begin{subfigure}[b]{0.4\textwidth}
    \includegraphics[width=60mm]{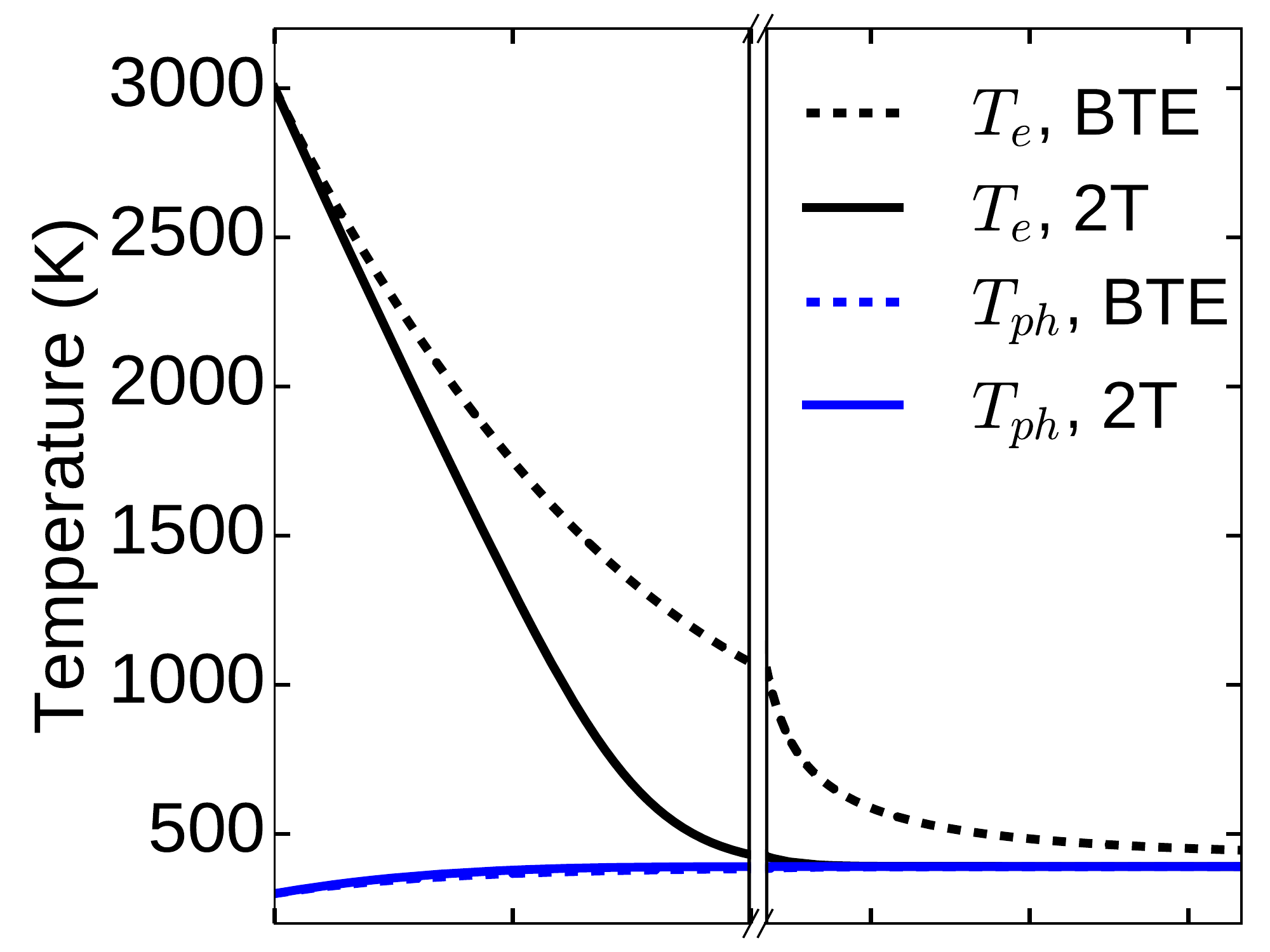}
    \caption{2T-GaP}
    \end{subfigure}\qquad\qquad
    \begin{subfigure}[b]{0.4\textwidth}
    \includegraphics[width=60mm]{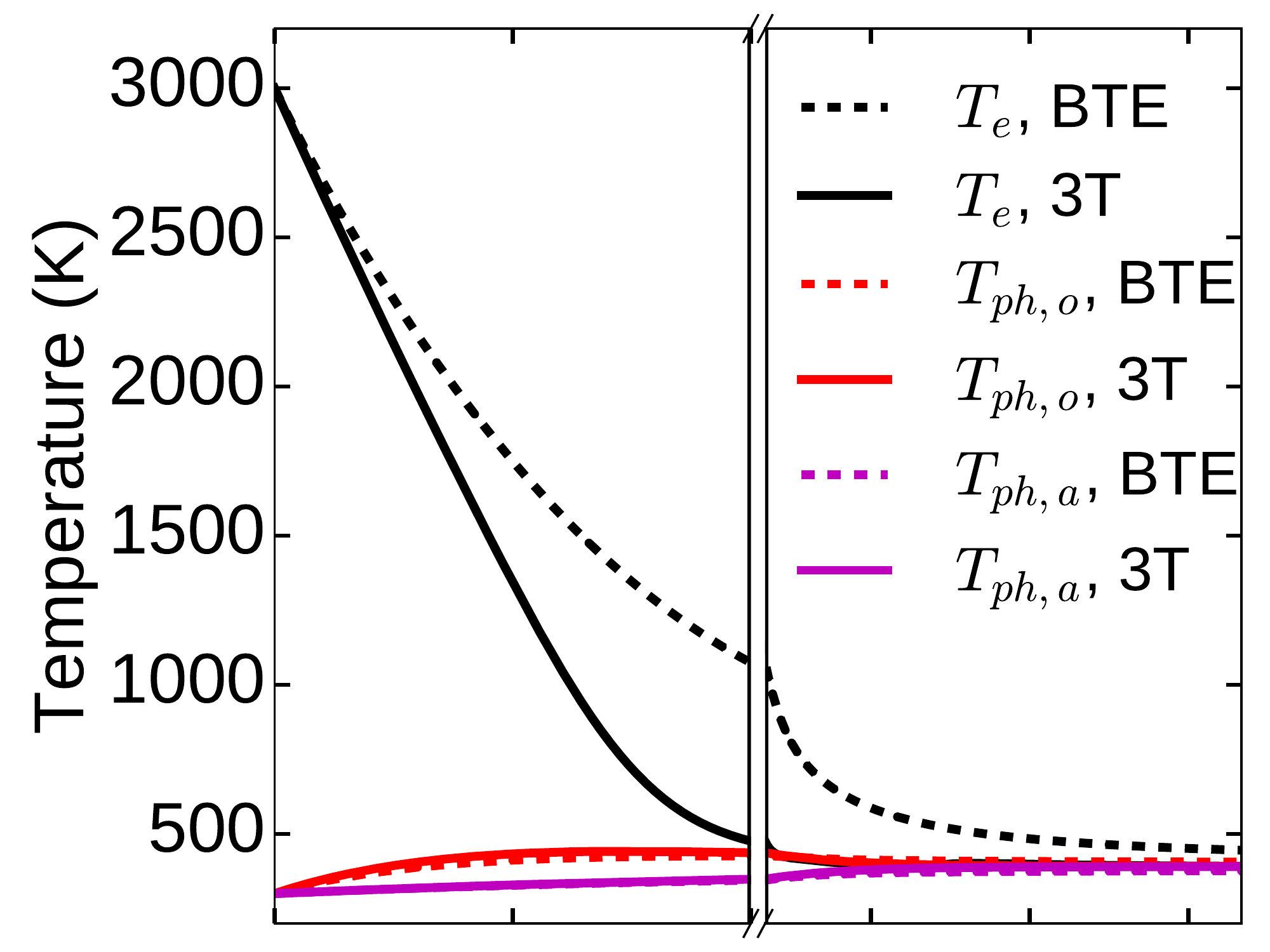}
    \caption{3T-GaP}
    \end{subfigure}\\
    \begin{subfigure}[b]{0.4\textwidth}
    \includegraphics[width=60mm]{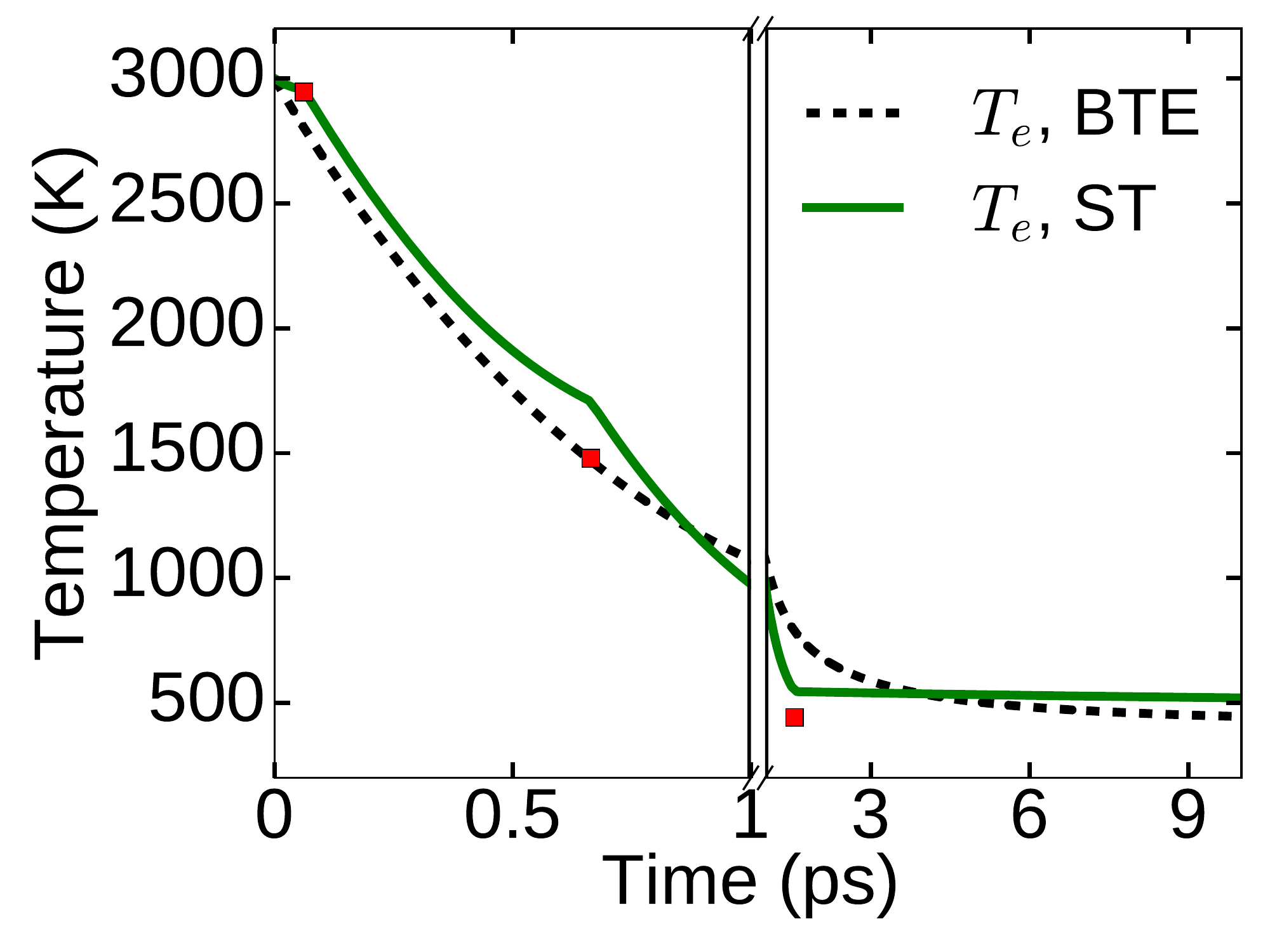}
    \caption{ST-GaP ($c=0.1$)}
    \end{subfigure}\qquad\qquad
    \begin{subfigure}[b]{0.4\textwidth}
    \includegraphics[width=60mm]{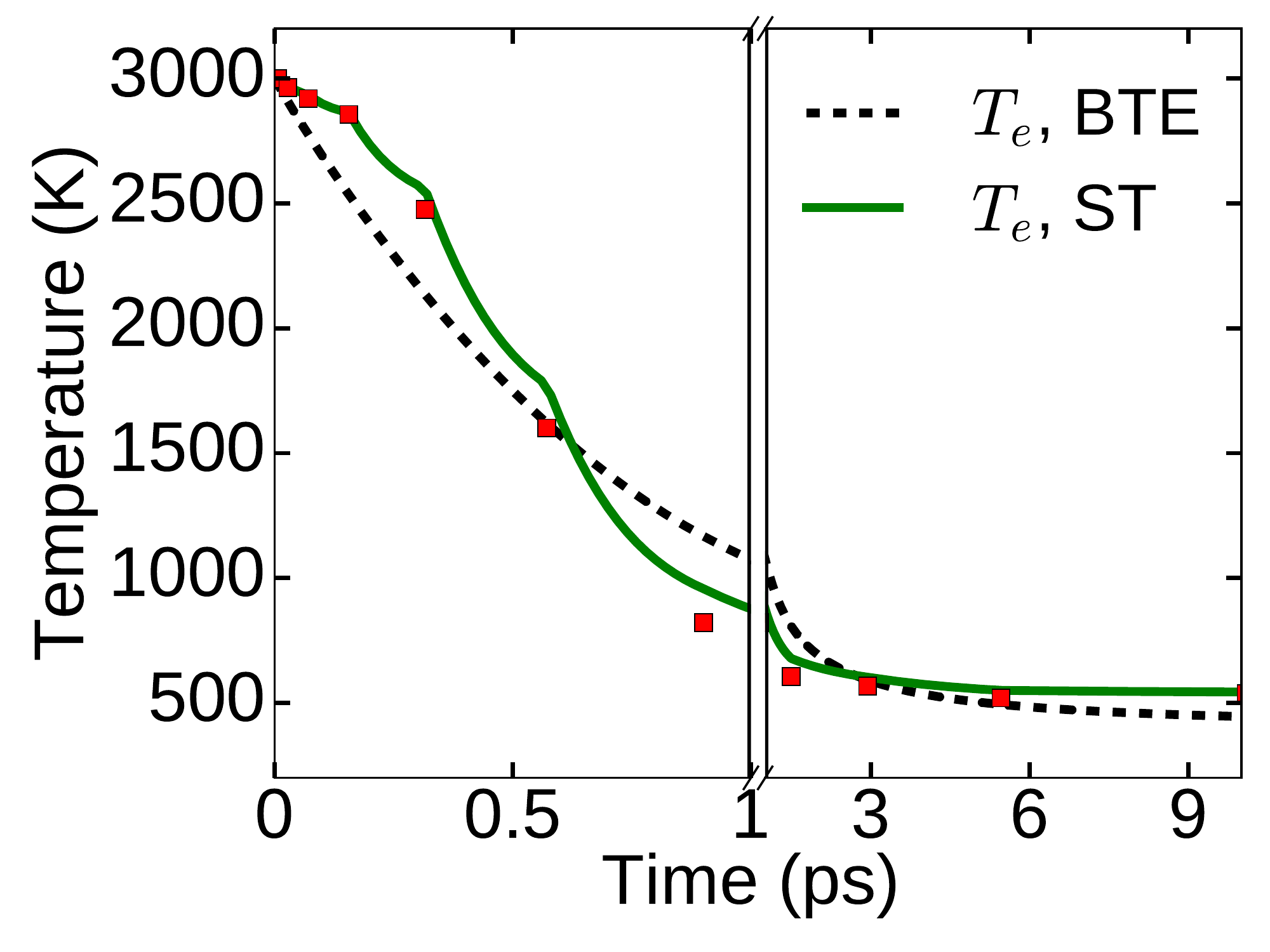}
    \caption{ST-GaP ($c=0.5$)}
    \end{subfigure}\\
    \begin{centering}
    \begin{subfigure}[b]{0.4\textwidth}
    \includegraphics[width=60mm]{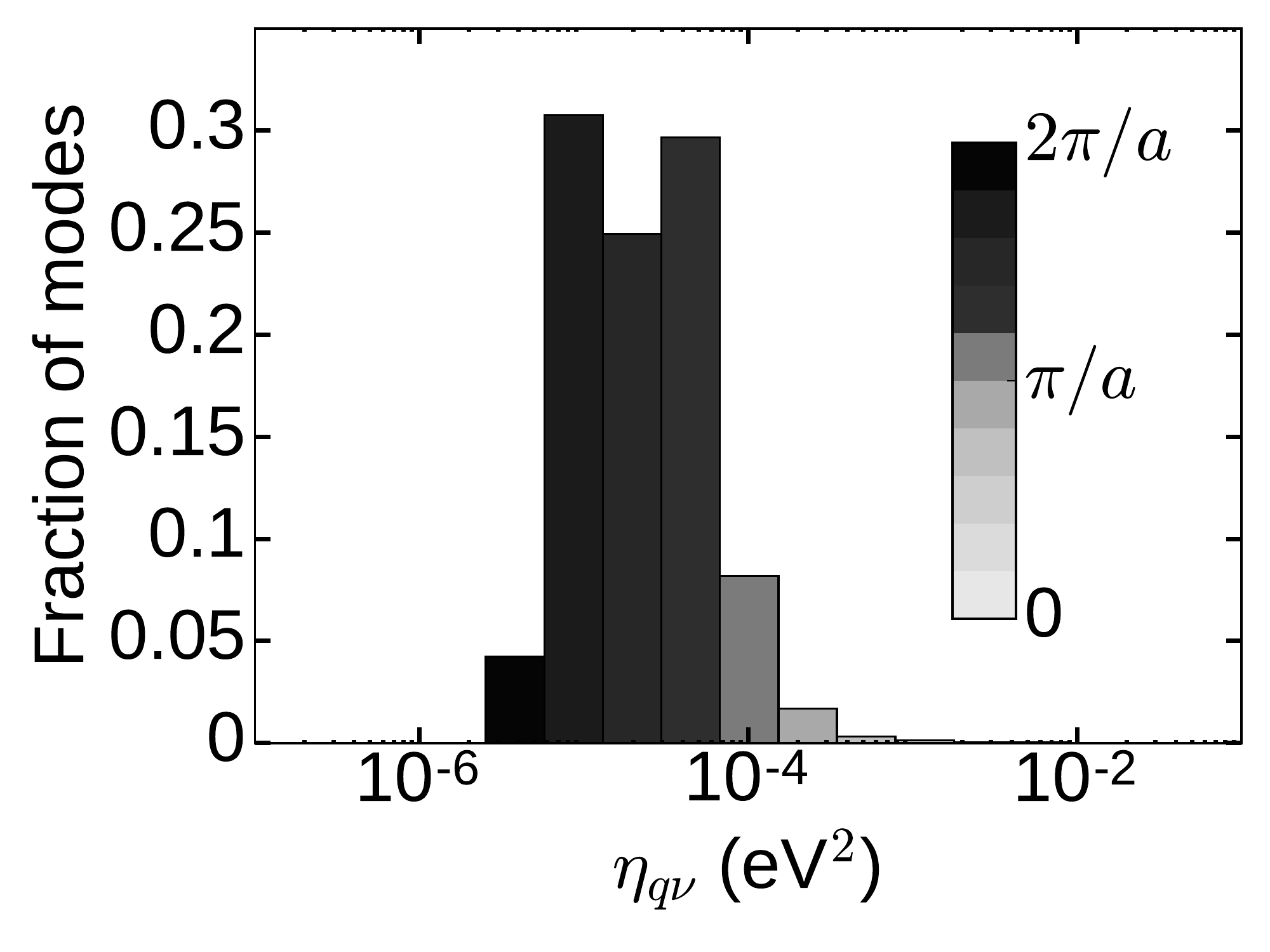}
    \caption{$\eta_{\textbf{q}\nu}$ - GaP}
    \end{subfigure}
    \end{centering}
    \caption{Comparison between the electronic temperature decay obtained from a full-BTE solution and the 2T (a), 3T (b), and successive thermalization (c,d) models for GaP. e) The distribution of phonon interaction strength $\eta_{\textbf{q}\nu}$ color-coded according to the average wavevector magnitude of phonons in each subset.}
\end{figure}

\newpage

\subsubsection{Gallium Arsenide (GaAs)}

\begin{figure}[h]
    \centering
   \begin{subfigure}[b]{0.4\textwidth}
    \includegraphics[width=60mm]{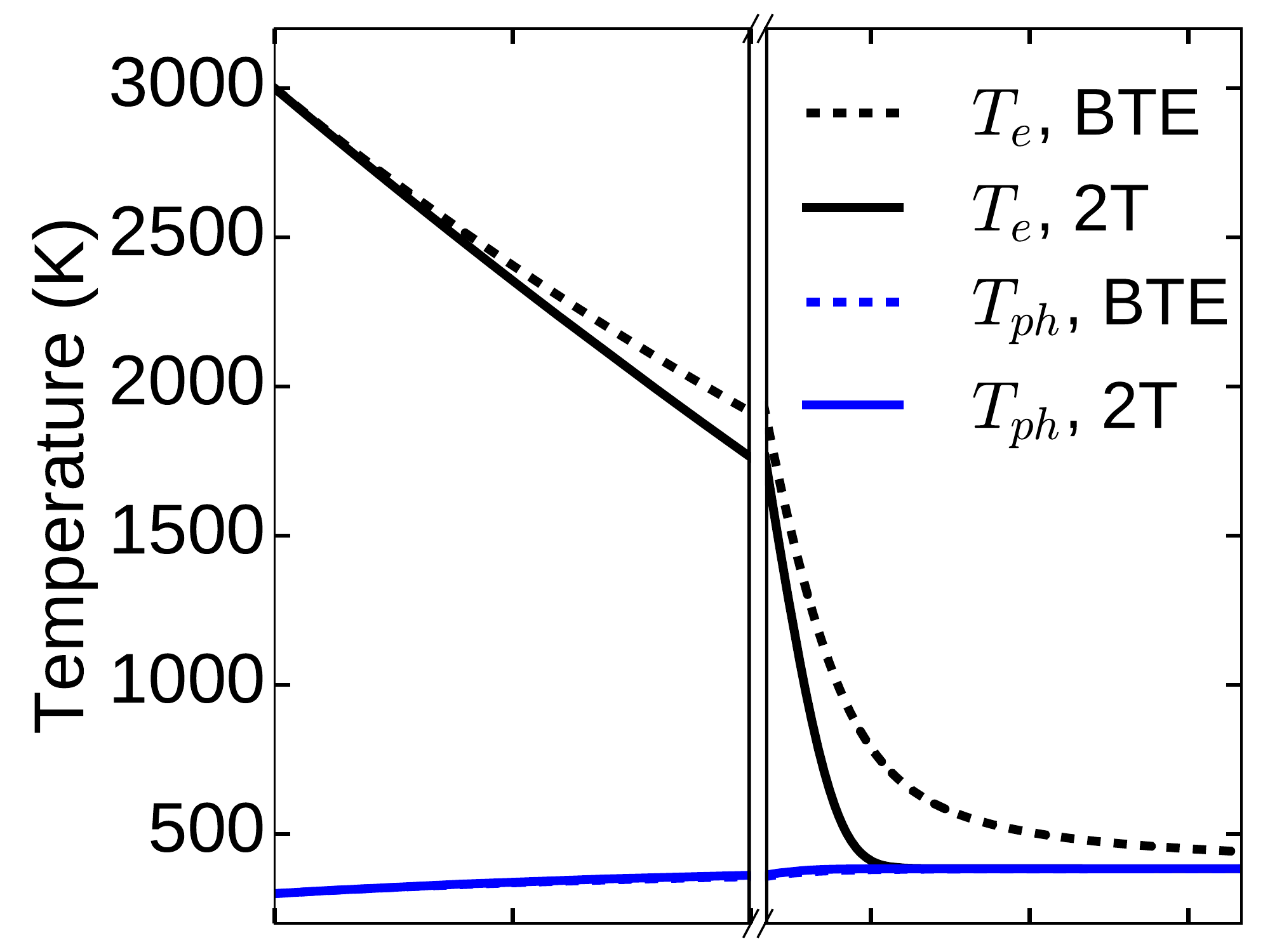}
    \caption{2T-GaAs}
    \end{subfigure}\qquad\qquad
    \begin{subfigure}[b]{0.4\textwidth}
    \includegraphics[width=60mm]{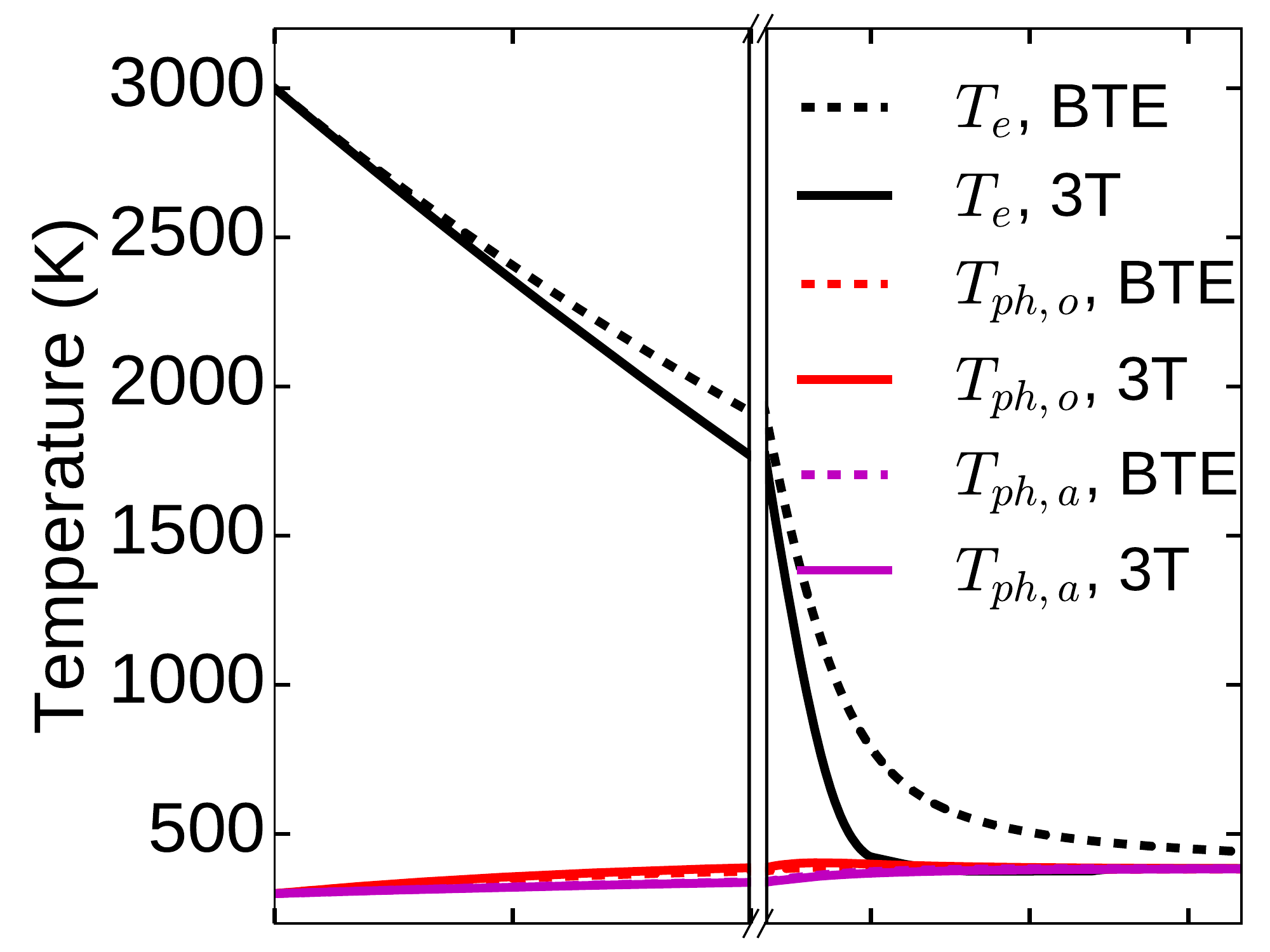}
    \caption{3T-GaAs}
    \end{subfigure}\\
    \begin{subfigure}[b]{0.4\textwidth}
    \includegraphics[width=60mm]{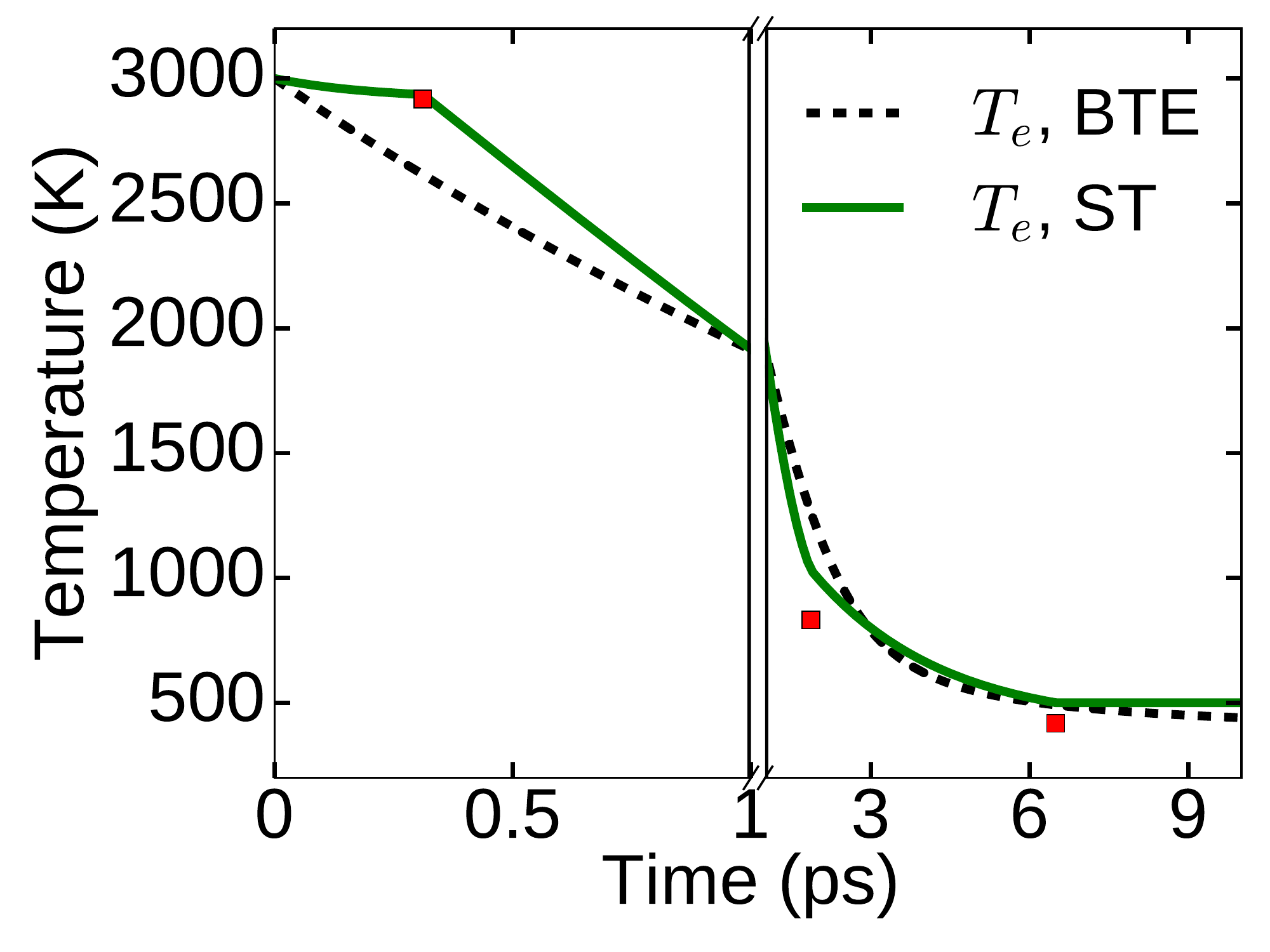}
    \caption{ST-GaAs ($c=0.1$)}
    \end{subfigure}\qquad\qquad
    \begin{subfigure}[b]{0.4\textwidth}
    \includegraphics[width=60mm]{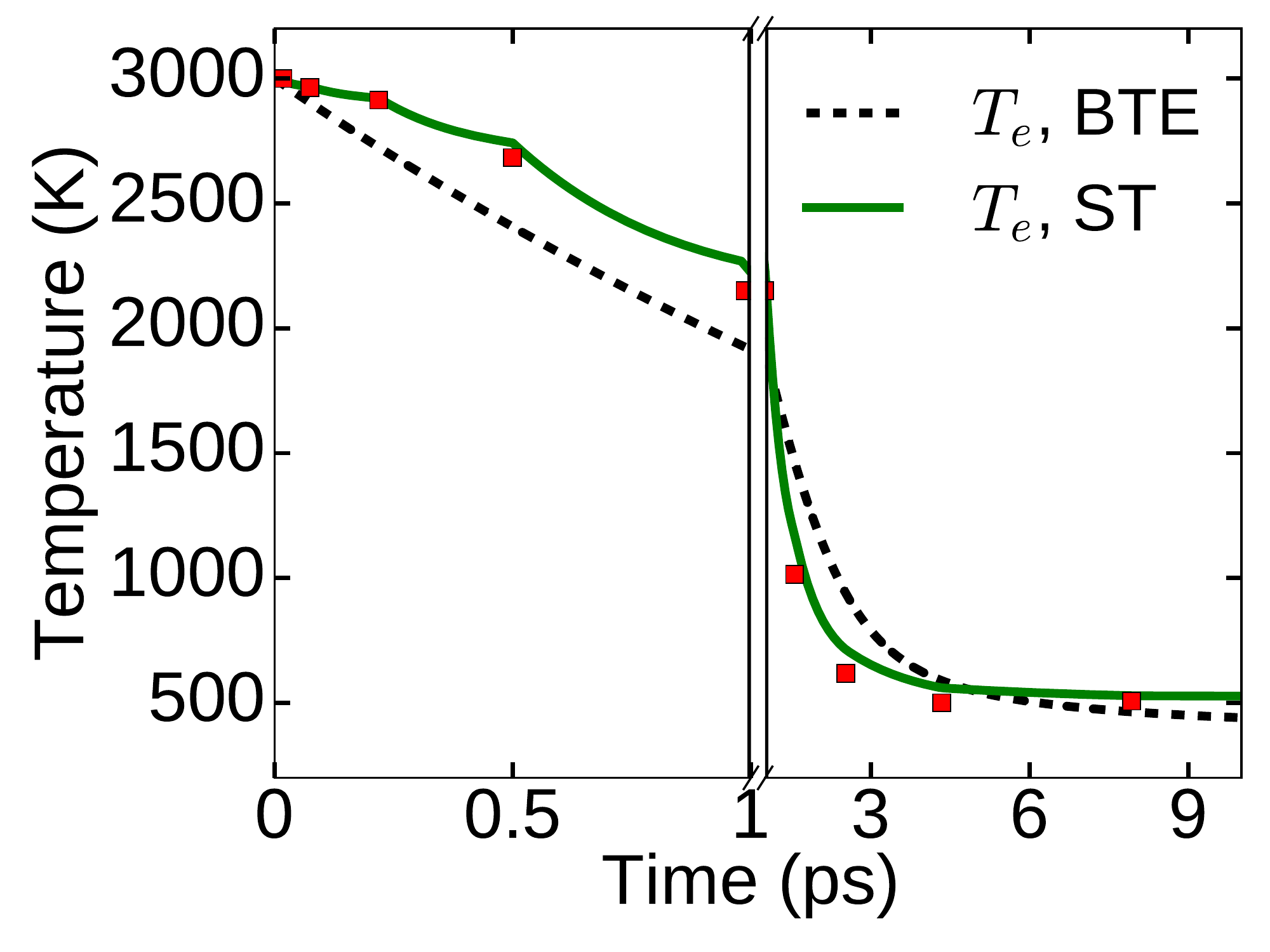}
    \caption{ST-GaAs ($c=0.5$)}
    \end{subfigure}\\
    \begin{centering}
    \begin{subfigure}[b]{0.4\textwidth}
    \includegraphics[width=60mm]{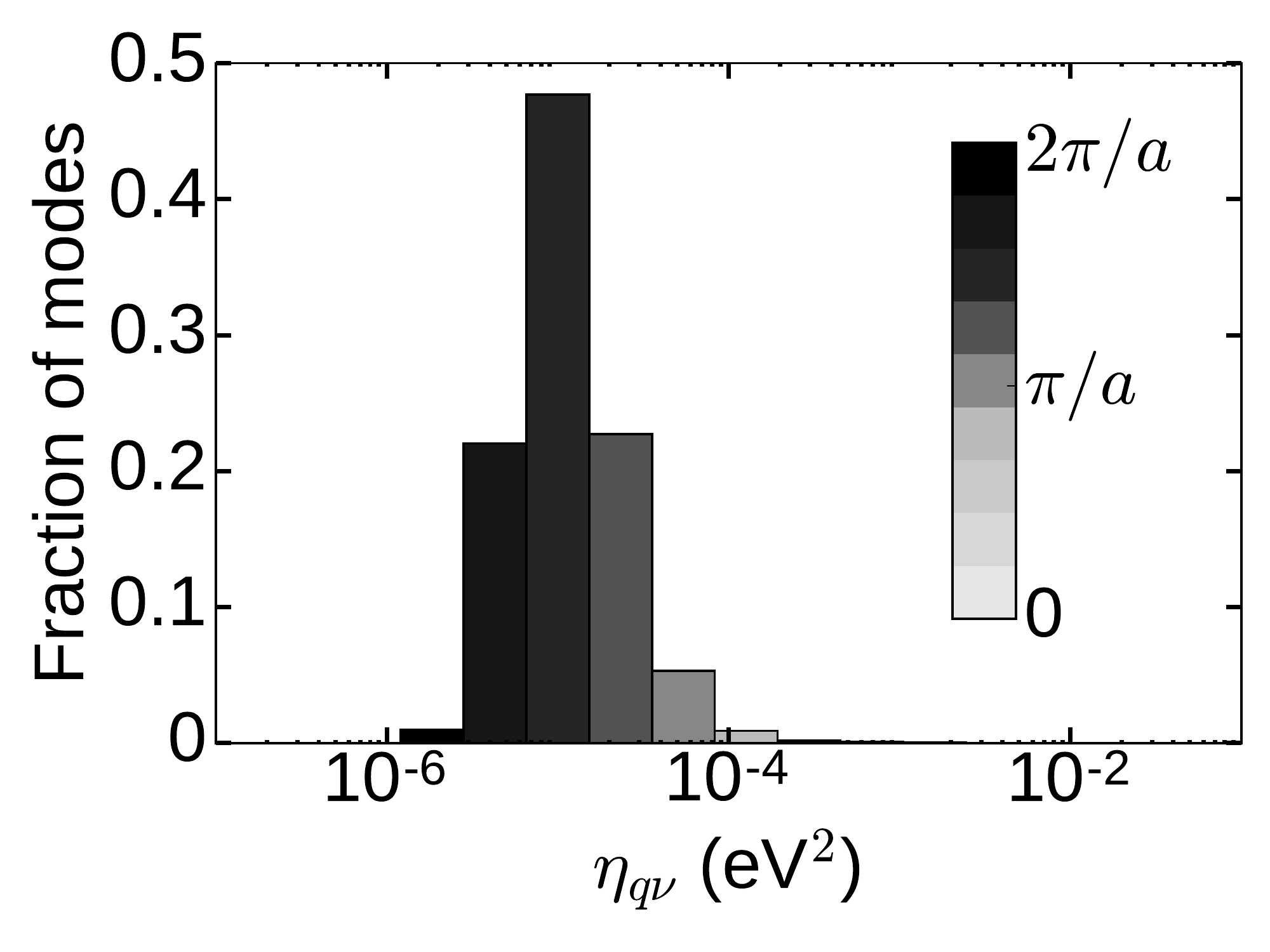}
    \caption{$\eta_{\textbf{q}\nu}$ - GaAs}
    \end{subfigure}
    \end{centering}
    \caption{Comparison between the electronic temperature decay obtained from a full-BTE solution and the 2T (a), 3T (b), and successive thermalization (c,d) models for GaAs. e) The distribution of phonon interaction strength $\eta_{\textbf{q}\nu}$ color-coded according to the average wavevector magnitude of phonons in each subset.}
\end{figure}

\newpage

\subsubsection{Diamond}

\begin{figure}[h]
    \centering
   \begin{subfigure}[b]{0.4\textwidth}
    \includegraphics[width=60mm]{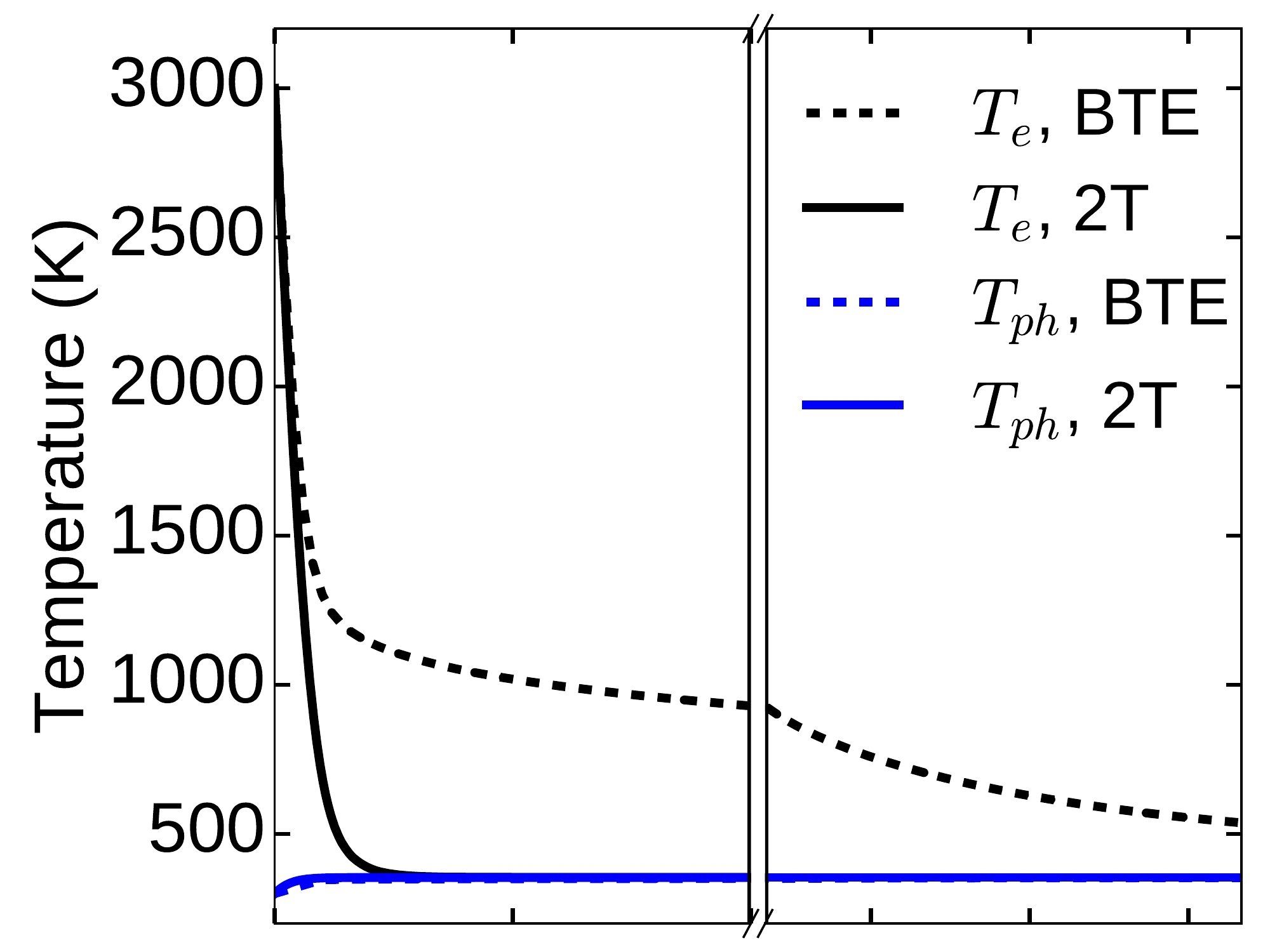}
    \caption{2T-Diamond}
    \end{subfigure}\qquad\qquad
    \begin{subfigure}[b]{0.4\textwidth}
    \includegraphics[width=60mm]{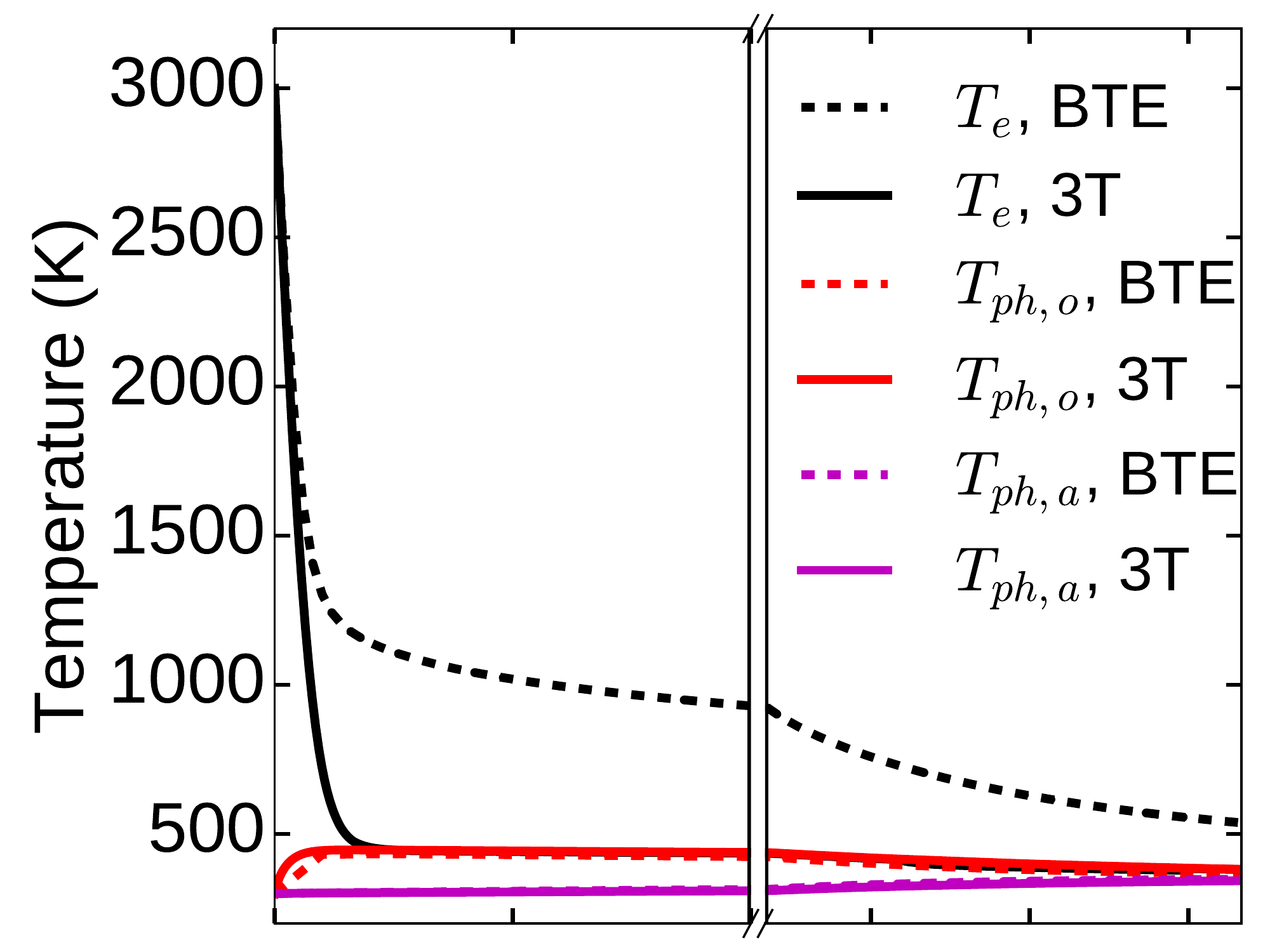}
    \caption{3T-Diamond}
    \end{subfigure}\\
    \begin{subfigure}[b]{0.4\textwidth}
    \includegraphics[width=60mm]{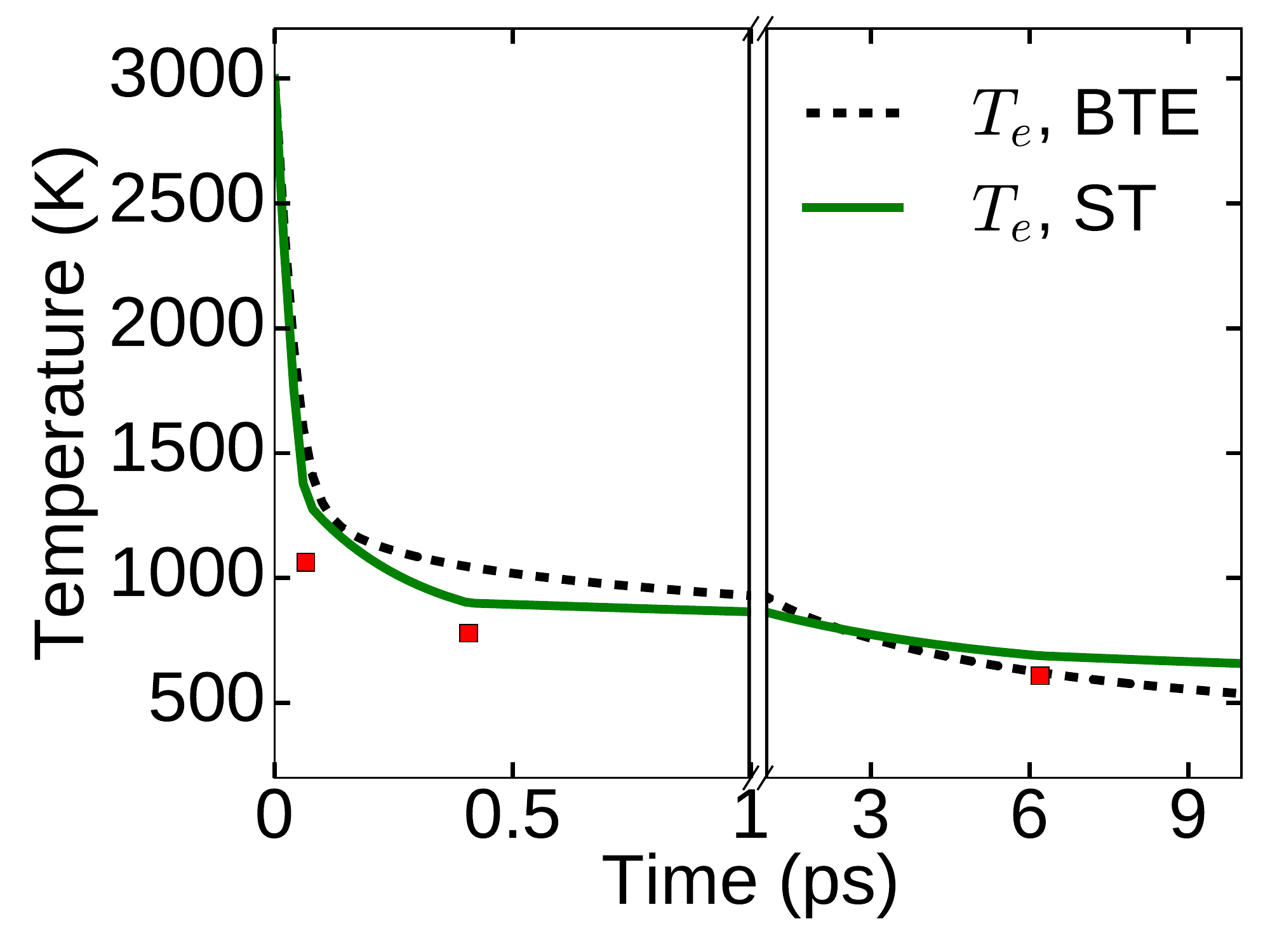}
    \caption{ST-Diamond ($c=0.1$)}
    \end{subfigure}\qquad\qquad
    \begin{subfigure}[b]{0.4\textwidth}
    \includegraphics[width=60mm]{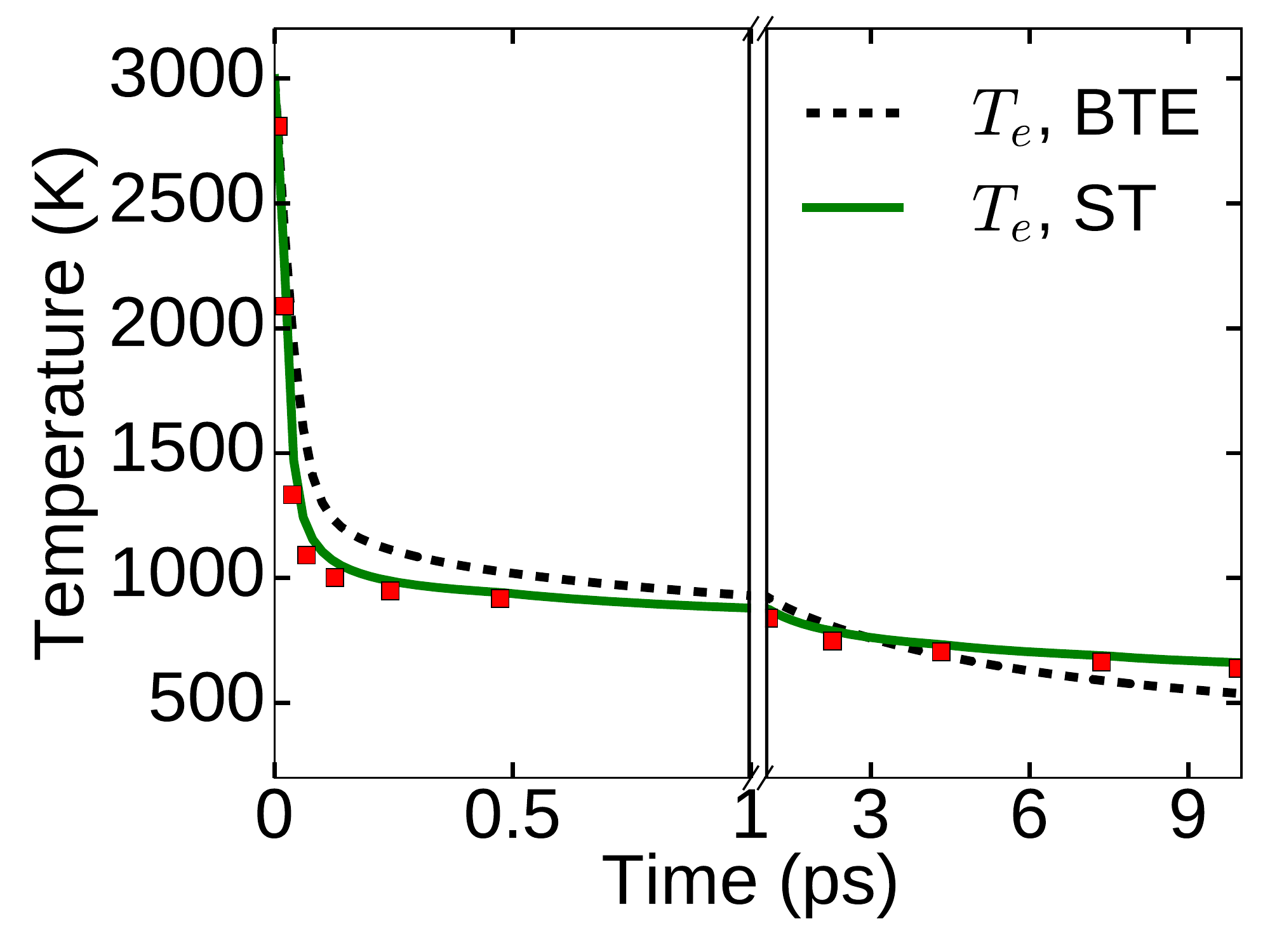}
    \caption{ST-Diamond ($c=0.5$)}
    \end{subfigure}\\
    \begin{centering}
    \begin{subfigure}[b]{0.4\textwidth}
    \includegraphics[width=60mm]{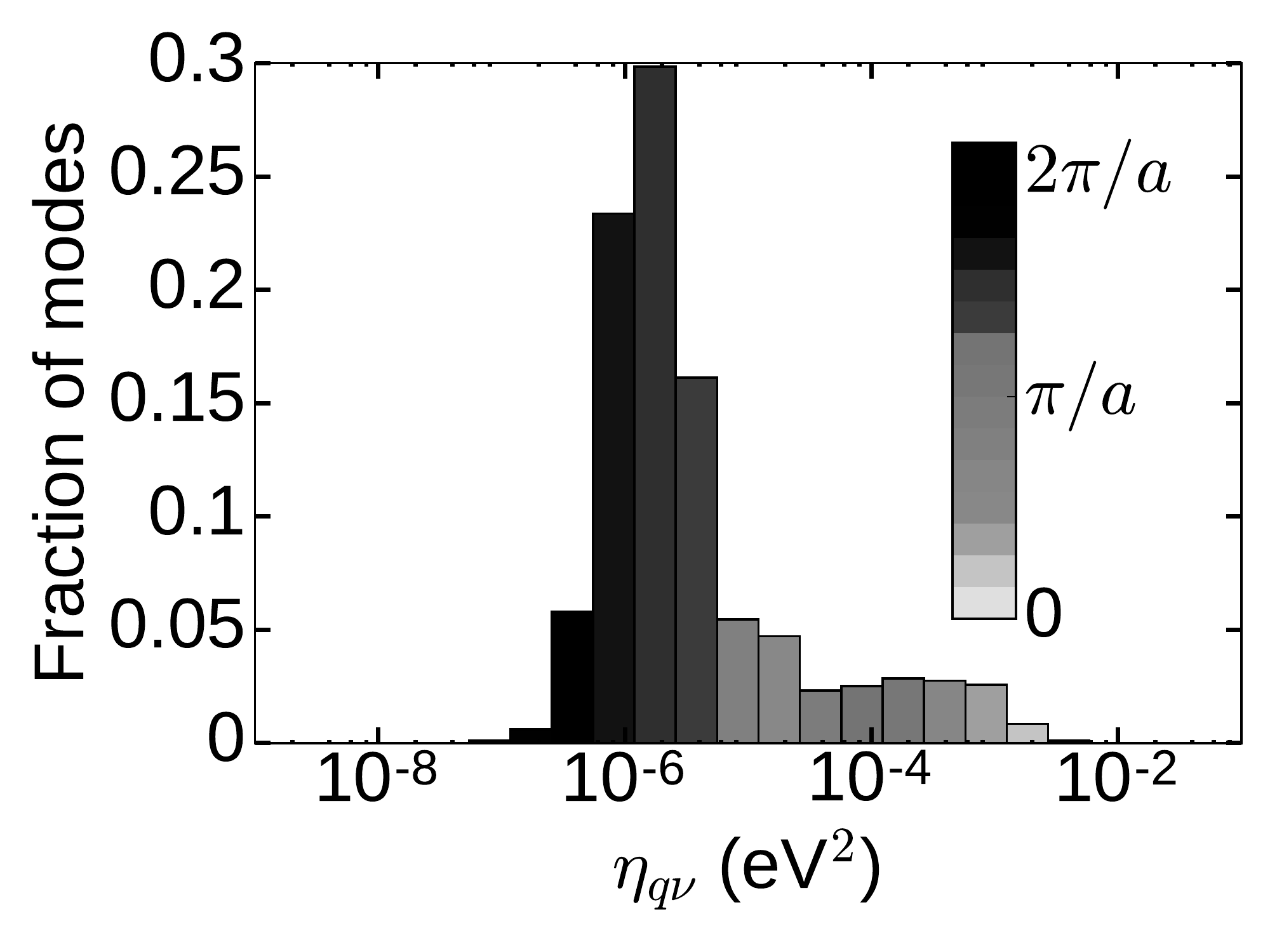}
    \caption{$\eta_{\textbf{q}\nu}$ - Diamond}
    \end{subfigure}
    \end{centering}
    \caption{Comparison between the electronic temperature decay obtained from a full-BTE solution and the 2T (a), 3T (b), and successive thermalization (c,d) models for Diamond. e) The distribution of phonon interaction strength $\eta_{\textbf{q}\nu}$ color-coded according to the average wavevector magnitude of phonons in each subset.}
\end{figure}

\newpage

\subsubsection{Silicon}
\begin{figure}[h]
    \centering
   \begin{subfigure}[b]{0.4\textwidth}
    \includegraphics[width=60mm]{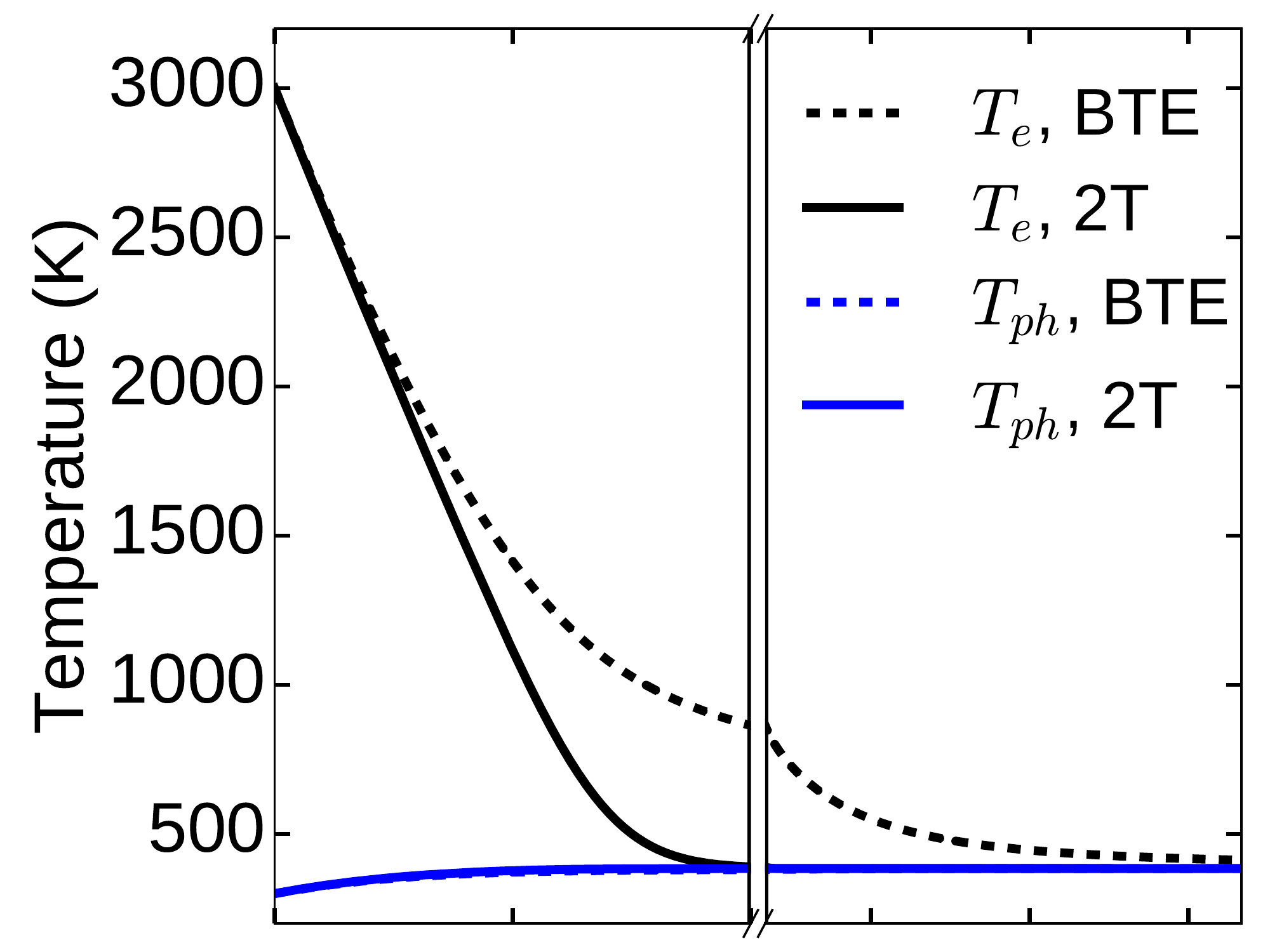}
    \caption{2T-Si}
    \end{subfigure}\qquad\qquad
    \begin{subfigure}[b]{0.4\textwidth}
    \includegraphics[width=60mm]{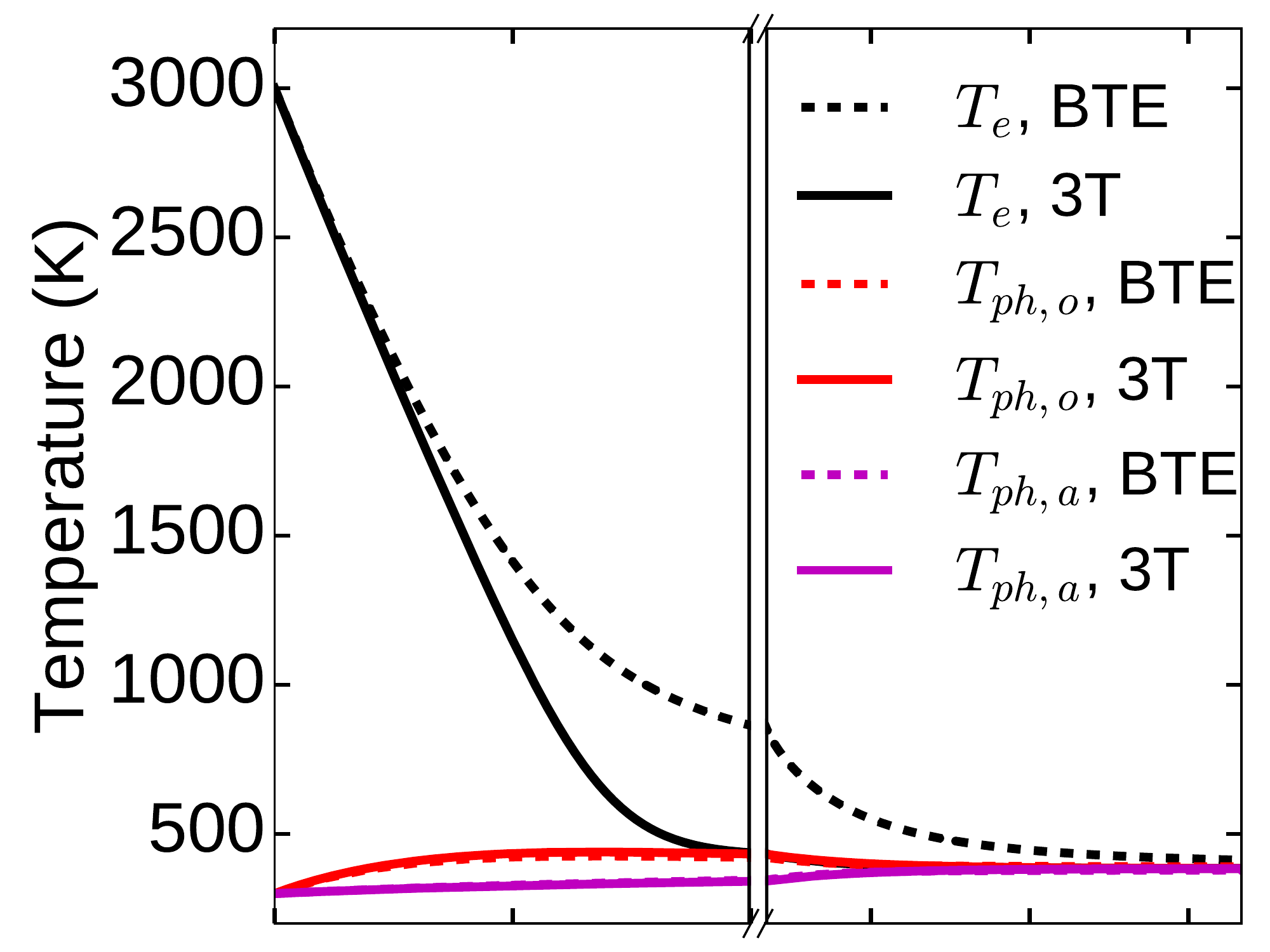}
    \caption{3T-Si}
    \end{subfigure}\\
    \begin{subfigure}[b]{0.4\textwidth}
    \includegraphics[width=60mm]{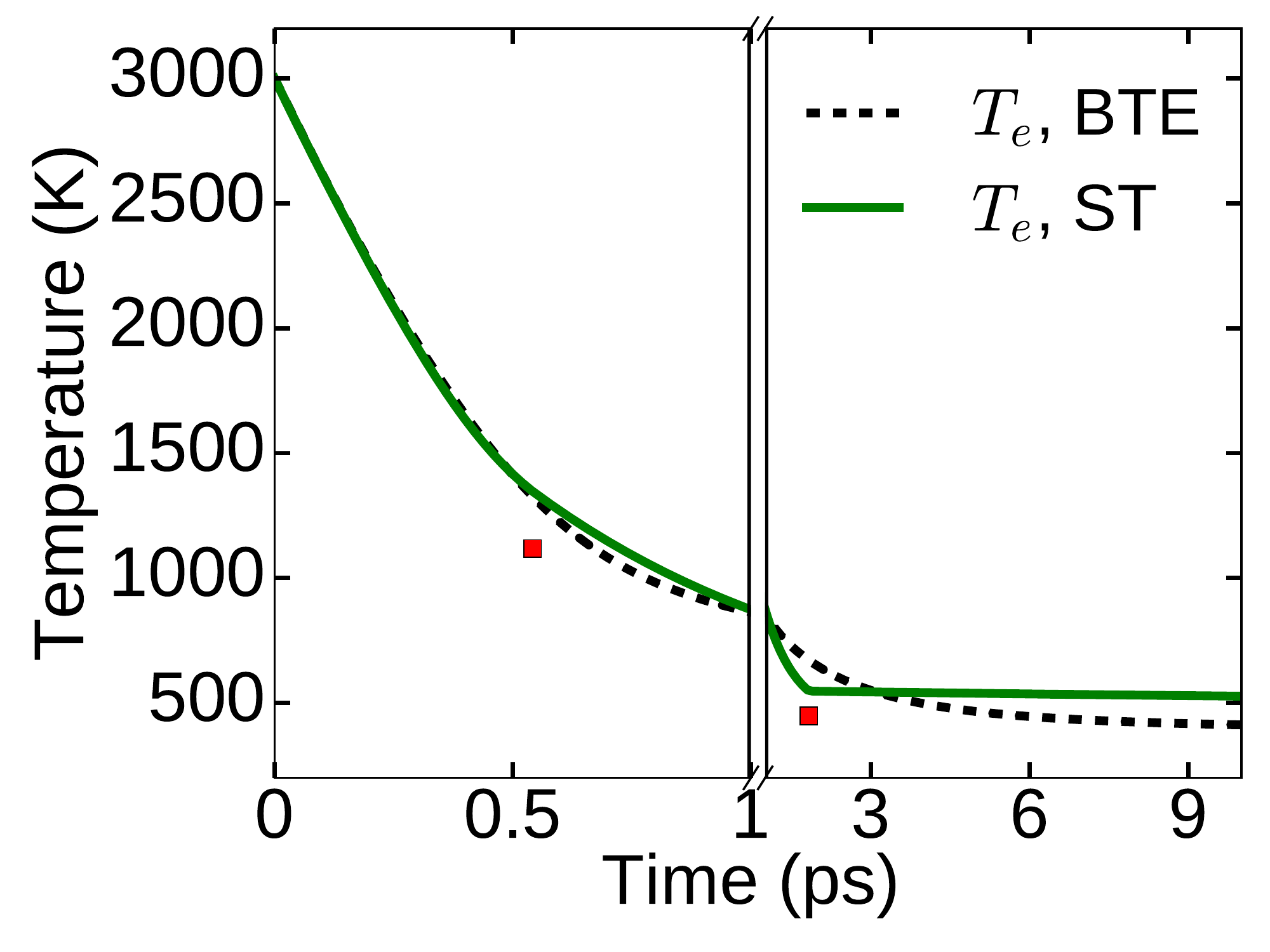}
    \caption{ST-Si ($c=0.1$)}
    \end{subfigure}\qquad\qquad
    \begin{subfigure}[b]{0.4\textwidth}
    \includegraphics[width=60mm]{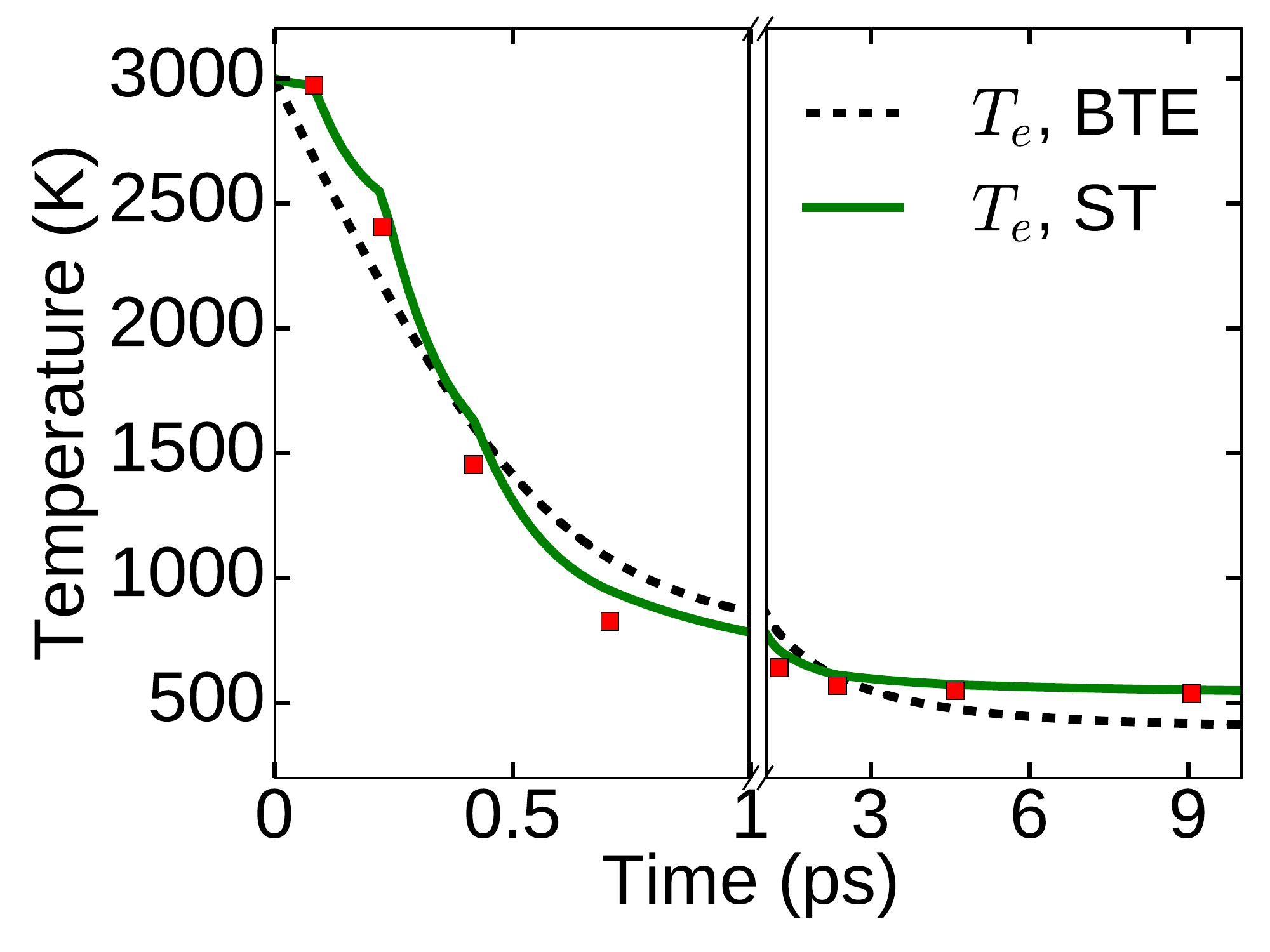}
    \caption{ST-Si ($c=0.5$)}
    \end{subfigure}\\
    \begin{centering}
    \begin{subfigure}[b]{0.4\textwidth}
    \includegraphics[width=60mm]{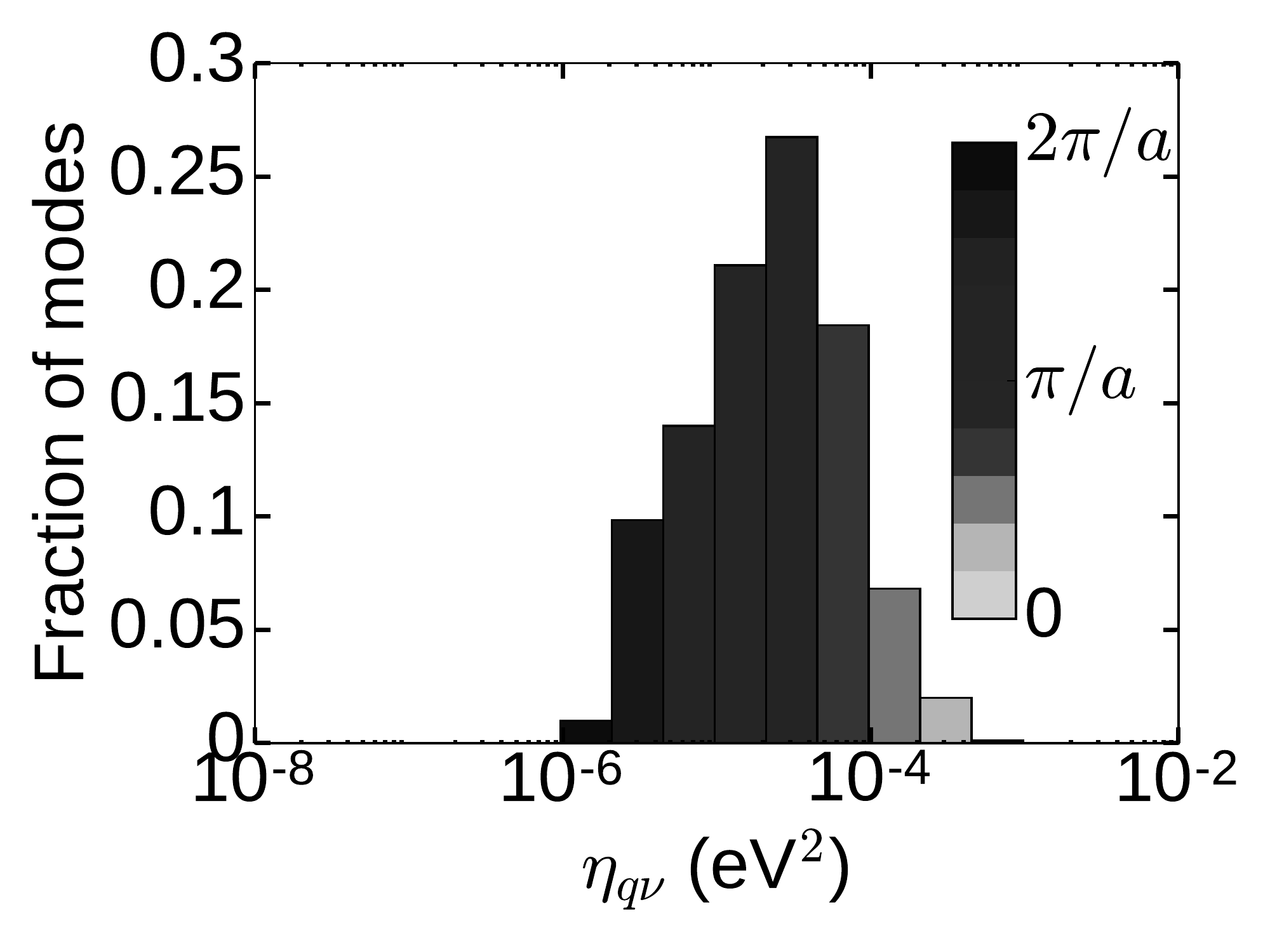}
    \caption{$\eta_{\textbf{q}\nu}$ - Si}
    \end{subfigure}
    \end{centering}
    \caption{Comparison between the electronic temperature decay obtained from a full-BTE solution and the 2T (a), 3T (b), and successive thermalization (c,d) models for Si. e) The distribution of phonon interaction strength $\eta_{\textbf{q}\nu}$ color-coded according to the average wavevector magnitude of phonons in each subset.}
   \label{Si_predictions}
\end{figure}

\subsection{Gold}
In this section, we present results from application of the simulation framework presented in this work to a simple metal gold. Our objective in this section is to compare predictions from a standard 2T model with BTE results for Au and contrast electron-phonon thermalization physics in a metal such as Au and most of the semiconductors considered in this work. 

As a sharp contrast to semiconductors such as BN and diamond with a heterogeneous distribution of $\eta_{\textbf{q}\nu}$ that span nearly four orders of magnitude, Au has a relatively homogeneous distribution of $\eta_{\textbf{q}\nu}$ as shown in Fig.~\ref{Au_results}a. Au does not have large heterogeneities in electron-phonon coupling due to lack of optical phonon branches, while the large number of bands crossing the Fermi surface (see Fig.~\ref{Au_results}c) implies that phonons with large momentum that connect two points on the Fermi surface are able to participate in electron-phonon scattering. This results in a much narrower distribution of phonon scattering strengths $\eta_{\textbf{q}\nu}$ and makes the physics of electron-phonon thermalization in Au fundamentally different from most of the semiconductors considered in this manuscript. Accordingly, the conventional two-temperature model predictions are in close agreement with the full-BTE results as shown in Fig.~\ref{Au_results}b. 

\begin{figure}[h]
    \centering
    \begin{subfigure}[b]{0.4\textwidth}
    \includegraphics[width=60mm]{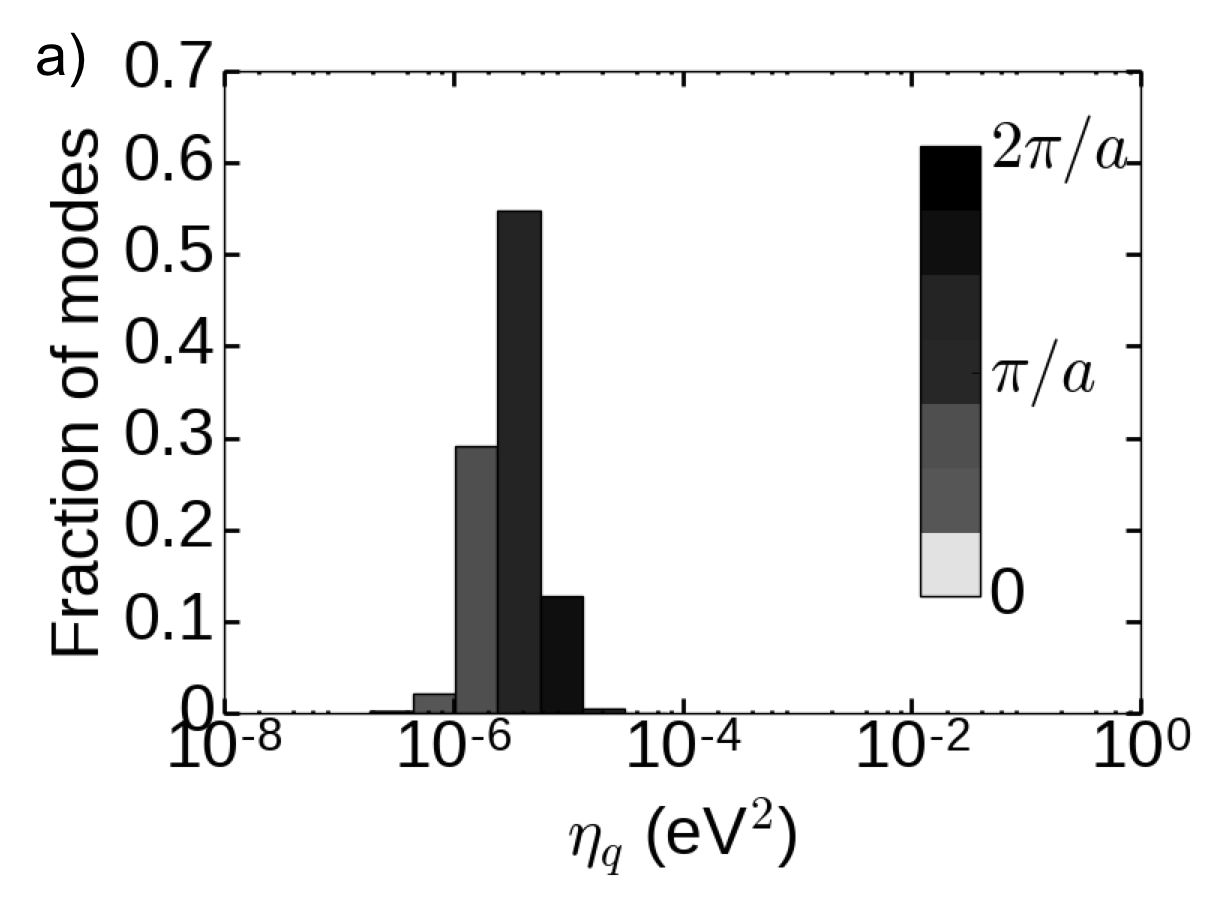}
    \end{subfigure}\qquad
    \begin{subfigure}[b]{0.4\textwidth}
    \includegraphics[width=60mm]{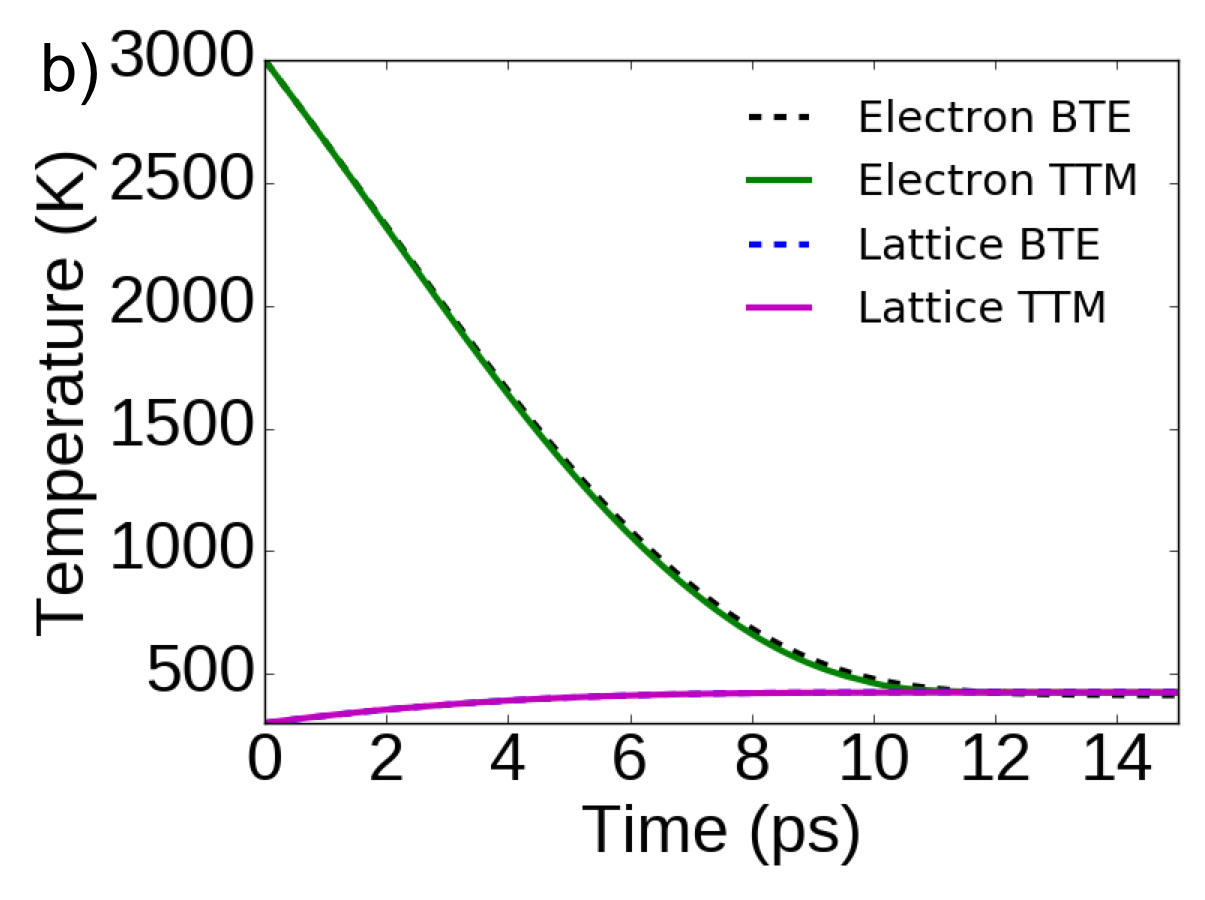}
    \end{subfigure}\qquad
     \begin{subfigure}[b]{0.4\textwidth}
    \includegraphics[width=60mm]{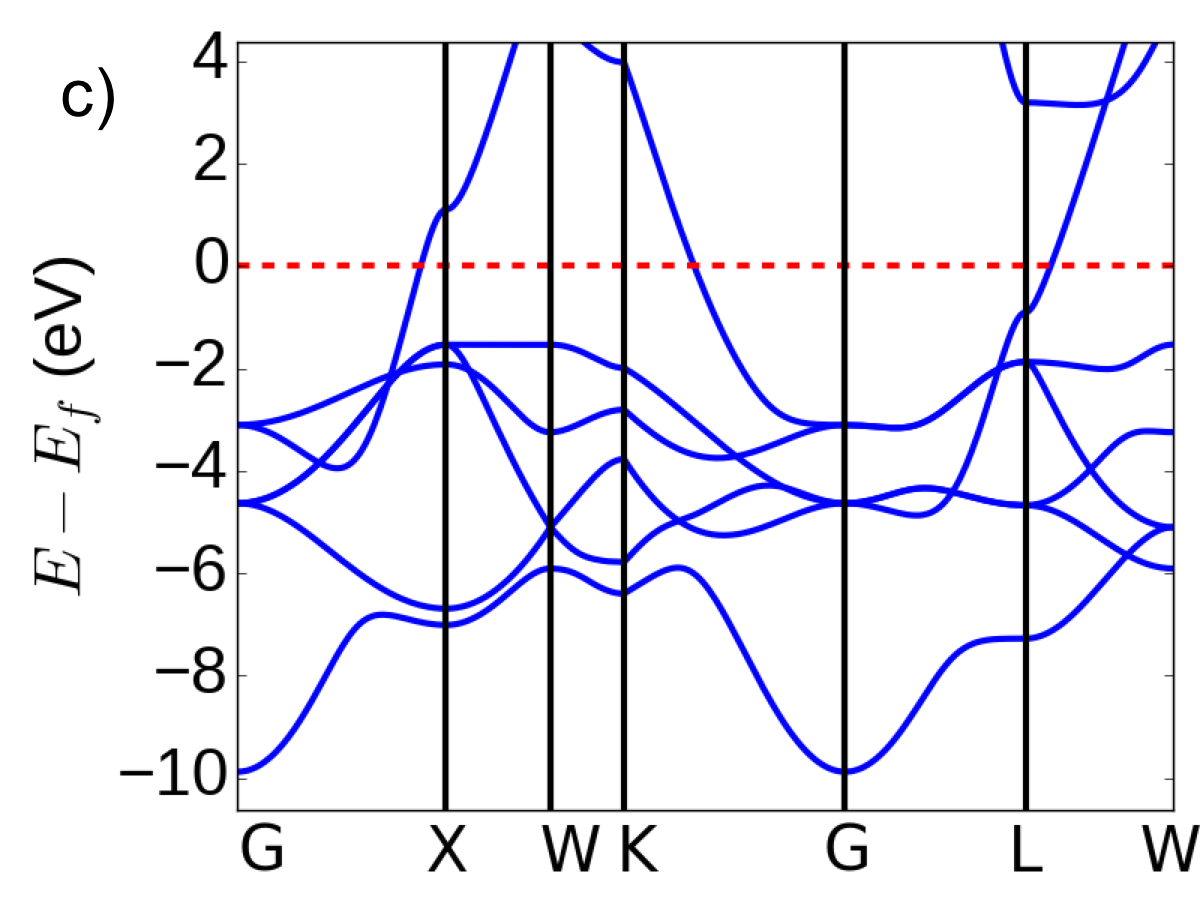}
    \end{subfigure}
    \caption{a) Distribution of $\eta_{\textbf{q}\nu}$ in Au. b) Electronic and lattice temperatures in Au from BTE and the standard 2T model. c) Electronic bandstructure of gold from DFT-LDA calculation.}
    \label{Au_results}
\end{figure}
\clearpage
\section{Generalized 2-Temperature Model}
\subsection{Heat Capacity and Electron-Phonon Coupling Accumulation Functions}

\subsubsection{Definitions}

We report here plots for the accumulation of lattice heat capacity $C_{\textrm{I}}(t)$ and electron-phonon coupling coefficient $G_{ep}(t)$ with respect to phonon thermalization timescale $t_{\textbf{q}\nu}$ for all the semiconductors considered in this manuscript. The heat capacity accumulation functions are evaluated at the lattice temperature of 300 K and the electron-phonon accumulation functions are evaluated at $T_{\textrm{el}}=3000$ K, $T_{\textrm{ph}}=300$ K. The timescale $t_{\textbf{q}\nu}$ for thermalization of each phonon mode is defined as the minimum of electron-phonon ($t_{\textbf{q}\nu,ep}$) and phonon-phonon ($t_{\textbf{q}\nu,pp}$) interaction timescales under the relaxation time approximation:
\begin{equation}
\frac{1}{t_{\textbf{q}\nu,ep}} = \frac{4\pi}{\hbar}   \sum\limits_{\textbf{k},m,n}(f_{m\textbf{k}+\textbf{q}}-f_{n\textbf{k}})|g_{\textbf{q}\nu}(m\textbf{k}+\textbf{q},n\textbf{k})|^2 \delta(E_{m\textbf{k}+\textbf{q}}-E_{n\textbf{k}}-\hbar\omega_{\textbf{q}\nu})
\end{equation}
\begin{equation}
\begin{split}
\frac{1}{t_{\textbf{q}\nu,pp}}  = \frac{2\pi}{\hbar^2n_{\textbf{q}\nu}(n_{\textbf{q}\nu}+1)}\sum\limits_{\textbf{q}'\nu'}\sum\limits_{\nu''}&\left\{|\Psi_{\textbf{q}\textbf{q}'\textbf{q}_1''}^{\nu\nu'\nu''}|^2n_{\textbf{q}\nu}n_{\textbf{q}'\nu'}(n_{\textbf{q}_1''\nu''}+1)\delta(\omega_{\textbf{q}\nu}+\omega_{\textbf{q}'\nu'}-\omega_{\textbf{q}_1''\nu''}) + \right. \\
&  \left.\frac{1}{2}|\Psi_{\textbf{q}\textbf{q}'\textbf{q}_2''}^{\nu\nu'\nu''}|^2(n_{\textbf{q}\nu}+1)n_{\textbf{q}'\nu'} n_{\textbf{q}_2''\nu''}\delta(\omega_{\textbf{q}\nu}-\omega_{\textbf{q}'\nu'}-\omega_{\textbf{q}_2''\nu''})\right\}
\end{split}
\end{equation}
The accumulation functions $C_{\textrm{I}}(t) = \sum\limits_{\textbf{q}\nu} C_{\textbf{q}\nu}\Theta(t-t_{\textbf{q}\nu})$ and $G_{ep}(t) = \sum\limits_{\textbf{q}\nu} G_{ep,\textbf{q}\nu}\Theta(t-t_{\textbf{q}\nu})$ are defined with respect to the thermalization timescale, and the mode resolved heat capacity $C_{\textbf{q}\nu}$ and electron-phonon coupling coefficient $G_{ep,\textbf{q}\nu}$ are given by:
\begin{equation}
C_{\textbf{q}\nu} = \frac{1}{V}\hbar\omega_{\textbf{q}\nu}\frac{\partial f_{BE}^o}{\partial T}
\end{equation}
\begin{equation}
\begin{split}
G_{ep,\textbf{q}\nu} = \frac{4\pi}{\hbar V(T_{\textrm{el}}-T_{\textrm{ph}})}  & \sum\limits_{\textbf{k},m,n}\hbar\omega_{\textbf{q}\nu}[f_{m\textbf{k}+\textbf{q}}(1-f_{n\textbf{k}})(n_{\textbf{q}\nu}+1)-\\&(1-f_{m\textbf{k}+\textbf{q}})f_{n\textbf{k}}n_{\textbf{q}\nu}]|g_{\textbf{q}\nu}(m\textbf{k}+\textbf{q},n\textbf{k})|^2 \delta(E_{m\textbf{k}+\textbf{q}}-E_{n\textbf{k}}-\hbar\omega_{\textbf{q}\nu})
\end{split}
\end{equation}
The total heat capacity $C(T_{\textrm{ph}}) = \sum\limits_{\textbf{q}\nu}C_{\textbf{q}\nu}$ and total electron-phonon coupling coefficient $G_{ep}(T_{\textrm{el}},T_{\textrm{ph}}) =  \sum\limits_{\textbf{q}\nu}G_{ep,\textbf{q}\nu}$ are the sum of mode-resolved heat capacity and electron-phonon coupling coefficient respectively. In the above equations for thermalization time-scale of a phonon mode, the equilibrium electron and phonon occupation functions are evaluated at the initial temperatures of electrons and phonons respectively. Hence, the definitions of these time-scales are heuristic and expected to only provide an approximate estimate of the actual equilibration timescale. 

\subsubsection{Computed accumulation functions}

\begin{figure}[!h]
    \centering
   \begin{subfigure}[b]{0.4\textwidth}
    \includegraphics[width=60mm]{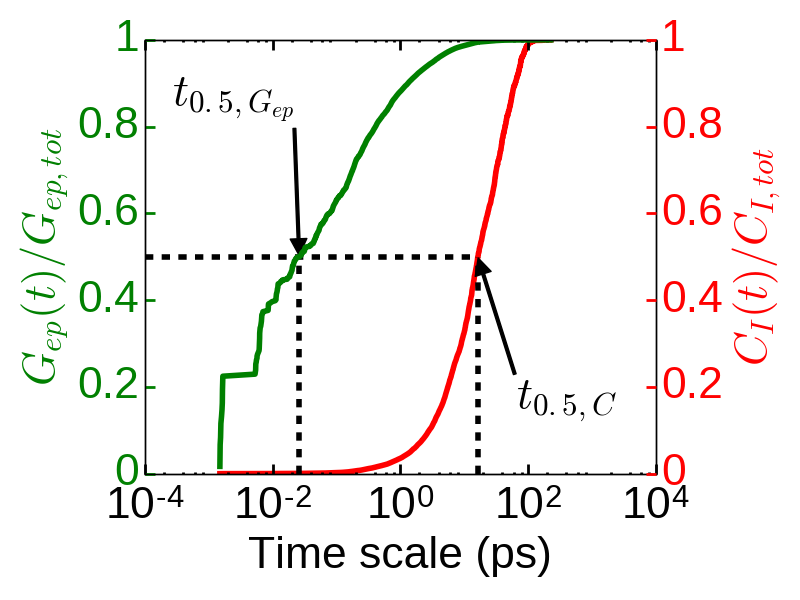}
    \caption{BN}
    \end{subfigure}\qquad\qquad
    \begin{subfigure}[b]{0.4\textwidth}
    \includegraphics[width=60mm]{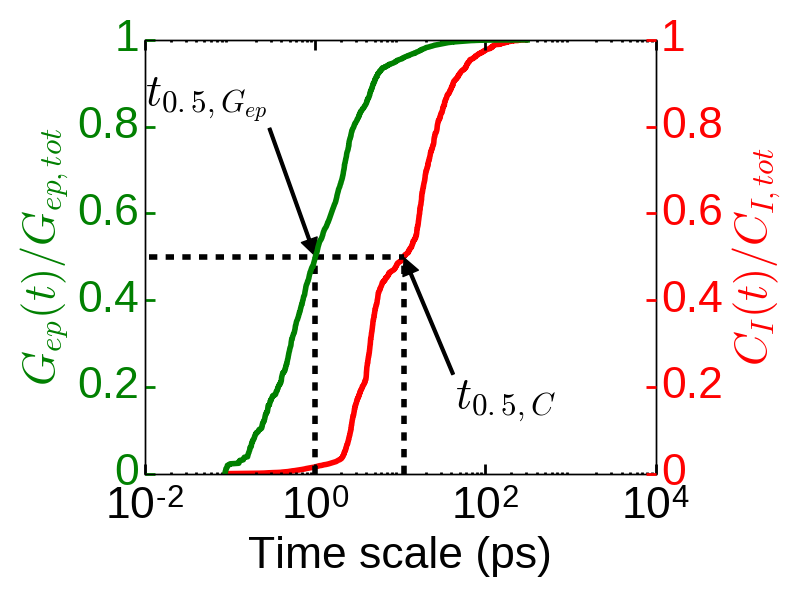}
    \caption{BP}
    \end{subfigure}\\
    \begin{subfigure}[b]{0.4\textwidth}
    \includegraphics[width=60mm]{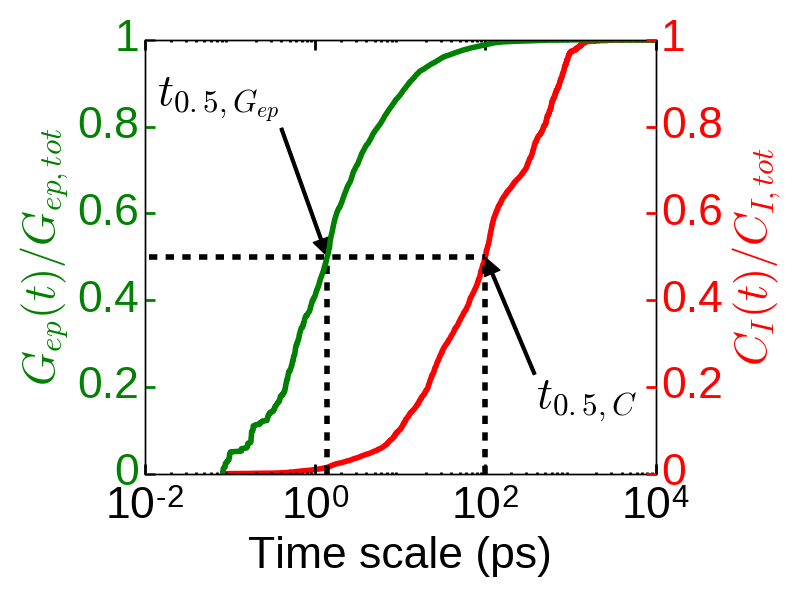}
    \caption{BAs}
     \end{subfigure}\quad\quad
    \begin{subfigure}[b]{0.4\textwidth}
    \includegraphics[width=60mm]{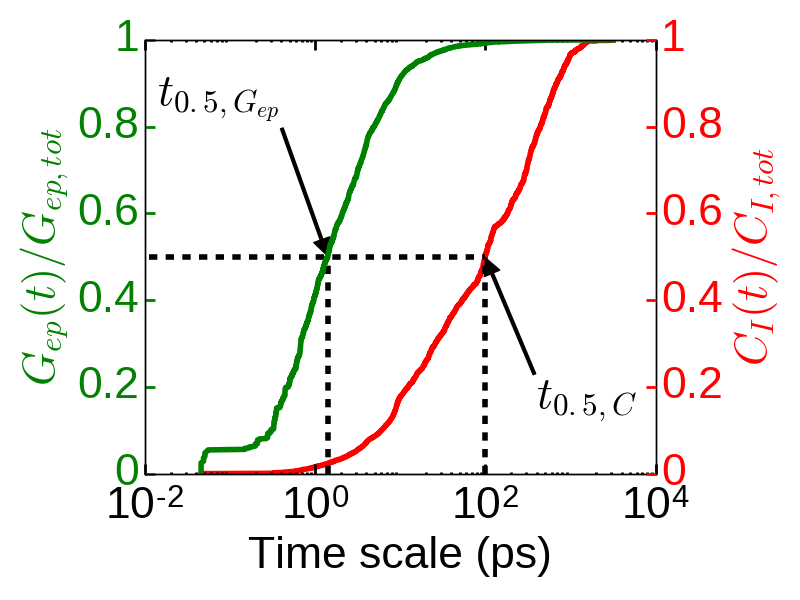}
    \caption{BSb}
    \end{subfigure}\\
        \caption{Accumulation of lattice heat capacity $C(t)$ and electron-phonon coupling coefficient $G_{ep}(t)$ for B series compounds.}
\end{figure}

\begin{figure}[!h]
    \centering
   \begin{subfigure}[b]{0.4\textwidth}
    \includegraphics[width=60mm]{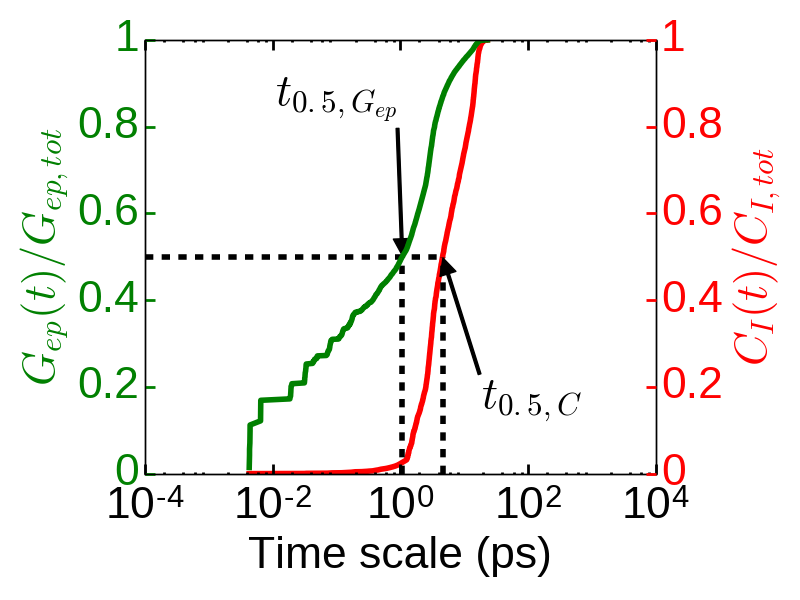}
    \caption{AlP}
    \end{subfigure}\qquad
    \begin{subfigure}[b]{0.4\textwidth}
    \includegraphics[width=60mm]{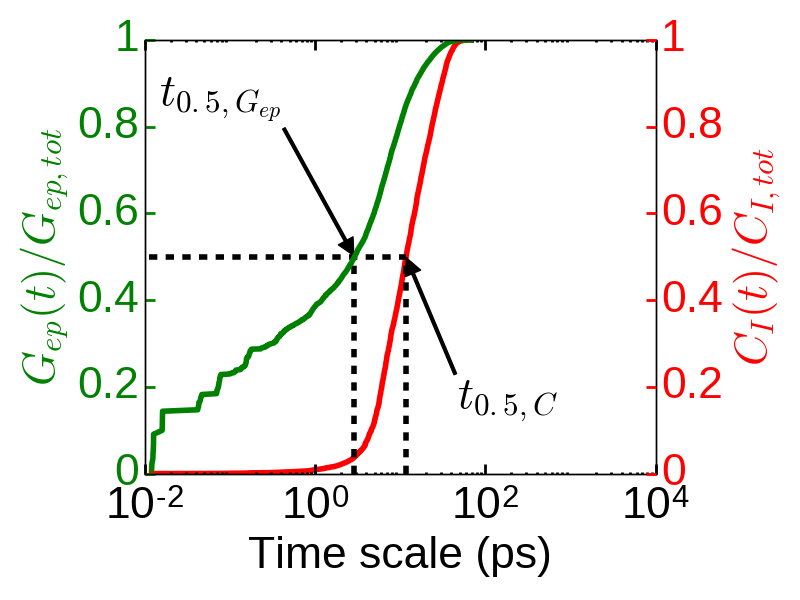}
    \caption{AlAs}
    \end{subfigure}\quad
    \begin{subfigure}[b]{0.4\textwidth}
    \includegraphics[width=60mm]{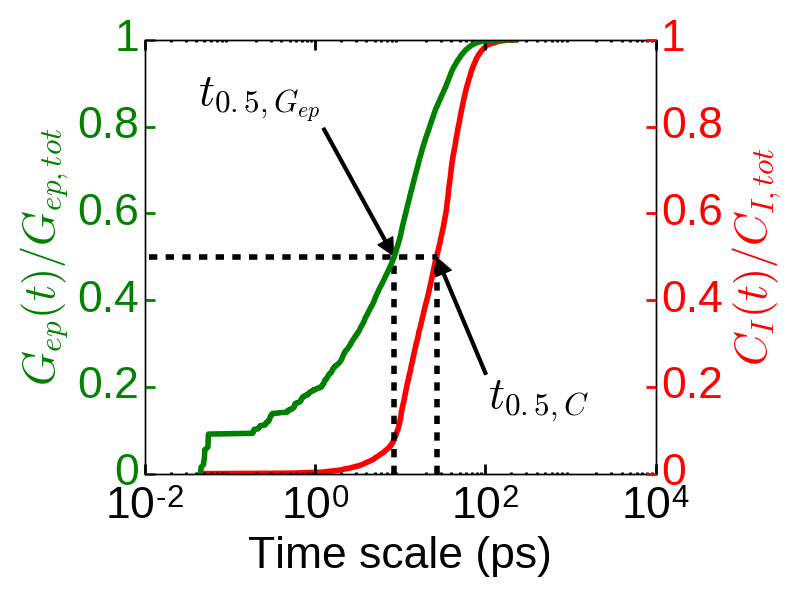}
    \caption{AlSb}
     \end{subfigure}
        \caption{Accumulation of lattice heat capacity $C(t)$ and electron-phonon coupling coefficient $G_{ep}(t)$ for Al series compounds.}
\end{figure}

\begin{figure}[h]
    \centering
   \begin{subfigure}[b]{0.4\textwidth}
    \includegraphics[width=60mm]{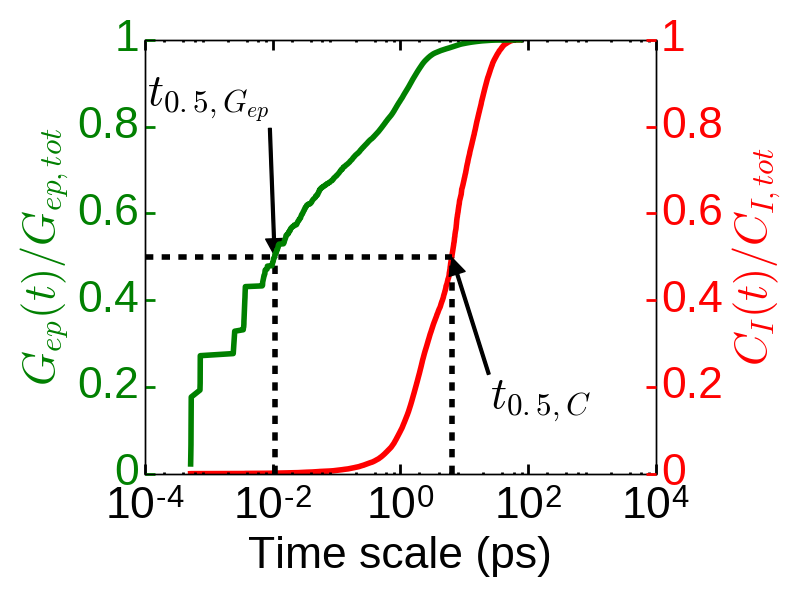}
    \caption{GaN}
    \end{subfigure}\qquad
    \begin{subfigure}[b]{0.4\textwidth}
    \includegraphics[width=60mm]{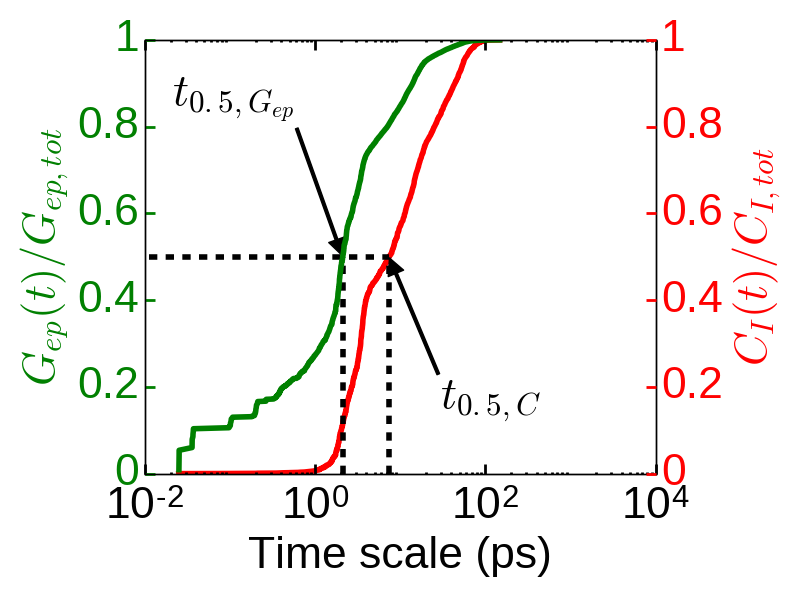}
    \caption{GaP}
    \end{subfigure}\quad
    \begin{subfigure}[b]{0.4\textwidth}
    \includegraphics[width=60mm]{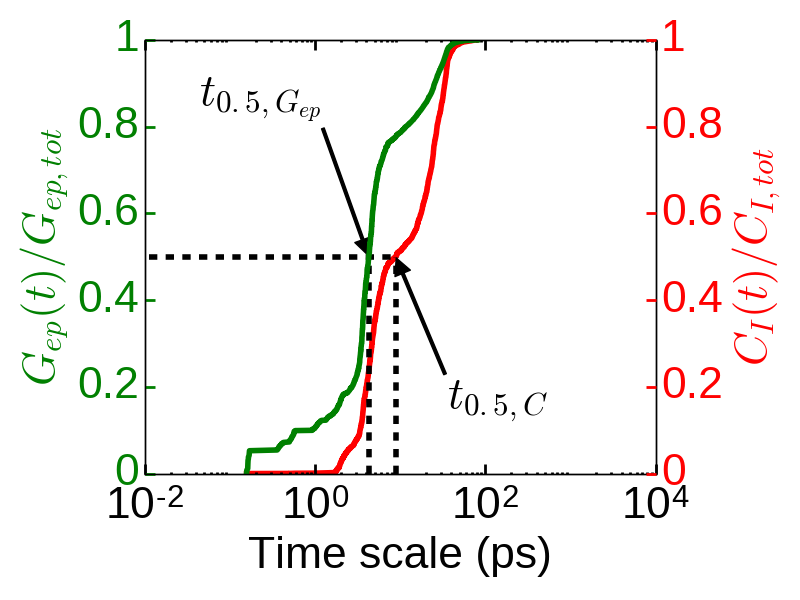}
    \caption{GaAs}
     \end{subfigure}
        \caption{Accumulation of lattice heat capacity $C(t)$ and electron-phonon coupling coefficient $G_{ep}(t)$ for Ga series compounds.}
\end{figure}

\begin{figure}[h]
    \centering
   \begin{subfigure}[b]{0.4\textwidth}
    \includegraphics[width=60mm]{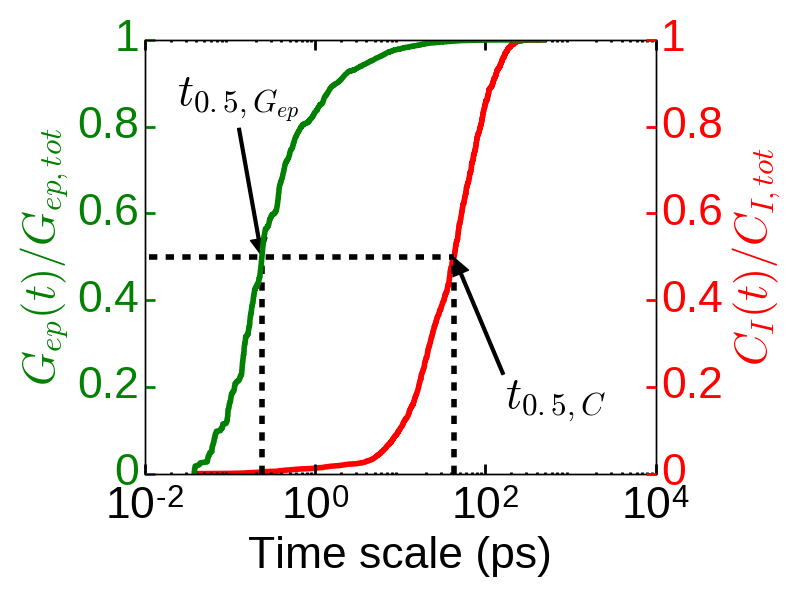}
    \caption{Diamond}
    \end{subfigure}\qquad
    \begin{subfigure}[b]{0.4\textwidth}
    \includegraphics[width=60mm]{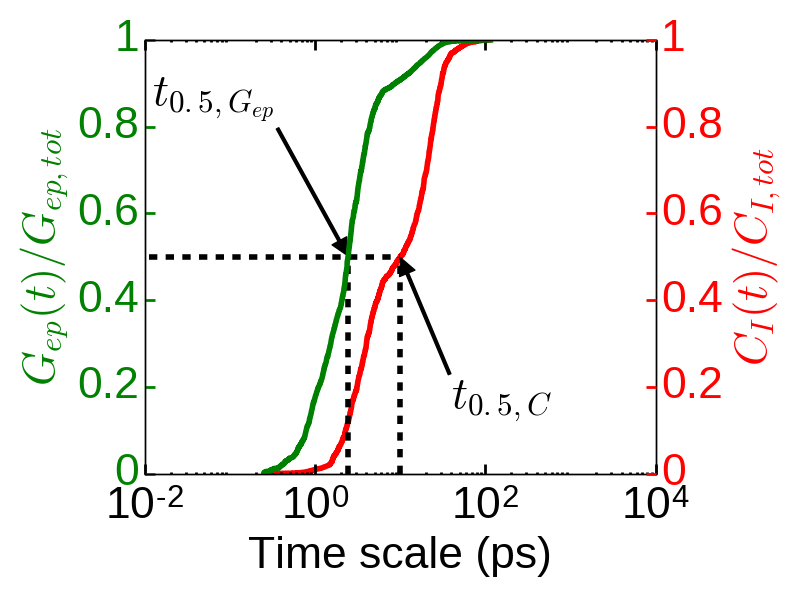}
    \caption{Si}
    \end{subfigure}
    \caption{Accumulation of lattice heat capacity $C(t)$ and electron-phonon coupling coefficient $G_{ep}(t)$ for diamond and Si.}
\end{figure}
\clearpage

\subsection{Comparison with BTE Results}
In this section, we compare predictions of the timescales of electronic cooling from the generalized 2T model with the full BTE results. Specifically, we compare the ratio of electronic temperature decay time constants $\alpha$ ($t\to 0$), $\beta$ ($t=10$ ps) for all the 12 semiconductors considered in this work.  At time $t=0$, the system I consists only of electrons and the decay rate $\alpha = \frac{d\log{T_\textrm{I}}}{dt}|_{t\to 0} = G_{ep,tot}/C_{\textrm{el}}$. At time $t=10$ ps, the decay rate $\beta$ in the generalized 2T model can be written as:
\begin{equation}
\beta = \frac{d\log{T_\textrm{I}}}{dt}\bigg\rvert_{t= 10\text{ ps}} = \frac{G_{ep,tot}-G_{ep,\textrm{I}}+G_{pp,\textrm{I}-\{\textrm{II},\textrm{III}\}}}{C_{\textrm{el}}+C_{\textrm{ph},\textrm{I}}}
\end{equation}
where $G_{ep,\textrm{I}}$ denotes the electron-phonon coupling coefficient for phonon modes thermalized with electrons in I. $G_{pp,\textrm{I}-\{\textrm{II},\textrm{III}\}}$ denotes the phonon-phonon coupling coefficient between phonon modes in I and the remaining phonon modes. The denominator contains the total heat capacity of subsystem I that contains electrons and a subset of phonons that are thermalized with electrons. At long times, the heat capacity of phonon modes that are thermalized with electrons ($C_{\textrm{ph},\textrm{I}}$) far exceeds the heat capacity of electrons ($C_{\textrm{el}}$) and the decay rate $\beta$ is significantly smaller than the initial decay rate $\alpha$. 

As seen in Fig.~\ref{alpha_by_beta_gttm}, the generalized 2T model captures the correct order-of-magnitude of the decay time constants in sharp contrast to the standard 2T model that would predict a single exponential decay ($\alpha/\beta \sim 1$) for all compounds. Such agreement is particularly noteworthy as, in the present work, the thermalization times of the generalized 2T model are not obtained from the BTE simulation, but, instead, obtained heuristically using the relaxation time approximation and the initial electronic, lattice temperatures. Further refinement of the definitions of these quantities is expected to produce better agreement between the BTE results and the generalized 2T model. 

\begin{figure}
    \includegraphics[width=60mm]{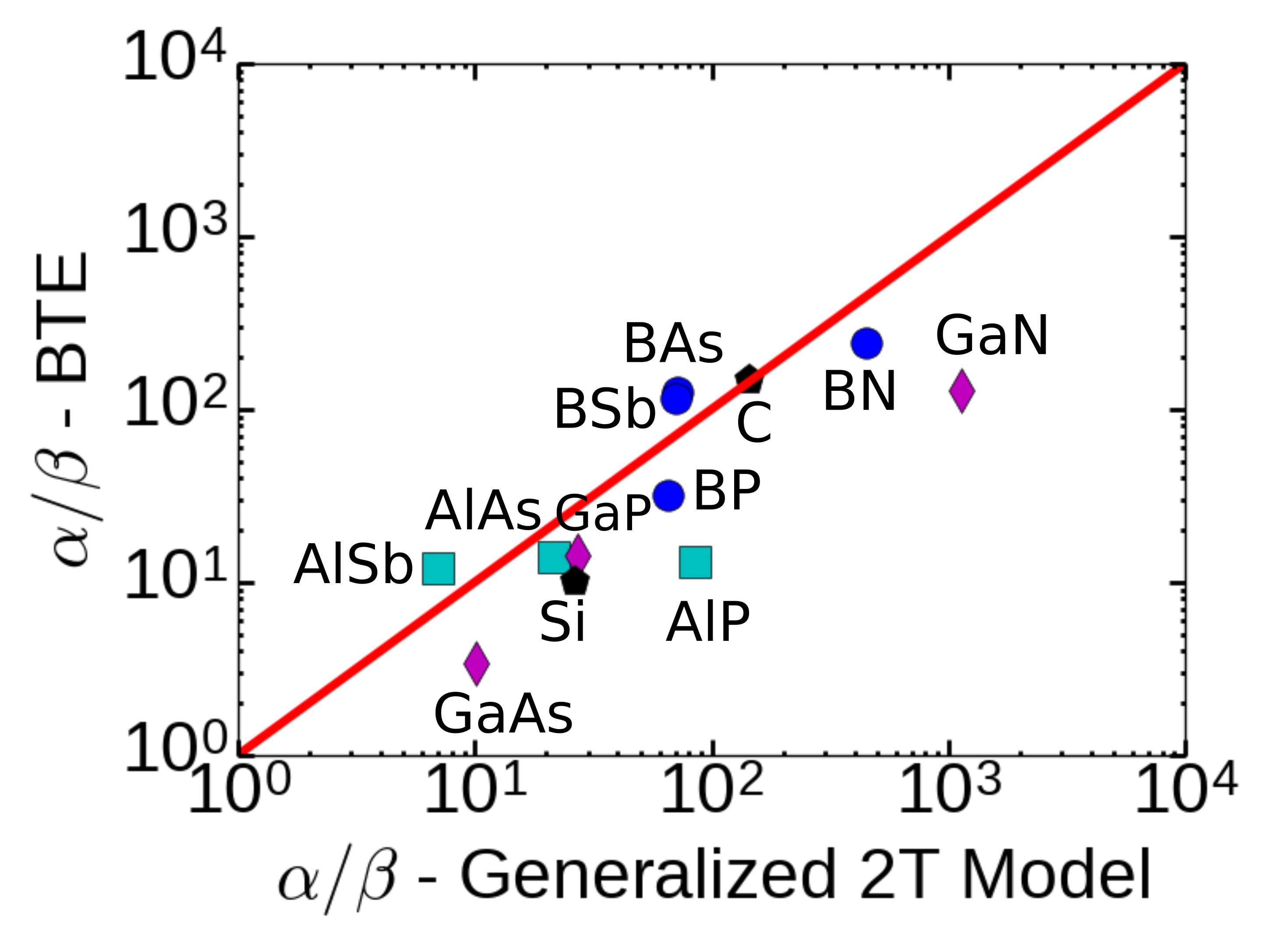}
    \caption{Comparison of the ratio of decay time constants $\alpha/\beta$ between the BTE and generalized 2T model for all 12 semiconductors considered in this work.}
        \label{alpha_by_beta_gttm}
\end{figure}
\section{Sensitivity of Results to Choice of Initial Conditions}
In this section, we confirm that the general conclusions reported in the manuscript of phonon non-equilibrium and significant deviations from the 2T model are not specific to the choice of initial electronic temperature (3000 K in the main manuscript) and Fermi level (0.3 eV below VBM in the main manuscript). Fig.~\ref{1500K} shows the electronic temperature decay for an initial electron temperature of 1500 K and Fig.~\ref{cbm} shows the electronic temperature decay with the Fermi level at 0.3 eV above the conduction band minimum for BN and BAs. For both the initial conditions, we observe a trend similar to the results in the main text where the 2T model under-predicts the equilibration time by an order-of-magnitude and a successive thermalization approach results in good quantitative agreement with the full BTE simulation.

\subsection{Electronic Temperature}

\begin{figure}[h]
    \centering
   \begin{subfigure}[b]{0.4\textwidth}
    \includegraphics[width=60mm]{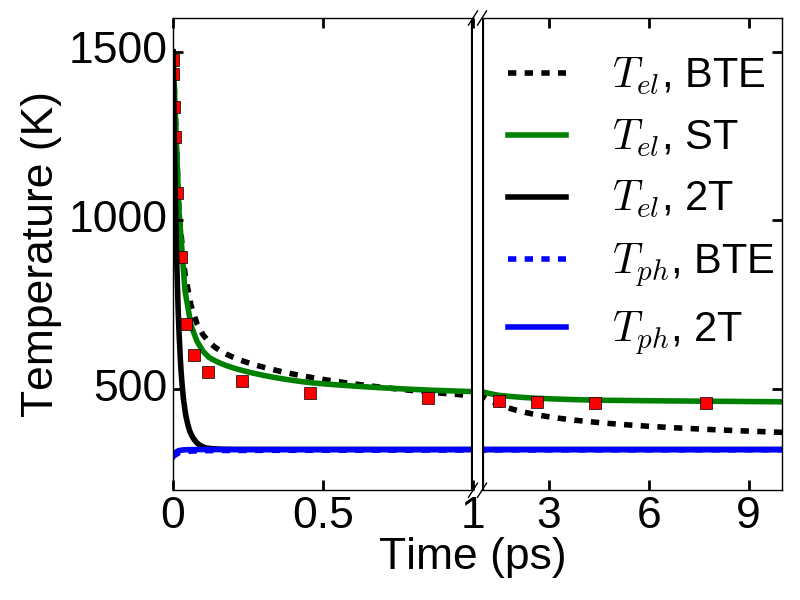}
    \caption{BN}
    \end{subfigure}\qquad
    \begin{subfigure}[b]{0.4\textwidth}
    \includegraphics[width=60mm]{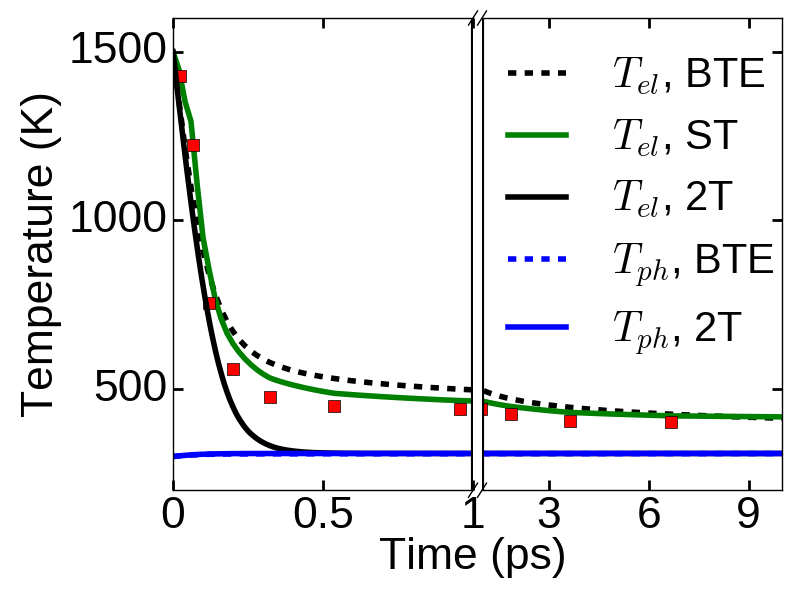}
    \caption{BAs}
    \end{subfigure}
    \caption{Electronic and lattice temperatures in BN (a) and BAs (b) obtained from the BTE, 2T model and a constrained successive thermalization (ST) simulation using the 2T model on a subset of phonons. The red squares indicate the times (and corresponding equilibration temperatures) at which subspace thermalization is achieved and a new set of modes is introduced in the ST simulation. This figure is similar to Figs.~2a,b of main text but with an initial electronic temperature of 1500 K.}
        \label{1500K}
\end{figure}

\subsection{Electron vs hole thermalization}

\begin{figure}
    \centering
   \begin{subfigure}[b]{0.4\textwidth}
    \includegraphics[width=60mm]{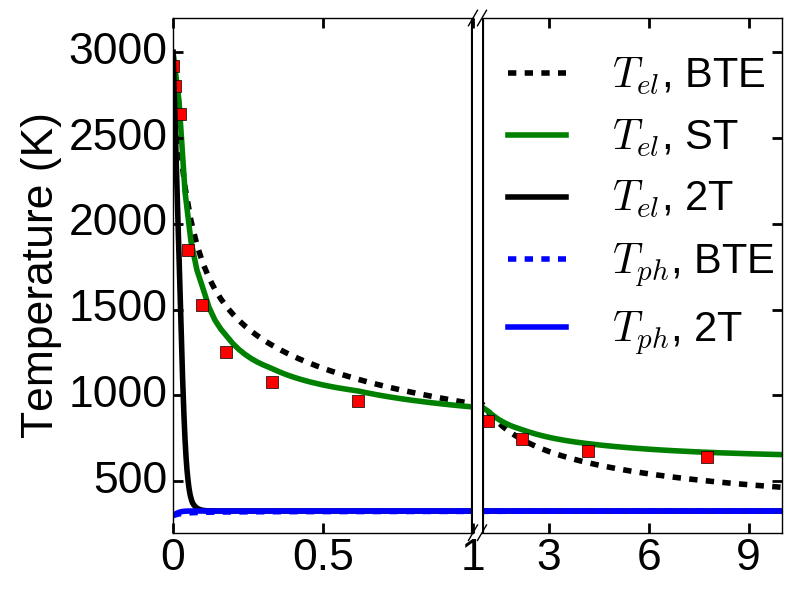}
    \caption{BN}
    \end{subfigure}\qquad
    \begin{subfigure}[b]{0.4\textwidth}
    \includegraphics[width=60mm]{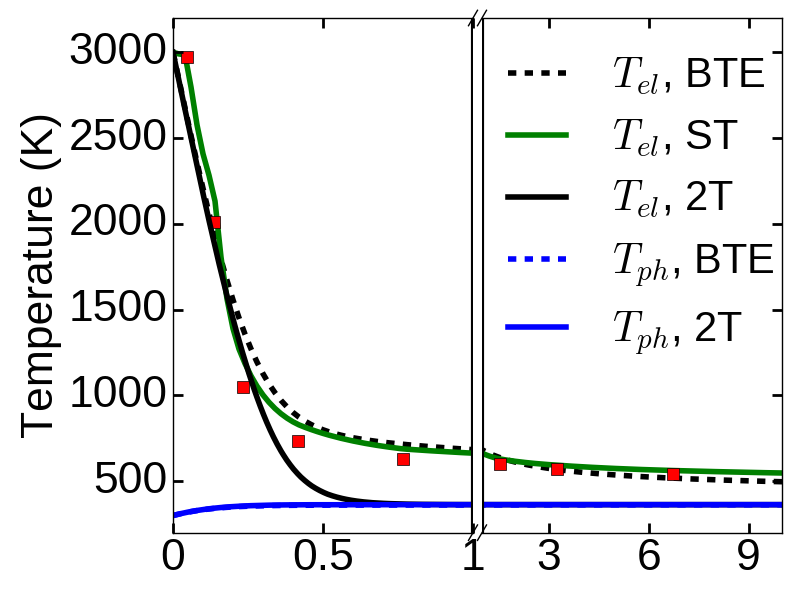}
    \caption{BAs}
    \end{subfigure}
    \caption{Electronic and lattice temperatures in BN (a) and BAs (b) obtained from the 2T model, BTE and a constrained successive thermalization (ST) simulation using the 2T model on a subset of phonons. The red squares indicate the times (and corresponding equilibration temperatures) at which subspace thermalization is achieved and a new set of modes is introduced in the ST simulation. This figure is similar to Figs.~2a,b of main text but the Fermi level is at 0.3 eV above the conduction band minimum (results in the main text involved relaxation of hot holes at 0.3 eV below VBM).}
        \label{cbm}
\end{figure}
\clearpage
\section{Lattice Constants}
Table~\ref{lattice_constants} reports the equilibrium lattice constants of all the semiconductors (in the zinc blende structure) obtained using LDA, norm-conserving pseudopotentials.
\begin{table}[h]
\centering
\caption{Lattice constants of the cubic compounds considered in the present work.}
\label{lattice_constants}
\begin{tabular}{|l|l|}\hline
Compound & Lattice constant ($\angstrom$) \\ \hline
BN       & 3.56                         \\ \hline
BP       & 4.46                           \\ \hline
BAs      & 4.72                           \\ \hline
BSb      & 5.19                           \\ \hline
AlP      & 5.40                           \\ \hline
AlAs     & 5.60                           \\ \hline
AlSb     & 6.09                           \\ \hline
GaN      & 4.42                           \\ \hline
GaP      & 5.33                           \\ \hline
GaAs     & 5.54                           \\ \hline
Diamond  & 3.52                           \\ \hline
Si       & 5.47                          \\ \hline
\end{tabular}
\end{table}
\bibliography{references}